\documentclass[preprint,3p]{elsarticle}

\usepackage{amssymb}

\usepackage{amsmath}
\usepackage{subeqnarray}
\usepackage{color}
\usepackage[]{algorithm2e}
\usepackage{psfrag}

\definecolor{orange}{RGB}{255,155,0}
\definecolor{darkgreen}{RGB}{0,130,0}

\newcommand\red[1]{\textcolor{red}{#1}}
\newcommand\ora[1]{\textcolor{orange}{#1}}
\newcommand\grn[1]{\textcolor{darkgreen}{#1}}

\journal{Computers \& Fluids}

\usepackage{psfrag}
\usepackage[breaklinks=true]{hyperref}

\begin{document}

\begin{frontmatter}



\title{Nonlinear model reduction: a comparison between POD-Galerkin
  and POD-DEIM methods}


\author{Denis Sipp}
\address{ONERA, 8 rue des Vertugadins, 92190 Meudon, France,
    Email: denis.sipp@onera.fr}

\author{Miguel Fosas de Pando}
\address{Dpto. Ingenier\'{\i}a Mec\'{a}nica y Dise\~{n}o Industrial,
  Escuela Superior de Ingenier\'{\i}a, Universidad de C\'{a}diz, 11519
  Puerto Real, Spain, Email: miguel.fosas@uca.es}

\author{Peter J. Schmid}
\address{Dept. of Mathematics, Imperial College London, London SW7 2AZ, United
  Kingdom, Email: pjschmid@imperial.ac.uk}

\begin{abstract}
  Several nonlinear model reduction techniques are compared for the
  three cases of the non-parallel version of the
  Kuramoto-Sivashinsky equation, the transient regime of flow
  past a cylinder at $Re=100$ and fully developed flow past a cylinder
  at the same Reynolds number. The linear terms of the governing
  equations are reduced by Galerkin projection onto a POD basis of
  the flow state, while the reduced nonlinear convection terms are
  obtained either by a Galerkin projection onto the same state basis,
  by a Galerkin projection onto a POD basis representing the
  nonlinearities or by applying the Discrete Empirical Interpolation
  Method (DEIM) to a POD basis of the nonlinearities. The quality of
  the reduced order models is assessed as to their stability, accuracy
  and robustness, and appropriate quantitative measures are introduced
  and compared. In particular, the properties of the reduced linear
  terms are compared to those of the full-scale terms, and the structure
  of the nonlinear quadratic terms is analyzed as to the conservation of
  kinetic energy. It is shown that all three reduction techniques
  provide excellent and similar results for the cases of the
  Kuramoto-Sivashinsky equation and the limit-cycle cylinder flow.
  For the case of the transient regime of flow past a cylinder, only
  the pure Galerkin techniques are successful, while the DEIM technique
  produces reduced-order models that diverge in finite time.
\end{abstract}

%


\end{frontmatter}


\section{Introduction}
\label{sec:intro}

Reduced-order models play an important role in many fluid applications.
Rapid evaluation of fluid systems in multi-query situations (such as
Monte-Carlo techniques or parameter sweeps) or the low-dimensional
representation of input-output behavior (such as in control design) are
but two of many applications where a more compact, yet accurate
description of fluid systems is a critical component in the overall
analysis or design process. Model reduction tries to capture the essential
features of the full-size system with far fewer degrees of freedom at a
fraction of the computational cost. While
the reduction of linear systems can be firmly based on a mathematical
framework involving linear-algebra techniques, the reduction of nonlinear
models is far less developed and understood.

Most commonly, model reduction for high-dimensional fluid systems relies
on a projection of the governing equations onto a proper, low-dimensional
basis. This basis is often extracted from a series of snapshots generated
by the governing equations, and thus contains coherent structures that are
deemed pertinent to the transport process one wishes to capture or model.
The proper orthogonal decomposition (POD)~\cite{berkooz1993proper}, producing
a hierarchy of flow fields that optimally (in the $L_2$-sense) express
the variance of fluctuations about a mean state, is quite popular among
numericists and experimentalists. Once the basis has been determined, the
governing equations are then recast, via a Galerkin projection, into a
nonlinear dynamical system for the coefficients of the expansion basis:
only the most energetic POD-modes are retained in the projection; modes
with negligible energy content are dismissed by truncating the Galerkin
expansion.

First efforts in this direction~\cite{berkooz1993proper,aubry1988dynamics,gillies1998low} revealed
that the use of a small number of POD modes required a modification of
the projection procedure, which accounts for the collective effect of the
neglected POD modes, in order to ensure stability of the resulting
dynamical system. In particular, the introduction of a
shift-mode~\cite{noack2003hierarchy} notably improved the stability and robustness
of the model-reduction method. Prompted by this observation, various
calibrations of the nonlinear Galerkin models were explored and
investigated~\cite{couplet2005calibrated,bergmann2009enablers,cordier2013identification}, while
the use of these models in flow control applications became increasingly
common~\cite{king2005nonlinear,bergmann2005optimal,bergmann2008optimal}.

The computational savings that are obtained in POD-Galerkin models greatly
depend on the underlying structure of the governing equations of the full-order
model. For instance, in the case of linear systems, the projection of the 
dynamics onto the subspace spanned by the selected POD modes leads to a reduced
system that can be evaluated at a negligible computational cost. In the case of general nonlinear dynamics, however, the evaluation cost of the projected nonlinear model is still comparable to that of the full-order model.

To circumvent this limitation, an alternative technique known as
POD-DEIM~\cite{chaturantabut2010nonlinear} advocates a different treatment
of linear and nonlinear terms. While linear terms are treated in exactly
the same fashion as in POD-Galerkin models, an additional POD basis is
introduced to represent the nonlinear terms in the reduced-order model. 
Then, nonlinear terms are incorporated into the reduced system according
to their values at selected \emph{interpolation} points. The location
of these \emph{interpolation} points is determined according to a greedy
algorithm, known as the Discrete 
Empirical Interpolation Method (DEIM), that minimizes the nonlinear residual. The effectiveness of this technique
relies then on the ability to cheaply evaluate nonlinear terms at the
interpolation points. The reader interested in error estimates is referred to~\cite{chaturantabut_state_2012,wirtz_posteriori_2014} for details.
Over the last decade, numerous extensions of the DEIM have been proposed
to either tailor this method to specific applications (see, for instance, the
\emph{matrix}~\cite{wirtz_posteriori_2014, 
negri_efficient_2015, bonomi_matrix_2017} or the
\emph{unassembled}~\cite{tiso_discrete_2013} variants) or to improve the 
quality of the reduced-order representation (see, for instance, the \emph{adaptive}~\cite{peherstorfer_online_2015, feng_adaptive_2017}, 
\emph{localized}~\cite{peherstorfer_localized_2014}, 
\emph{trajectory-based}~\cite{tan_trajectory-based_2019},
\emph{non-negative}~\cite{amsallem_energy_2016}, 
\emph{QR-factorization based}~\citep{drmac_new_2016} and 
\emph{weighted}~\cite{drmac_discrete_2018} variants).
In the context of fluid flows, this technique has been applied successfully to
incompressible~\cite{xiao_non-linear_2014} and compressible~\cite{fosasde2016nonlinear} cases, as well as reacting
flows~\cite{huang_exploration_2018}. 

Despite notable success over the past years, the field of nonlinear
model reduction for fluid systems is marked by empiricisms and heuristics
for the choice of basis, the treatment of nonlinearities, the enforcement
of physical constraints, or the selected model order. It is the objective
of this paper to compare the performance of POD-DEIM models with the more traditional Galerkin methods.
More specifically, we will study the convergence of the models as the
number of POD modes increases; we are, however, not interested in the
calibration of models of very small size or in data-driven regression techniques \citep{peherstorfer2016data}, based on sparsity-promoting techniques \citep{loiseau2018constrained} or on linear models \citep{brunton2017chaos,alomar2020reduced}. We also propose a new
reduction technique for the nonlinear terms that takes advantage of
a supplementary POD basis for the representation of the nonlinear
terms: however, instead of using interpolation to determine the
coefficients (like in the DEIM technique), we proceed straightforwardly
by projecting the nonlinear terms onto this additional basis.

The outline of the article is as follows. Section \ref{sec:testcases} is devoted to the
presentation of two test-cases: (i) cylinder flow at $Re=100$,
and (ii) a model problem, the non-parallel Kuramoto-Sivashinsky equations. For each case, we will introduce different trajectories, a transient initialized by the fixed-point solution, a limit-cycle solution and a transient initialized by a mean-flow solution.
These trajectories may be considered both for the building of the projection bases and for evalution of the models.
Section \ref{sec:ModelRed} introduces the three model reduction techniques: (i) Galerkin
projection with a single POD basis, (ii) Galerkin projection with two POD
bases, and (iii) the POD-DEIM technique. In the same section, we will
analyze the various properties of the reduced-order models, in particular
in view of the energy-preservation of the nonlinear convective terms.
Sections \ref{sec:KuramotoS}, \ref{sec:cyl_TR} and \ref{sec:CYL_LC} then apply the introduced techniques to the different
test-cases, with various choices for the trajectories used for building the projection bases. Section \ref{sec:Concl} offers a summary of our main results and concluding
remarks.

\section{Test cases, trajectories, numerical discretization} \label{sec:testcases}

The objective of this article is to generate low-order models that
accurately, stably and robustly reproduce the full nonlinear dynamics
of the underlying governing equations on given trajectories.
We choose two systems that
undergo a Hopf bifurcation, settling into a finite-amplitude
limit-cycle behavior. First, as a fluid system of this type,
we consider the incompressible flow past a cylinder at a supercritical
Reynolds number of $Re = 100.$ 
Second, we consider a simpler, one-dimensional model problem: the
non-parallel version of the Kuramoto-Sivashinsky equation. This model
equation mimics the behavior of cylinder flow, but allows a simpler and
more straightforward analysis of various model-reduction techniques. We
describe in the next two subsections (\S~\ref{sec:CYL},
and \S~\ref{sec:KS}) the respective governing equations, various considered trajectories and the adopted (spatial and temporal) numerical discretization.

\subsection{Flow past a cylinder}
\label{sec:CYL}

We consider flow past a cylinder at a Reynolds number of $Re=100,$ which is governed by the incompressible 2D Navier-Stokes equations, made non-dimensional with the upstream velocity and the cylinder diameter. Below, we first (\S \ref{sec:geq}) describe the different equivalent formulations in perturbative form around the fixed point $ w_b$ or the time-averaged flow around the limit-cycle $ \overline{w} $ and the spatial and numerical discretization details.

\subsubsection{Governing equations} \label{sec:geq}

Considering finite elements, the semi-discretized form of the Navier-Stokes equations governing the composite velocity-pressure variable $ w=[u,v,p] $
can be written as \citep{sipp2007global}:
\begin{equation} \label{eq:NS}
    Q\frac{dw}{dt}=r(w),
\end{equation}
with
\begin{equation}
  \label{eq:eq2}
  Q = \left( {\begin{array}{*{20}c}
       M & 0 & 0\\
       0 & M & 0\\
       0 & 0 & 0\\
       \end{array} } \right),
\end{equation}
and matrix $M$ designating the mass matrix linked to the finite element discretization of one velocity component. Throughout our study, we use the finite-element package {\tt{FreeFEM++}} \citep{Hecht2012} to implement all computations.  The unknown $w$ is discretized using $[P1b,P1b,P1]$ finite elements \citep{arnold1984stable} on a triangular mesh. The mesh contains 33586 triangles and extends from $x=-10$ to $ x=25$ in the streamwise direction and $ y=-10$ to $ y=10 $ in the cross-stream direction, with the cylinder located at $ (x,y)=(0,0) $.
Uniform Dirichlet boundary conditions are imposed at the inlet boundary, no-slip conditions at the cylinder surface, symmetric boundary conditions at the lateral boundaries and no-stress outflow conditions at the outlet boundary.
A $ w=[u,v,p]$ unknown holds 71499 degrees of freedom, the region in the vicinity of the cylinder exhibiting an isotropic mesh with triangles of size $\Delta x=0.1$. This corresponds to a rather coarse mesh: yet, all relevant features of cylinder flow are sufficiently captured, the flow becoming unstable for $ 47.3 < Re < 47.4 $ with a marginal eigenvalue appearing at $\lambda=0.8025\mathrm{i}$. These are classical values for such a configuration \citep{barkley2006linear}.

In supercritical cylinder flow, there are two specific flow fields of interest, around which we may consider perturbations:
\begin{subeqnarray}
w(t) &=& w_b+w'(t) \\
&=& \overline{w}+w''(t),
\end{subeqnarray}
where $w_{b} = \left(u_{b}, v_{b},
p_{b} \right)$ is a fixed point of the Navier-Stokes equations (base-flow) and $\overline{w}=[\bar{u},\bar{v},\bar{p}]$ the time-averaged flow (mean-flow):
\begin{subeqnarray}
r(w_b)&=&0 \\
 \overline{w}&=&\lim_{T \rightarrow \infty} \frac{1}{T} \int_0^T w(t)\; dt.
\end{subeqnarray}
The base-flow solution $w_b$ is computed in a classical manner \citep{sipp2007global} using Newton's method, based on a direct Lower-Upper (LU)-solver (in our case the MUMPS package \cite{MUMPS:1}).

These two changes of variables will provide two alternative governing equations in either the $w'$ or $w''$ variables, with a different split between linear and nonlinear operators. Depending on the case, and as shown below, the linear operator is either unstable when selecting the formulation involving $ w' $, or nearly marginal if the formulation based on $ w'' $ is chosen. As the modelling strategies may be different for the linear and the nonlinear operators, deciding on either of the two formulations will not be equivalent.

\subsubsection{Perturbative form around the base-flow (BF formulation)}

 When considering a trajectory of the flow that starts in the vicinity of the base-flow, it may seem important to accurately capture the linear dynamics close to the base-flow and hence explicitly introduce the linearized operator around the base-flow in the equations that will be reduced. In doing so, we separate the dynamics into two distinct parts: a linearized part, which exhibits the well-known unstable global mode, and a nonlinear part which imposes a stabilizing effect. In so far, we aim at structurally reproducing the dynamics of a Stuart-Landau amplitude equation\cite{sipp2007global}:
 \begin{equation}
\frac{dA}{dt}=\lambda A -\mu A|A|^2.
\end{equation}
In such an equation, the first term on the right-hand side represents the exponential instability and induces perturbation growth; the second term exerts a restoring stabilizing force, and the equilibrium between these two effects results in a saturated limit-cycle of a given amplitude.

The semi-discretized form of the perturbed equations
can then be written as (BF formulation in the following):
\begin{equation}
  \label{eq:eq1}
  Q \frac{d w'}{dt} = A' w' + n(w',w'),
\end{equation}
where matrix $A'$ is the Jacobian of $ R $ around $ w_b $,
\begin{equation}
A'=\left.\frac{\partial r}{\partial w}\right|_{w_b},
\end{equation}
and the term $n(w',w')$ refers to the quadratic convection term. The first term on the right-hand side corresponds to the (destabilizing) linear dynamics around the base-flow, the second one to the (stabilizing) nonlinear term that maintains the flow on the limit-cycle.

In the context of finite elements, weak forms should be favored wherever possible for the spatial discretization, as done in the definition of the residual $r$ (eq. \eqref{eq:NS}). We adopt a classical weak formulation for the linear dynamics $Qdw'/dt=A'w'$.
Yet, for computational efficiency when time-marching the equations, the convection term can also be discretized as   
\begin{equation}
  n(w',w') = -\left(
  {\begin{array}{*{20}c}
      u' \otimes (D_{x} u')  \\
      u' \otimes (D_{x} v') \\
      0 \\ \end{array} } \right)
- \left(
  {\begin{array}{*{20}c}
      v' \otimes (D_{y} u')  \\
      v' \otimes (D_{y} v')  \\
      0\\ \end{array} } \right),
\end{equation}
where $D_{x}$ and $D_{y}$ are the weak forms of the derivative matrices and
$\otimes$ denotes the element-wise (Hadamard) product.
In view of the DEIM reduction, we note that this choice of discretization of the quadradic term is not pointwise, i.e., the evaluation of the nonlinearity at a given location $n_i(w') $ does not correspond to the evaluation of a nonlinear function $ \tilde{n} $ at that location $\tilde{n}(w'_i)$. 
Yet, it still provides a quick and efficient evaluation of the nonlinear term since
only four sparse matrix-vector products and two element-wise products are required. For computational efficiency of the DNS solver, the various sparse matrices $M$, $D_x$, $D_y$ and $A'$ should be evaluated and stored in a preliminary stage of the DNS computations.

As mentioned in the introduction, a specific projection basis for the nonlinear term
will be considered here (in addition to the projection basis for the state $w'$). To this end, instead of considering the weak nonlinear term $n(w',w')$ for the snapshot series, we prefer to choose a physically more meaningfull nonlinear field, $f(w',w') $, such that:
\begin{equation} \label{eq:eq1b}
 n(w',w')=Qf(w',w').
\end{equation}
For a quick evaluation of the nonlinear term with a DEIM reduced order modelling technique \citep{chaturantabut2010nonlinear}, 
we finally choose the following (symmetrized, see advantages below for the implied structure of the nonlinear reduced order model) discretization scheme for $f$:
\begin{equation}
  \label{eq:eq3}
  f(w'_1,w'_2) = -\frac{1}{2}\left(
  {\begin{array}{*{20}c}
      u_{1}' \otimes (M^{-1} D_{x} u_{2}') + u_{2}' \otimes (M^{-1} D_{x} u_{1}') \\
      u_{1}' \otimes (M^{-1} D_{x} v_{2}') + u_{2}' \otimes (M^{-1} D_{x} v_{1}') \\
      0 \\ \end{array} } \right)
  - \frac{1}{2}\left(
  {\begin{array}{*{20}c}
      v_{1}' \otimes (M^{-1} D_{y} u_{2}') + v_{2}' \otimes (M^{-1} D_{y} u_{1}') \\
      v_{1}' \otimes (M^{-1} D_{y} v_{2}') + v_{2}' \otimes (M^{-1} D_{y} v_{1}') \\
      0\\ \end{array} } \right).
\end{equation}
The evaluation of this nonlinear term involves additional inverses of the mass matrix $M$, which may be efficiently handled by a conjugate-gradient solver using a diagonal preconditioner.
As mentioned earlier, this choice of discretization of the convection term stands as an approximation of the true weak-form discretization used in $R$. We have verified that we obtain the same trajectories with both discretizations (the time-evolution of various signals in the flow field are indistinguishable), validating our choice. 
In what follows, we have considered, for all simulations, this implementation of the nonlinear quadratic terms. 

For the time integration, we use a second-order semi-implicit scheme, with the linear operator in $ w=[u,v,p]^T $ being inverted by the direct LU solver at each time-step \citep{cerqueira2014eigenvalue}. The time-step is set to $\Delta t=0.01$, which ensures a maximum CFL number (based on the base-flow velocity) of 0.13.

\subsubsection{Perturbative form around the mean-flow (MF formulation)}

The evolution equation for the perturbation
$w''$ is given as (MF formulation):
\begin{equation}
  \label{eq:eq4}
  Q \frac{d w''}{dt} = b + A'' w'' + Q f(w'',w''),
\end{equation}
where
\begin{subeqnarray}
b &=& r(\overline{w}) \\  
A''&=&\left.\frac{\partial r}{\partial w}\right|_{\overline{w}},
\end{subeqnarray}
 are, respectively, the residual of the discretized Navier-Stokes equations taken at the mean-flow $\overline{w}$ and
the Jacobian of the residual around this flow.
We will use the same discretization choices
for $ f(w'',w'')$ as those presented in the previous section.

These are the governing equations that naturally arise when the trajectory is taken along the limit-cycle \citep{bergmann2009enablers,fosasde2016nonlinear}.
It is also the natural choice if only the mean-flow $\overline{w}$ is known (and not the
base flow $w_b).$ 
We remark that this equation is now inhomogeneous with a constant term $b$ and that the linear operator now corresponds to the Navier-Stokes equations linearized around the mean-flow, which exhibits stability properties different from the ones involving the base-flow.
It is expected that the flow snapshots on the limit-cycle are close to the features of the mean-flow marginal eigenvector.   
For a harmonic flow (which is a reasonable approximation for cylinder flow), the frequency of the flow field and its Fourier mode correspond to a marginal-eigenvalue and eigenvector of the linearized operator around the mean-flow \citep{mezic2013analysis,turton2015prediction}. Therefore, eq. \eqref{eq:eq4} likely represents the best choice to reproduce the dynamics around the mean-flow, since most of the features of the limit-cycle are already captured by  the linear operator, while the nonlinear operator only needs to correct small defects of the linear representation (for example, the eigenvalue might be slightly unstable).
Also, this equation is at the heart of all successfull mean-flow-based resolvent studies, which justify the validity of the approach by different arguments (dominant singular value condition \citep{beneddine2016conditions}, white noise approximation of the forcing \citep{towne2018spectral} or small-amplitude assumption of the fluctuation field $ w''$ \citep{leclercq2019linear}).  
Of course, in the present case, the dynamics close to the base-flow, that is the frequency shift of the dynamics and the different wavelengths of the perturbations, needs to be entirely modelled by the nonlinear terms. Hence, it is likely that the linear dynamics close to the base-flow will be more poorly represented, since nonlinear model reduction generally performs poorer than linear reduction. 
As in the previous section, we will consider snapshots $f(w'',w'')$ for building the nonlinear projection basis. 

\subsubsection{Transient (TR), Limit-Cycle (LC) and Mean-Flow Transient (MFTR) trajectories} \label{sec:traj_cyl}

\begin{figure}[htbp]
\centerline{(a)}
\psfrag{x}{$x$}
\psfrag{y}{$y$}
\psfrag{u}{$u$}
  \centerline{\includegraphics[width=0.7\textwidth]{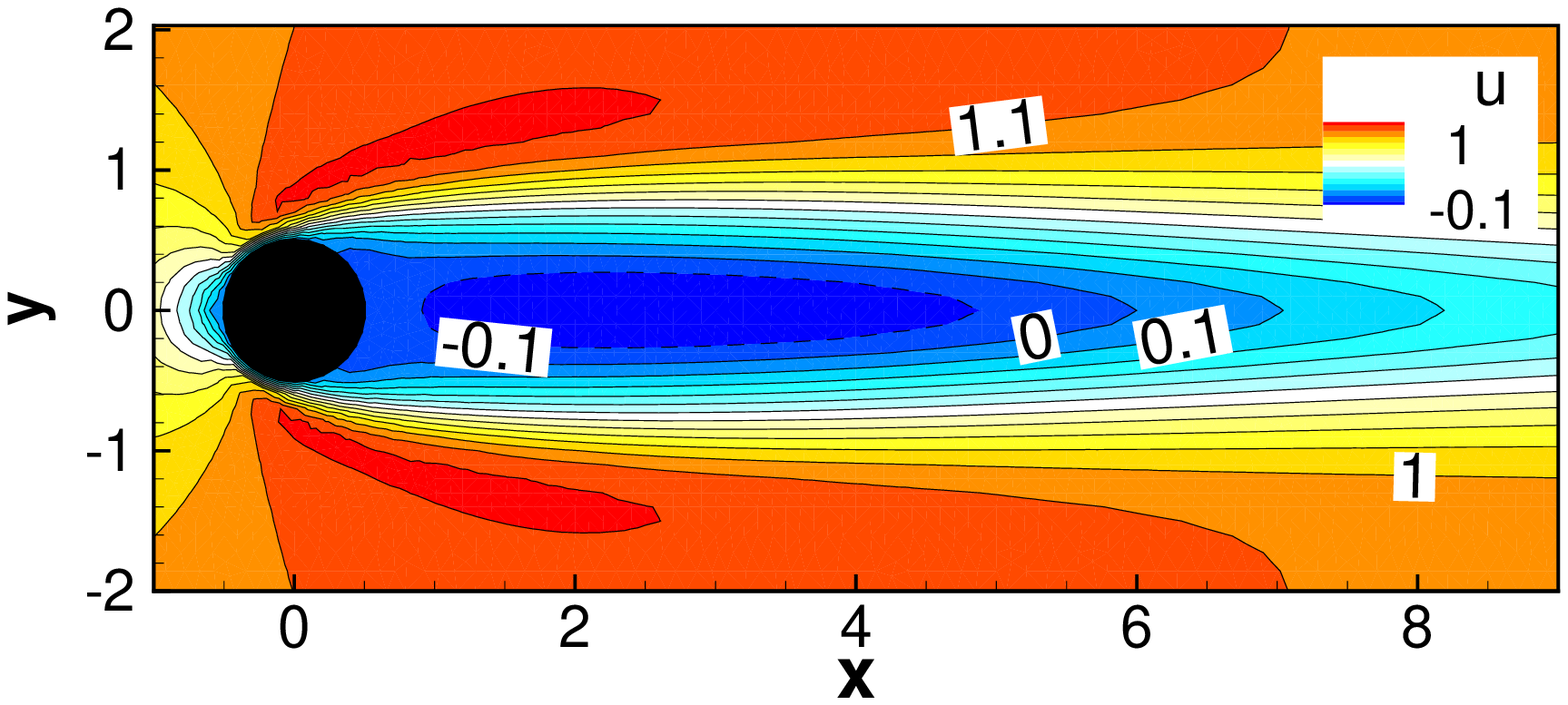}}
 \centerline{(b)}
  \psfrag{x}{$x$}
\psfrag{y}{$y$}
\psfrag{u}{$u$}
  \centerline{\includegraphics[width=0.7\textwidth]{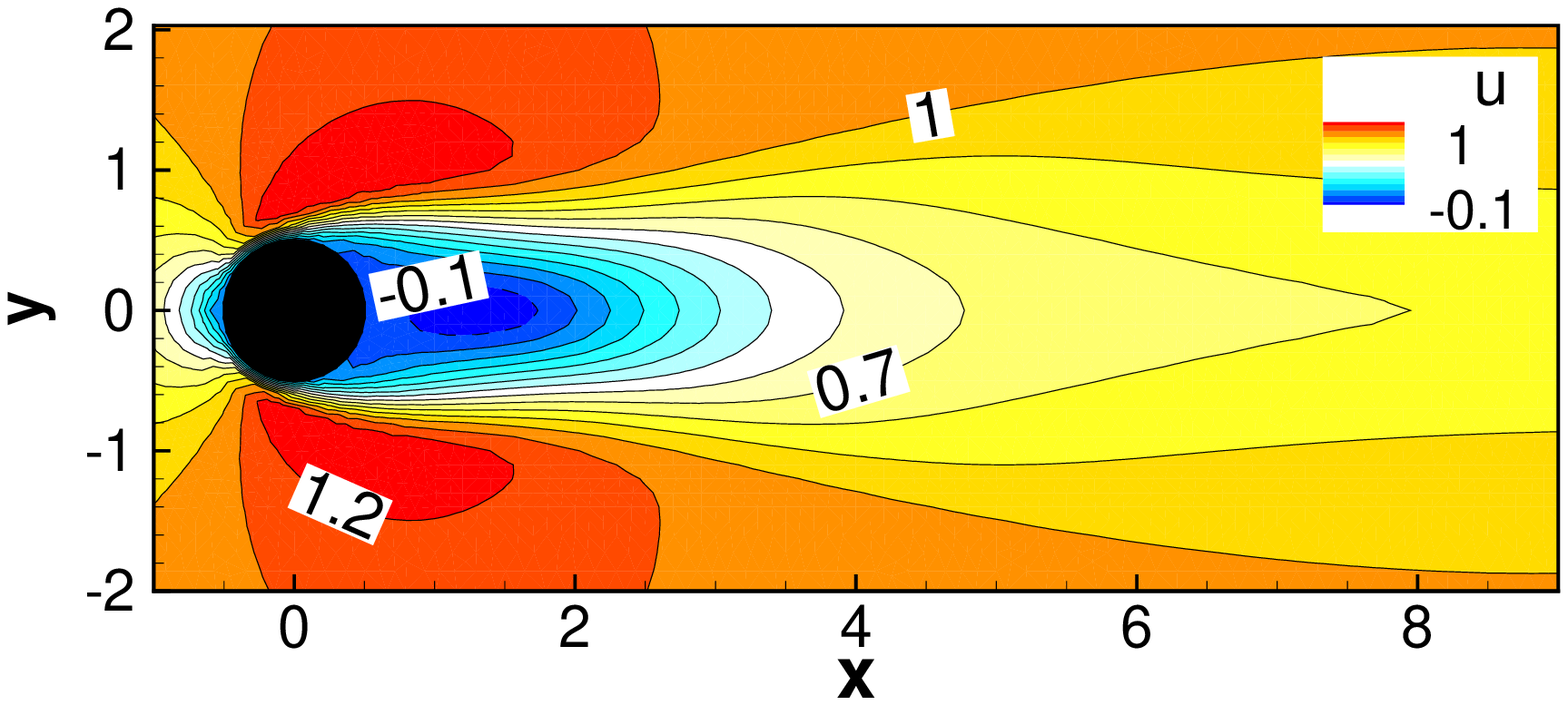}}
  \caption{\label{fig:baseandmeanflow} (a): Base- and (b): mean-flow solutions for flow past a
  cylinder at a supercritical Reynolds number of $Re=100,$ visualized
  by contours of the streamwise velocity component.}
\end{figure}

 The base- and mean-flows at $Re=100$ are  represented in figures~\ref{fig:baseandmeanflow}(a,b) with iso-values of streamwise velocity.
The mean-flow solution shows a markedly reduced recirculation
zone behind the cylinder, when compared to the base-flow solution.

\begin{figure}[htbp]
  \centering
  \begin{tabular}{cc} (a) & (b) \\
      \psfrag{s}{$\sigma$}
    \psfrag{w}{$\omega$}
    \psfrag{a}{BF}
    \psfrag{b}{MF}
    \psfrag{c}{$\omega^{DNS}$}
    \includegraphics[width=0.5\textwidth]{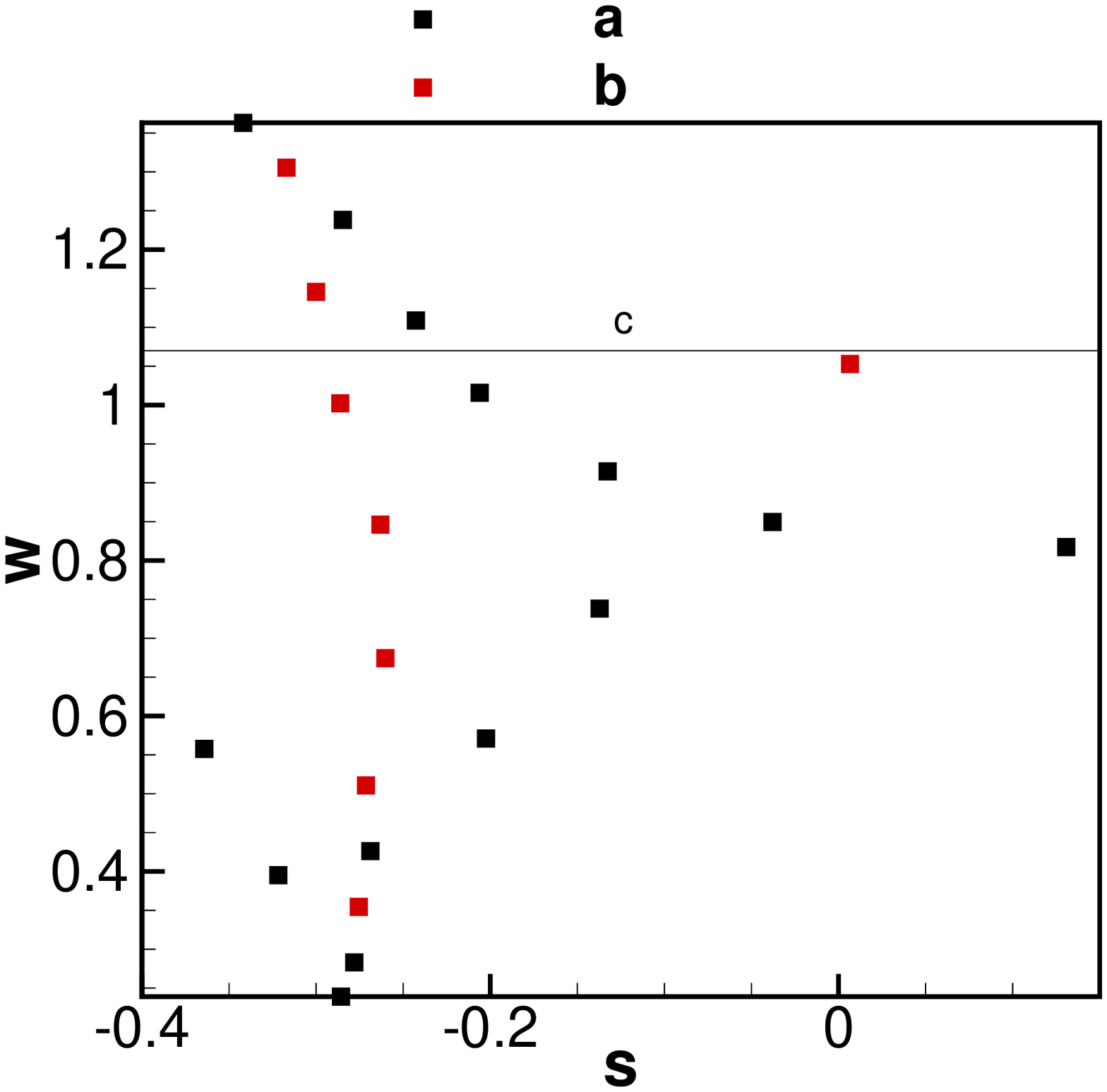}&
    \psfrag{x}{$x$}
    \psfrag{v}{$\mbox{Re}(\hat{u})$}
    \psfrag{a}{BF}
    \psfrag{b}{MF}
    \includegraphics[width=0.5\textwidth]{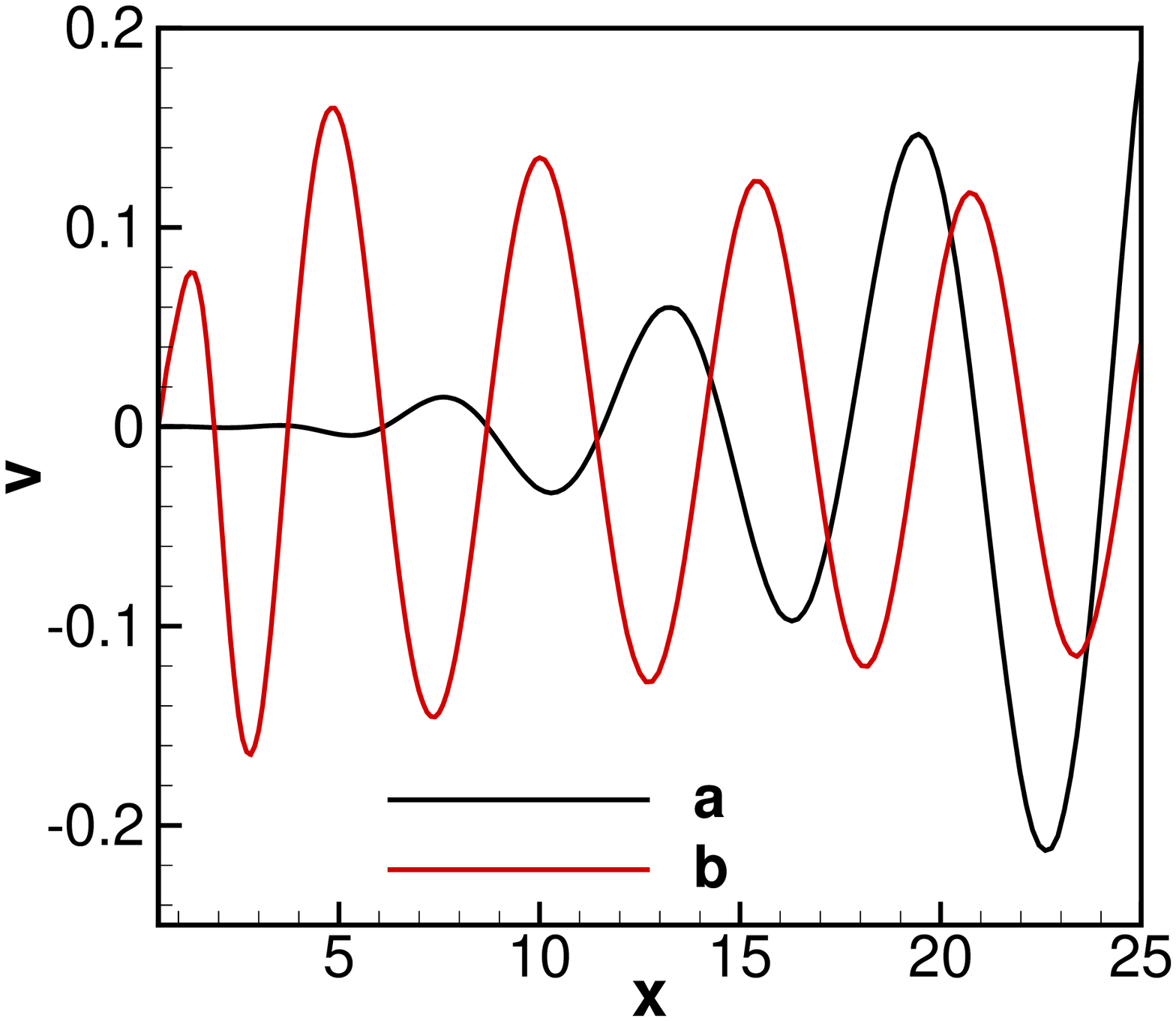}
    \end{tabular}
      \caption{\label{fig:CYLeig} (a): Eigenspectrum of the linearized operator around the base-flow (black symbols) and mean-flow (red symbols). The horizontal solid line depicts the frequency of the limit-cycle obtained by DNS. (b): Real part of the streamwise component of the associated eigenvectors.}
\end{figure}

Normal modes of the linear operators can be formulated as:
$ w'=e^{\lambda^{BF} t} \hat{w}^{BF}$
and $ w''=e^{\lambda^{MF} t} \hat{w}^{MF}$.
These are solution of the eigen-problems 
\begin{subeqnarray}
  A'\hat{w}^{BF}&=&\lambda^{BF} Q\hat{w}^{BF} \\
  A''\hat{w}^{MF}&=&\lambda^{MF} Q\hat{w}^{MF},
\end{subeqnarray}
which can be treated by Krylov methods linked to a shift-invert strategy based on the direct LU-solver \citep{sipp2007global}.
As expected, we obtain a pair of unstable eigenvalues $\lambda^{BF} = 0.131 \pm 0.817\mathrm{i}$ for $A'$
and a pair of (nearly-) marginal eigenvalues
$\lambda^{MF} = 0.00195 \pm 1.06\mathrm{i}$ for $A''$.  These compare well with those in the literature \citep{barkley2006linear}. Note also that the value of the frequency $1.06$ is very close to the frequency of the limit-cycle, as obtained from the DNS solution ($\omega^{DNS}\approx1.07$ here).
A close-up view of both spectra in the vicinity of the two previously mentioned eigenvalues is shown in figure \ref{fig:CYLeig}(a), with black symbols for $\lambda^{BF}$ and red symbols for $\lambda^{MF}$. Also, the horizontal black solid line depicts the frequency $\omega^{DNS}.$ The real part of the streamwise velocity of the corresponding eigenvectors are shown with lines in corresponding colors in figure \ref{fig:CYLeig}(b).
The shapes of the two eigenvectors are very different: the base-flow eigenvector exhibits a gradual growth of the oscillation amplitude in the downstream direction, while the mean-flow eigenvector exhibits a peak around $ x=5$, before slowly decreasing. The last pattern is representative of oscillations of the flow on the limit-cycle. These observations are reminiscent of the property that an eigenvalue/eigenvector of a mean-flow solution exactly reproduces the frequency and spatial structure of the unsteady solution, if the unsteady solution exhibits a harmonic behavior \citep{barkley2006linear,sipp2007global,mezic2013analysis,turton2015prediction}. This property is well satisfied in the case of cylindar flow \citep{turton2015prediction}.

\begin{figure}[htbp]
  \centering
  \begin{tabular}{cc} (a) & (b) \\
    \psfrag{T}{$t$}
    \psfrag{K}{$2K'$}
    \psfrag{a}{TR}
    \psfrag{b}{LC}
    \psfrag{c}{MFTR}
    \includegraphics[width=0.5\textwidth]{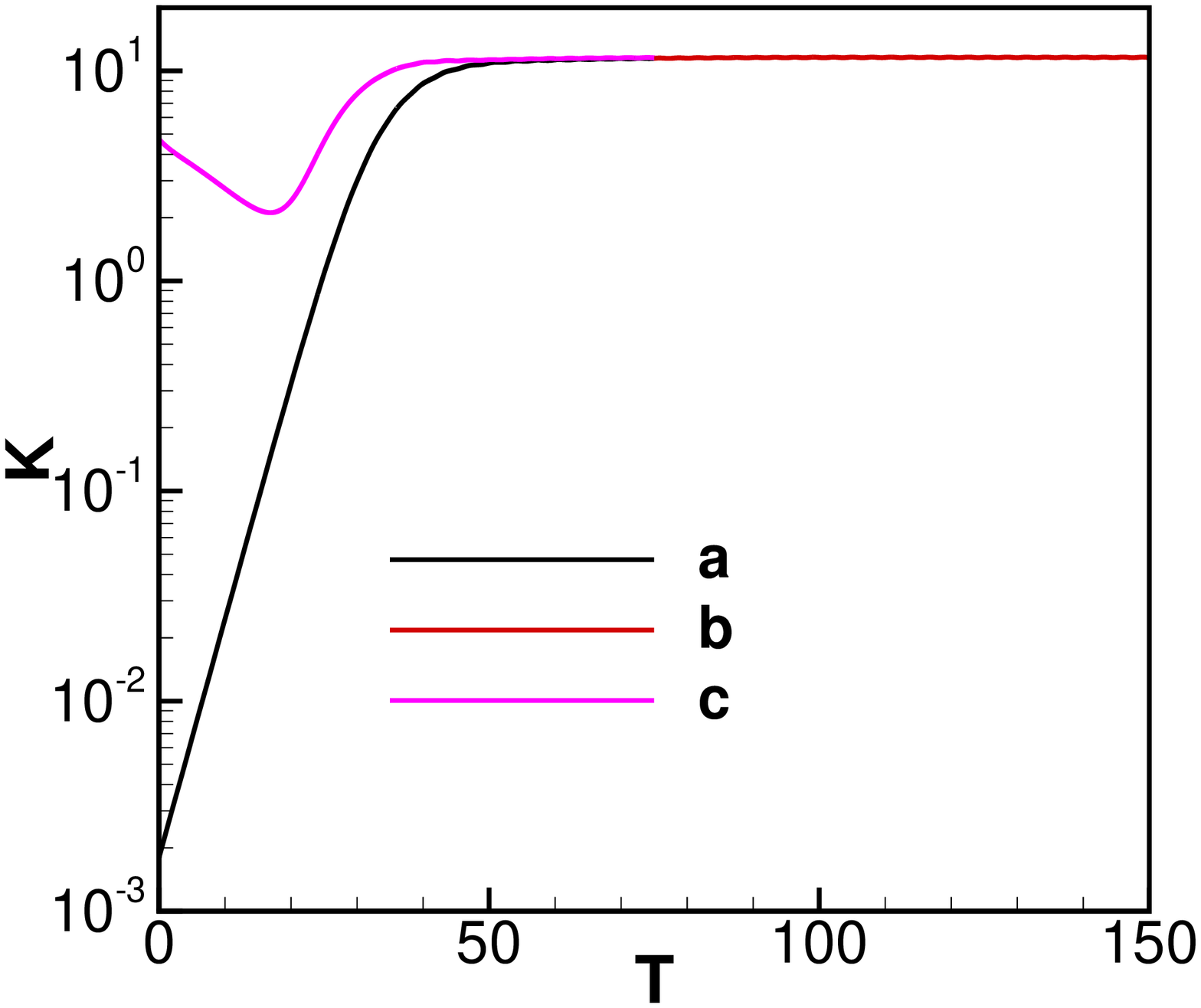} &
    \psfrag{A}{TR}
    \psfrag{B}{\textcolor{red}{LC}}
    \psfrag{a}{TR}
    \psfrag{b}{LC}
    \psfrag{c}{MFTR}
    \psfrag{T}{$t$}
    \psfrag{U}{$u'(5,0)$}
\psfrag{u}{$u'(5,0)$}
    \psfrag{v}{$u''(5,0)$}
    \includegraphics[width=0.5\textwidth]{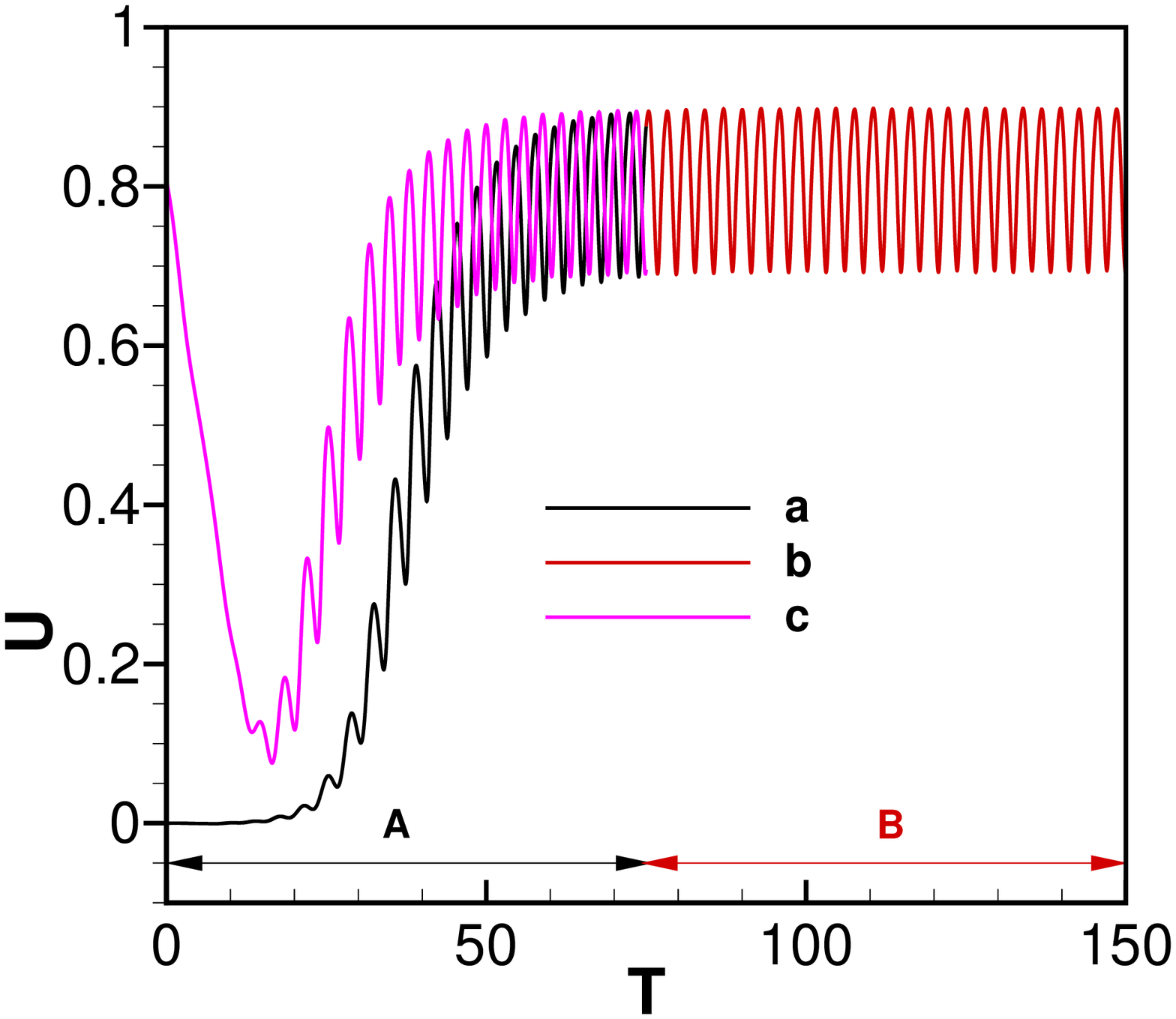}
  \end{tabular}
  \caption{\label{fig:evoEnergy} (a) For
  flow past a cylinder at $Re=100$, time evolution of the kinetic
  energy $2K'=w'^H Q w'$ of the perturbation about the base-flow for the transient TR from the base-flow (black line), for the limit-cycle solution (red line) and the transient MFTR from the mean-flow (magenta line). (b) Time trace of
  the streamwise velocity extracted at $(x=5,y=0)$
  for each trajectory TR, LC, MFTR. The TR and MFTR trajectories span the time-range $ 0 \leq t \leq 75$ and the LC trajectory the range $ 75 \leq t \leq 150$.}
\end{figure}

A direct numerical simulation has been carried out, starting with the small-amplitude unstable
global mode as an initial condition,
$ w'=\alpha \mbox{Re}(\hat{w}')$, with $ \alpha$ standing for a small amplitude.
The evolution of the perturbation
towards the limit-cycle is best visualized by the
time-evolution of the kinetic energy given by the quadratic form
$2K' = w'^H Q w'.$ Figure~\ref{fig:evoEnergy}(a) clearly displays (see the black solid line labelled TR) 
an exponential instability over about $30$ time-units (at an amplification rate corresponding to twice the amplification rate of the unstable global mode), until saturation
sets in, as the elimit-cycle behavior is reached (red solid line labelled LC). We also have shown with a solid magenta line the transient MFTR trajectory, which is initialized by the mean-flow solution about the limit-cycle: after a quick decrease of the kinetic energy, the perturbation grows again showing a near exponential growth rate, before saturation on the limit-cycle sets in. We will use this particular trajectory to assess the robustness of the reduced-order models, by evaluating their performance on a trajectory that was not considered for the building of the model.   
A more detailed view of the
same evolutions is given by the time-trace of the streamwise velocity
component $w'$ at point $(x=5,y=0)$ in the wake of the cylinder
(see figure~\ref{fig:evoEnergy}(b)). Again,
an exponential instability is clearly discernible for the black solid line, before convergence
towards a limit-cycle behavior sets in (red solid line). The vortex shedding frequency
is visible in the velocity trace.

\begin{figure}[htbp]
\centerline{(a)}
\psfrag{x}{$x$}
\psfrag{y}{$y$}
\psfrag{u}{$u'$}
  \centerline{\includegraphics[width=0.8\textwidth]{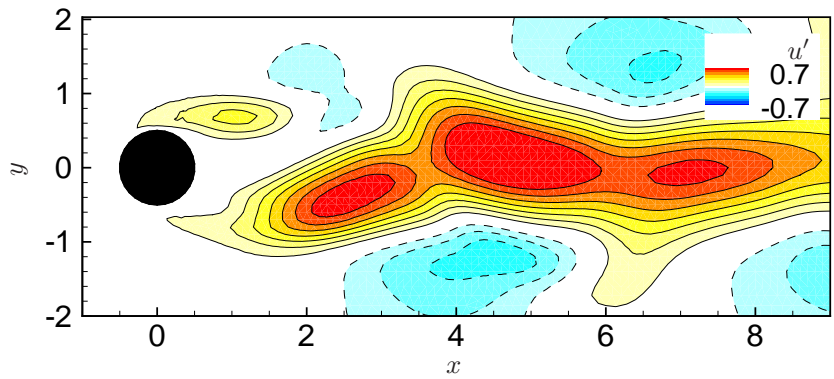}}
\centerline{(b)}
    \psfrag{x}{$x$}
\psfrag{y}{$y$}
\psfrag{u}{$u''$}
  \centerline{\includegraphics[width=0.8\textwidth]{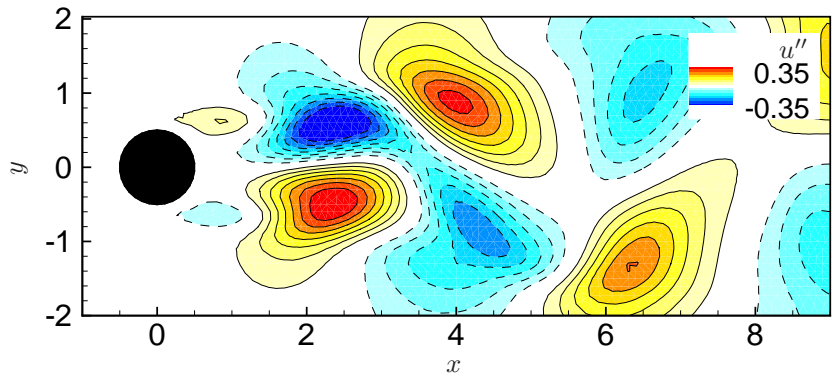}}
  \caption{\label{fig:LCsnapshot} (a): Representative snapshot of the
  perturbation field $w'$ at $t=400$ from the limit cycle of flow
  past a cylinder at Reynolds number $Re=100,$ visualized by
  iso-contours of the streamwise velocity component.
  (b): Same snapshot but represented with the $ w''=w'+w_b-\overline{w}$ variable.}
\end{figure}

In addition, we show representative
snapshots of the perturbation field $w'$ and $w''$ on the limit cycle
(see figures~\ref{fig:LCsnapshot}(a,b)), visualized by iso-contours of the
streamwise velocity component. In the left snapshot, the dominant red colors in the central part of the wake represent the mean-flow deformation with respect to the base-flow (the recirculation bubble shortens, inducing positive mean-values of $ u'$), and the antisymmetric streamwise large-scale modulations depict the vortex-shedding mode. In the right snapshot, only the vortex shedding mode is visible, since the mean-deformation was subtracted by the change of variables.
These differences will yield different projection bases depending on the considered formulation, since
the $w'$ snapshots are used in the BF formulation and the $w''$ snapshots in the MF formulation.

\subsection{Non-parallel Kuramoto-Sivashinsky equation}
\label{sec:KS}

This equation is commonly used as a model equation for complex fluid
motion as it contains many features in a one-dimensional setting that
have equivalents in higher dimensions. It is thus a valuable proxy for
investigating analytical and computational techniques and for
quantifying the influence of its ingredients on user-specified
performance measures. We will use this equation for assessing the
accuracy, stability and robustness of model-reduction methods in a simpler case. To this
end, we consider the non-parallel version of the Kuramoto-Sivashinsky
equation in the form

\begin{equation}
  \label{eq:eq7}
  \partial_t u' +
  u_b \partial_x u' + u' \partial_x u'
  = -\mu(x) \partial_{xx} u' -
  \gamma \partial_{xxxx} u'
\end{equation}
with
\begin{equation}
  \label{eq:eq8}
  \mu(x) = \mu_0 \exp\left(-\frac{x^2}{d^2}\right)
\end{equation}
and $u'$ as a real function defined on the interval
$\left[ -x_\infty, \  x_\infty \right].$ To complete the problem,
we provide the boundary conditions on $u'$ according to
\begin{subeqnarray}
  \label{eq:eq9}
  &u'=\partial_{xx} u' = 0 \qquad &\mathrm{at}
  \quad x=-x_\infty, \\
  &\partial_{x} u' =\partial_{xxx} u'
  = 0 \qquad &\mathrm{at} \quad x=x_\infty.
\end{subeqnarray}
The term $u_b\partial_x u'$ models uniform convection at a
prescribed speed of $u_b,$ and $u'\partial_{x} u'$ constitutes
the nonlinear, quadradic convection term which is also present in the
full Navier-Stokes equations. The expression
$-\mu(x)\partial_{xx} u'$ in the Kuramoto-Sivashinky equation
models an instability with a strength of $\mu_0,$ its origin located
about $x=0$ and its spatial extent governed by the width parameter $d.$  This term mimics the streamwise-localized instability mechanism acting in the recirculation bubble of cylinder flow.
The final term $-\gamma \partial_{xxxx} u'$ provides a
stabilizing hyper-diffusion which damps high-wavenumber (small-scale)
structures.

In view of spatial discretization with classical finite-element methods, we consider the auxiliary variable $v'=\partial_{xx} u'$ to render the system second-order, so that:
\begin{eqnarray}
  \partial_t u' +
  u_b \partial_x  u' + u' \partial_x u'
  &=& -\mu(x) v' -
  \gamma \partial_{xx} v' \\
  v' &=& \partial_{xx} u'
\end{eqnarray}
with $u'=v' = 0$ at $ x=-x_\infty$ and
  $\partial_{x} u' =\partial_{x} v'
  = 0$ at $ x=x_\infty$.
  The second equation can be seen as a constraint on the state $ w '$, reminiscent of the divergence-free constraint in the Navier-Stokes equations.

Semi-discretization of the above
equation with finite elements yields an expression whose structure is similar to
(\ref{eq:eq1}) for the state-variable $w'=[u',v']$ . We obtain
\begin{equation}
  \label{eq:eq11}
  Q \frac{d w'}{dt} = A' w' + Q f\left( w',w' \right),
\end{equation}
with
\begin{equation}
   Q=\left( \begin{array}{cc}
        M & 0 \\
        0 & 0
   \end{array}\right),
\end{equation}
$A'$ as the weak-form of
\begin{equation}
\left( \begin{array}{cc}
        -u_b \partial_x & -\mu(x) -\gamma\partial_{xx} \\
        -\partial_{xx} & 1
   \end{array}\right),
\end{equation}
and the symmetrized bilinear term:
\begin{equation}
  f(w'_1,w'_2) = -\frac{1}{2}\left( \begin{array}{c}
      u_{1}' \otimes (M^{-1} D_{x} u_{2}') + 
      u_{2}' \otimes (M^{-1} D_{x} u_{1}')    \\
       0 
  \end{array}
       \right),
\end{equation}
where, again, the point-wise discretization scheme of the nonlinear term has been chosen.

We use second-order $ [P2,P2] $ elements for the discretization of $ w'=[u',v']$. For the mesh, we choose $ x_\infty=100$ and
the elements are of size $ \Delta x=0.05$, which yields 4000 elements.
The time-step for the simulation is $ \Delta t=0.01$. The same semi-implicit time integration strategy as described for the Navier-Stokes equations is used here.

\begin{figure}[htbp]
  \centering
  \begin{tabular}{cc} (a) & (b) \\
    \psfrag{s}{$\sigma$}
    \psfrag{w}{$\omega$}
    \psfrag{a}{BF}
    \psfrag{b}{MF}
    \psfrag{c}{$\omega^{DNS}$}
    \includegraphics[width=0.45\textwidth]{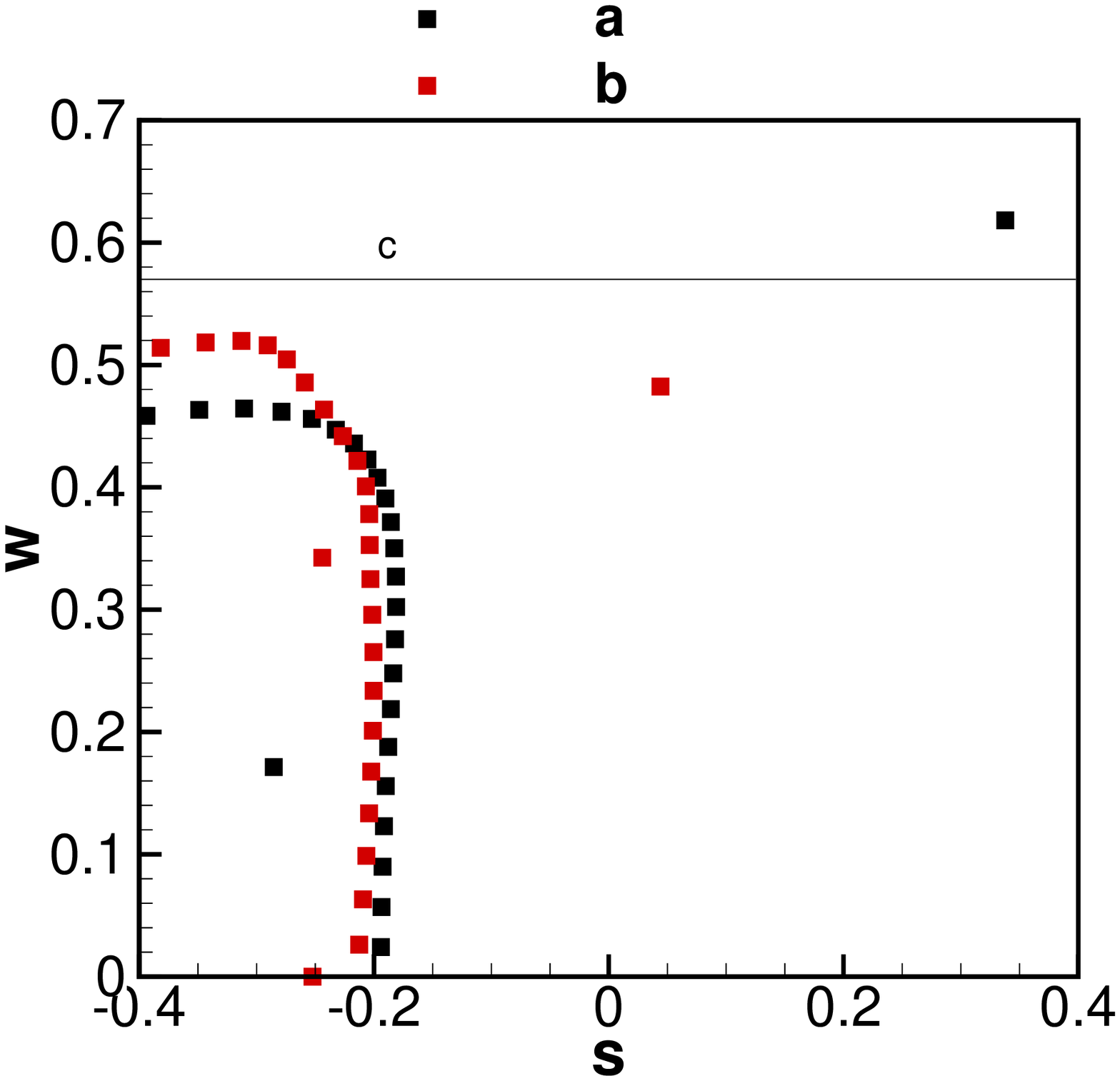}&
    \psfrag{x}{$x$}
    \psfrag{v}{$\mbox{Re}(\hat{u})$}
    \psfrag{a}{BF}
    \psfrag{b}{MF}
    \includegraphics[width=0.45\textwidth]{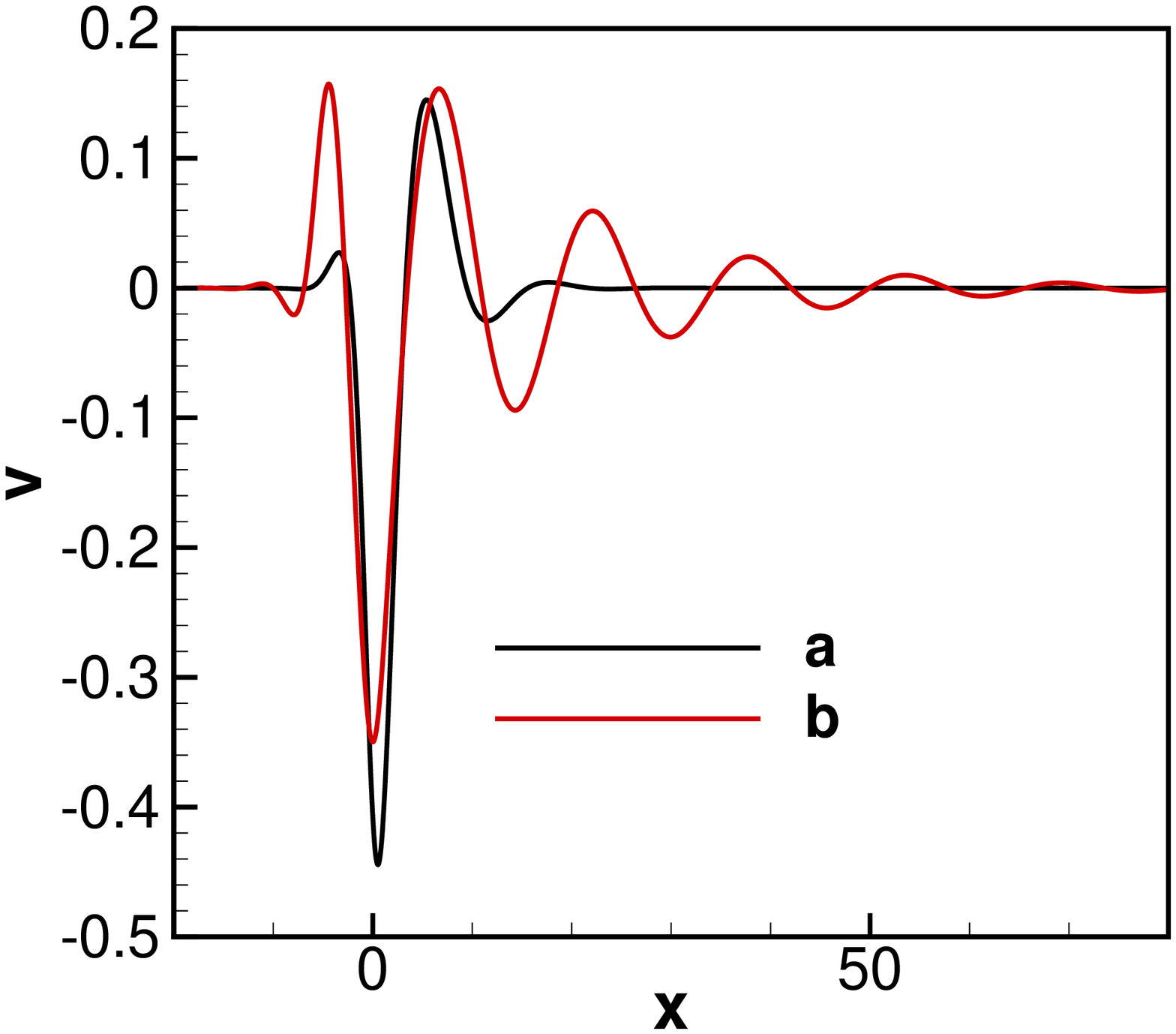} \\ (c) & (d) \\
    \psfrag{T}{$t$}
    \psfrag{K}{$2K'$}
    \psfrag{a}{TR}
    \psfrag{b}{LC}
    \psfrag{c}{MFTR}
    \includegraphics[width=0.45\textwidth]{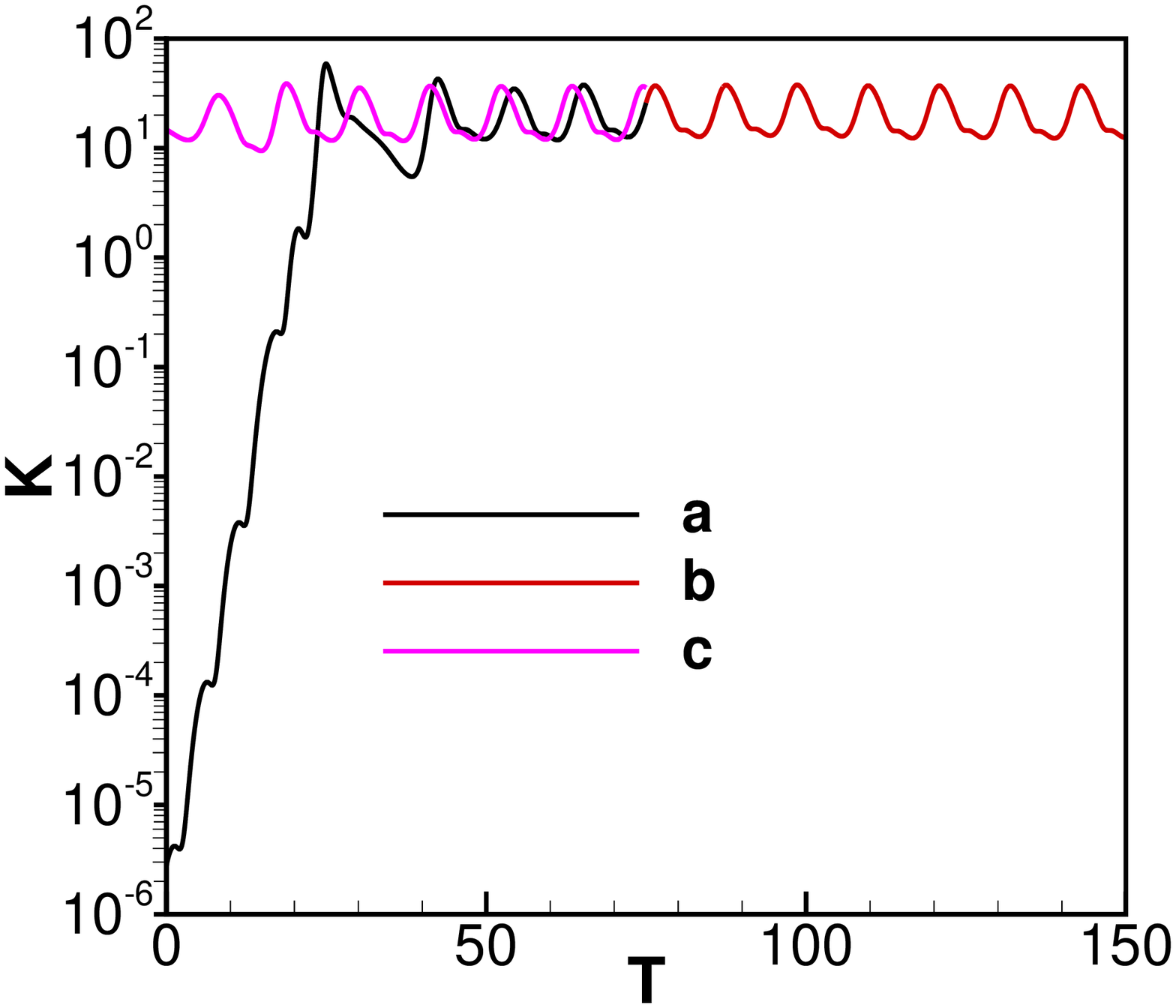}
    &        \psfrag{T}{$t$}
    \psfrag{W}{$u'(x=10,t)$}
        \psfrag{A}{TR}
    \psfrag{B}{\textcolor{red}{LC}}
    \psfrag{a}{TR}
    \psfrag{b}{LC}
    \psfrag{c}{MFTR}
    \includegraphics[width=0.45\textwidth]{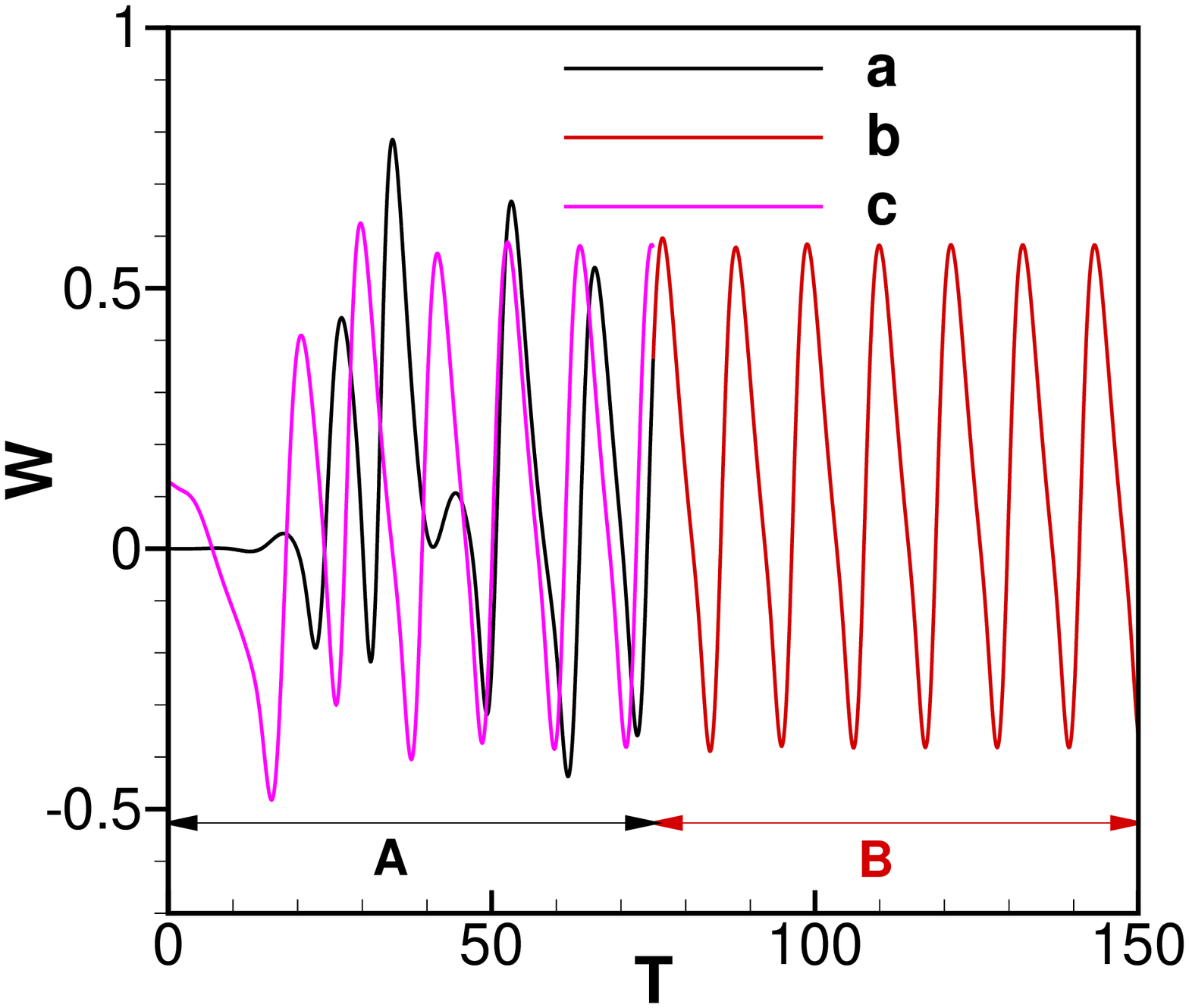}
    \\ (e) & (f) \\
 \psfrag{x}{$x$}
    \psfrag{w}{$u'(x,t)$}
    \psfrag{a}{$t=141.61$}
    \psfrag{b}{$t=144.41$}
    \psfrag{c}{$t=147.23$}
    \psfrag{d}{$t=150.01$}
    \includegraphics[width=0.45\textwidth]{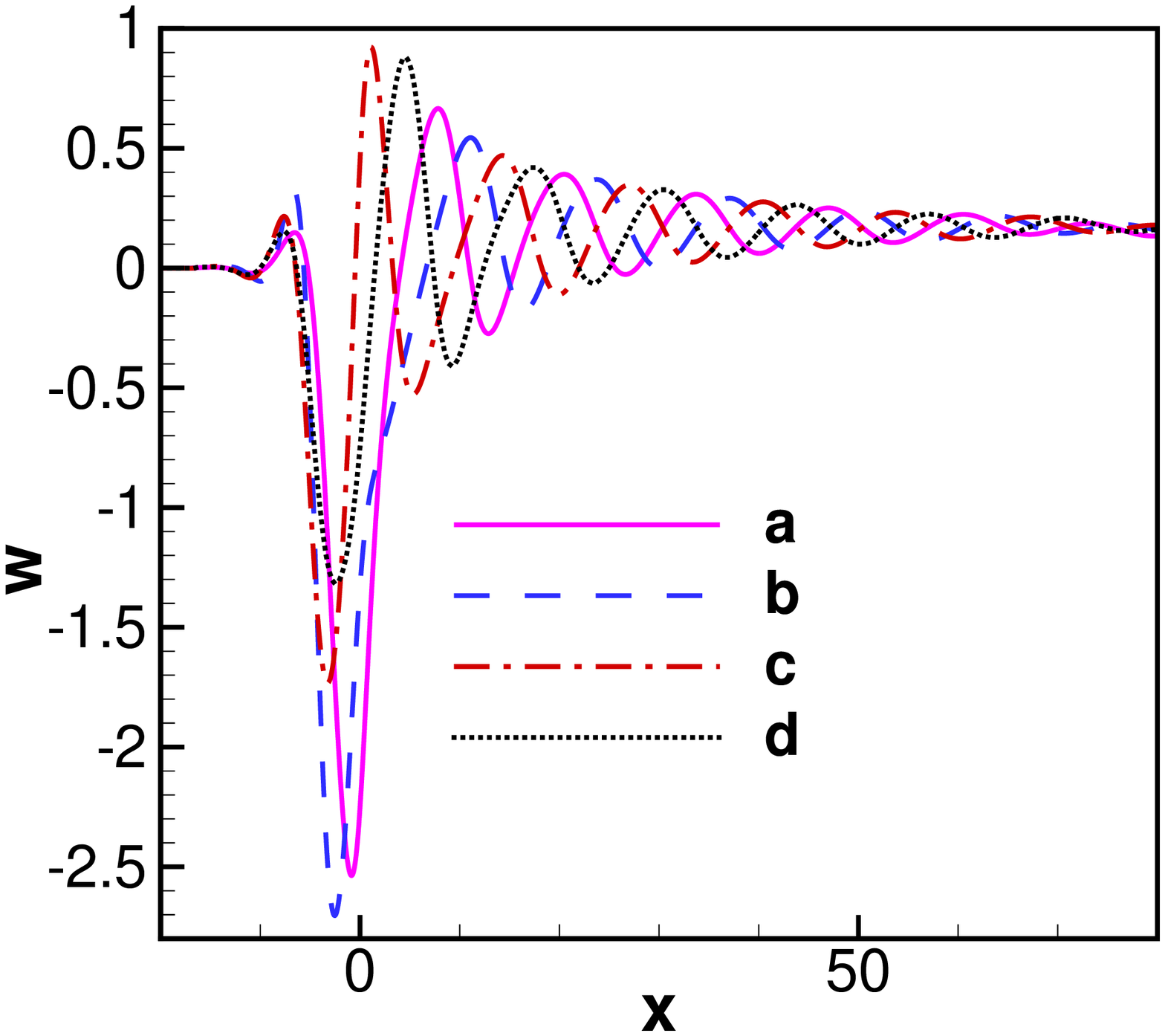}
    & \psfrag{x}{$x$}
    \psfrag{w}{$\overline{u'}$}
    \includegraphics[width=0.45\textwidth]{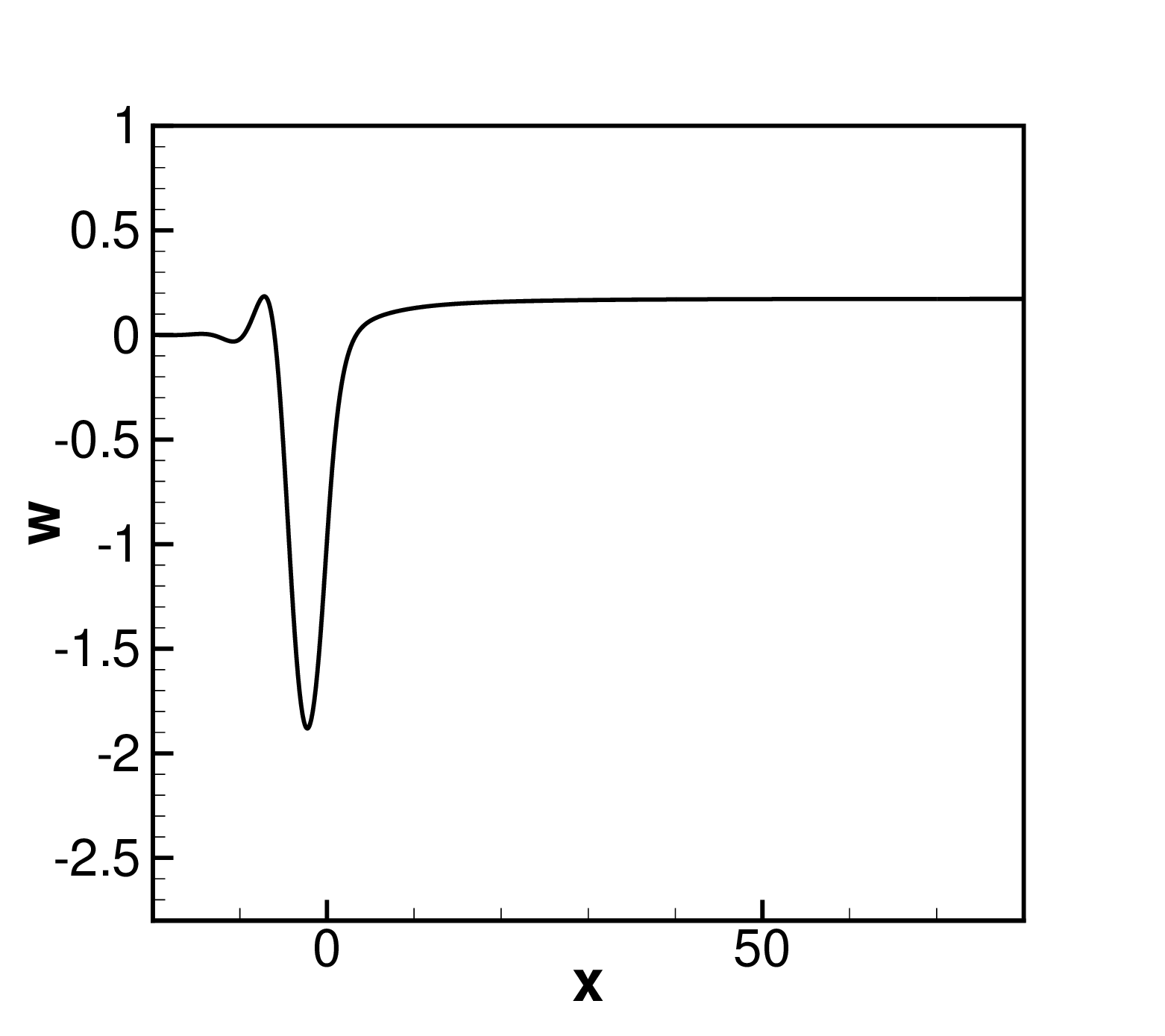}
  \end{tabular}
  \caption{\label{fig:KS4} Kuramoto-Sivashinsky equation.
  (a): Eigenvalue spectrum of linear operator in BF formulation ($ w'$ variable) and MF formulation ($w''$ variable). The horizontal solid line depicts the frequency of the limit-cycle obtained by DNS. (b): Corresponding real parts of leading eigenvectors. (c): Temporal
  evolution of kinetic perturbation energy $2 K = w'^H Q w'.$
  The initial condition corresponds to a small-amplitude
  global mode (see subfigure (b), in black). (d): Temporal evolution
  of $u'$ evaluated at $x=10.$ TR  and LC refer, respectively, to the time intervals $0\leq t \leq 75$ and $75 \leq t \leq 150$. Trajectory MFTR (magenta line) has been initialized by the mean-flow solution about the limit-cycle. (e): Four snapshots $u'$
  within one period of the limit-cycle regime. (f): Mean-flow solution about the limit-cycle.}
\end{figure}

For our
analysis, we choose the following constants
\begin{equation}
  \label{eq:eq12}
  u_b=1,\qquad \gamma =1, \qquad d=1,\qquad \mu_{0}=3.95,
\end{equation}
which leads to an unstable linearized operator (about the base
flow $w'=0$) with a single pair of unstable eigenvalues at
$\lambda^{BF} =0.338\pm 0.618\mathrm{i}.$
The eigenvalue spectra for the linearized operators associated to the base-flow (BF) and mean-flow (MF) formulations are shown with red and black symbols in figure ~\ref{fig:KS4}(a),
and the (real part of the) global mode associated
with the leading eigenvalue in figure~\ref{fig:KS4}(b) with the same color.
Figures~\ref{fig:KS4}(c,d,e) depict the TR, MF and MFTR trajectories in a similar
manner than in subsection \S~\ref{sec:traj_cyl},
i.e., the evolution of the perturbation kinetic energy
(figure~\ref{fig:KS4}(c)) for the three trajectories TR, LC and MFTR, the time-trace of the variable
$u'$ at $x=10$ (figure~\ref{fig:KS4}(d)) and four
representative snapshots from the limit cycle
(figure~\ref{fig:KS4}(e)). 
We have also shown the mean-flow solution corresponding to the limit-cycle in figure~\ref{fig:KS4}(f). Coming back to the mean-flow eigenvalue  $\lambda^{MF}=0.0440\pm 0.482\rm{i}$, we observe that, similarly to the case of cylinder flow, the amplification rate of the linearized operator around the mean-flow (MF) is closer to marginality. Yet, the frequency of the mode has decreased well below the frequency of the limit-cycle, which is $ \omega^{DNS}\approx0.57 $. The eigenvector of the mean-flow unstable eigenmode is now closer to the oscillations of the limit-cycle. The observed discrepancies are due to the fact that the limit-cycle for the KS equation exhibits many harmonics and is therefore far from harmonic.
The "harmonic property" is better satisfied in the case of cylinder flow.
To conclude, despite these differences, we can state that the behavior of
this model problem is qualitatively rather close
to the case of flow past a cylinder: the solution undergoes
a Hopf bifurcation and progresses towards a limit-cycle
behavior.

\section{Model reduction methodology}
\label{sec:ModelRed}

In this section, we develop a mathematical framework for the
projection-based model reduction of semi-discretized evolution
problems of the form

\begin{equation}
  \label{eq:eq13}
  Q \frac{d w}{dt} = b + A w + Q f(w,w),
\end{equation}
where $f(w,w)$ is a bilinear symmetric term. We express the
perturbation $w$ as $w = W z$ where the matrix $W$ contains
$p$ basis vectors as its columns and $z$ denotes the
vector of $p$ expansion coefficients, and project the resulting
equation onto the subspace spanned by $W$. The reduced-order model
then reads
\begin{equation}
  \label{eq:eq14}
  \frac{dz_i}{dt} = c_i + \sum_j L_{ij} z_j +
  \sum_{j,k} N_{ijk} z_j z_k
\end{equation}
with
\begin{subeqnarray}
  \label{eq:eq15}
  W^H Q W &=& I, \\
  N_{ijk} &=& N_{ikj}.
\end{subeqnarray}
The last relation stems from the symmetry of the nonlinear operator.
The subscripts ${}_{ijk}$ indicate the $i,j,k$-th component of the corresponding vectors, matrices
and tensors. The values of the coefficients $c_i$, $L_{ij}$ and $N_{ijk}$ will be given in the next sections.
 The operation count associated with this reduced-order model scales as $p^3$, since for each of the $p$ degrees of freedom, we have a double-summation over $ p^2 $ terms. Yet, as shown below, in the case of the DEIM method, the scaling of the operation count linked to the evaluation of the nonlinear term can be drastically reduced in the case of a point-wise nonlinearity (as is the case here). 

Based on the above approach, the building of the reduced-order
model involves three steps: the computation of the POD basis $W$ for the representation of the state
(section \S\ref{sec:POD}), the modelling of the constant
term (section \S\ref{sec:const}), of the linear term (section
\S\ref{sec:lin}) and of the nonlinear term (section
\S\ref{sec:nonlin}).
Section \S\ref{sec:mathprop} discusses the mathematical properties of the model while \S \ref{sec:crit} presents all quality and error measures that will serve to assess the models.

\subsection{The orthonormal basis $W$}
\label{sec:POD}

The orthogonal basis contained in the columns of the matrix $W$ are
computed from snapshots gathered from numerical simulations of the
full system. With $X$ denoting the snapshot matrix, consisting of
columns that constitute the flow fields at equispaced instants in
time, we form the temporal correlation matrix and determine its eigenvalues
and eigenvectors according to
\begin{equation}
  \label{eq:eq16}
  X^H Q X T = T \mathrm{\Sigma}^2.
\end{equation}
The matrix $\mathrm{\Sigma}$ contains the non-negative eigenvalues
along its diagonal, ranked in decreasing order, and the columns of $T$ provide the corresponding
eigenvectors. The POD-modes, and thus the basis matrix $W,$ is then
formed following
\begin{equation}
  \label{eq:eq17}
  W = X T \mathrm{\Sigma }^{-\frac{1}{2}}.
\end{equation}
By construction, the columns of $W$ are orthonormal; the corresponding
matrix $W$ satisfies the orthonormality condition $W^H Q W = I.$ The POD-modes associated to the largest eigenvalues $ \Sigma $ form an optimal basis to represent the snapshots contained in $X$. 

\subsection{The constant term}
\label{sec:const}

The constant term $c$ in the reduced-order model can easily be
determined from the constant term of the full-scale
equation~(\ref{eq:eq13}). It is simply obtained by a left-multiplication
of the full-scale constant term $b$ by $W^H$ which is equivalent to a
Galerkin projection of $b$ onto the basis spanned by the columns of
$W,$ i.e. the POD-modes. Mathematically, we have
\begin{equation}
  \label{eq:eq18}
  c = W^H b.
\end{equation}

\subsection{The linear term}
\label{sec:lin}

The projection of the linear term onto the reduced-order basis $W$ yields
a reduced system matrix $L,$
that may be computed following:
\begin{equation}
  \label{eq:eq19}
  L = W^H A W.
\end{equation}
The number of POD modes in $W$ (i.e., the number of columns of $W$) may
be chosen such that the spectrum of the reduced matrix $L$ displays
unstable eigenvalues close to the unstable eigenvalues of the
large-scale generalized eigenproblem defined by $(A,Q)$. 

\subsection{The nonlinear term}
\label{sec:nonlin}

While there is little choice for the reduced expression of the constant
and linear terms, the treatment of the nonlinear terms allows more choice
and flexibility. In this article, three different techniques are
investigated: the traditional Galerkin projection of the nonlinear terms
onto the reduced basis $W$ (method 1, see section \S~\ref{sec:Gal1}), the Galerkin
projection of the nonlinear terms onto a new dedicated basis $F$
representing only the nonlinearities (method 2, see section \S~\ref{sec:Gal2})
and the Discrete Empirical Interpolation Method (DEIM) applied to the
dedicated basis $F$ (method 3, see section~\S~\ref{sec:DEIM}).

\subsubsection{Method 1: traditional Galerkin projection}
\label{sec:Gal1}

This method follows the traditional derivation of a nonlinear
dynamical system for the coefficients $z.$ In it, the
arguments of the bilinear function $f$ are expressed in terms of
their Galerkin expansion, after which the nonlinear expression
is left-multiplied by the matrix $W^H$ which is equivalent to a
projection onto our orthonormal basis. We obtain the nonlinear
coefficients $N_{ijk}$ as
\begin{equation}
  \label{eq:eq20}
  N_{ijk} = W_{:,i}^H Q f\left( W_{:,j}, W_{:,k} \right)
\end{equation}
where we have used the common notation $W_{:,i}$ indicating
the $i$-th column of $W.$ Since $f$ is symmetric in its arguments,
we have $N_{ijk} = N_{ikj}$ for all indices ${}_{ijk}.$

\subsubsection{Method 2: Galerkin projection with an additional
nonlinear basis}
\label{sec:Gal2}

Besides the common basis extracted
from the snapshot sequence $X,$ the second method introduces a second basis intended to
represent the nonlinear terms $f(w,w)$ and thus nonlinear effects.
For this, we evaluate the nonlinearities for all snapshots in $X$
to form a second snapshot sequence which we refer to as $Y = f(X,X).$
Analogous to the first method, the correlation matrix based on
this second sequence $Y$ is then decomposed into its eigenvalues
and eigenvectors according to
\begin{equation}
  \label{eq:eq21}
  Y^H Q Y U = U {\mathrm{\Gamma}}^{2},
\end{equation}
and the corresponding POD modes $F$ are determined as
\begin{equation}
  \label{eq:eq22}
  F = Y U {\mathrm{\Gamma}}^{-\frac{1}{2}}.
\end{equation}
As before, the structures contained in the columns of $F$ are
orthonormal by construction, which is expressed mathematically as
$F^H Q F = I.$

The nonlinear terms may then be projected onto the $F$ basis
\begin{equation}
  \label{eq:eq23}
  f(Wz,Wz) = F \hat{f}
\end{equation}
with $\hat{f} = F^H Q f(Wz,Wz)$ as the coefficient
vector. Using this new basis to express the nonlinearities in
the governing equations, we arrive at the nonlinear terms of our
reduced-order model as
\begin{equation}
  \label{eq:eq24}
  N_{ijk} = W_{:,i}^H Q F F^H Q f(W_{:,j},W_{:,k}).
\end{equation}
Again, the symmetric nature of $f$ leads to $N_{ijk} = N_{ikj}.$

\subsubsection{Method 3: Discrete Empirical Interpolation Method
with an additional nonlinear basis}
\label{sec:DEIM}

The basis $F$ introduced in section \S~\ref{sec:Gal2} for the
representation of the nonlinear term is considered again, but the
coefficients $\hat{f}$ are obtained differently. If $F$ contains
$q$ columns representing $q$ structures onto which we
project the nonlinear term, we enforce equation $f(Wz,Wz) = F \hat{f}$ not in
a least-squares sense as above, but instead by enforcing equality
at $q$ selected interpolation points. Mathematically, we
premultiply the above equation by a row-selector matrix $P^H$
which yields
\begin{equation}
  \label{eq:eq25}
  P^H f(Wz,Wz) = P^H F \hat{f}.
\end{equation}
The matrix $P$ contains $q$ columns, each displaying a single
unit value at some row with the remaining entries as zero.
The choice of these $q$ columns and interpolation points
follows a greedy algorithm and is given below. The
premultiplication by $P^H$ ensures that $P^H F$ is invertible
and, as a consequence, the above equation can be solved for
the coefficient vector $\hat{f}.$ We thus have $\hat{f} = (P^H
F)^{-1} P^H f(Wz,Wz)$ and invoking the bilinearity of the nonlinear operator:
\begin{equation}
P^H f(Wz,Wz)=P^HF\left(\sum_j W_{:,j}z_j,\sum_k W_{:,k}z_k\right)=\sum_{j,k} z_jz_kP^Hf(W_{:,j}, W_{:,k}),
\end{equation}
we obtain the following representation of the nonlinear term
\begin{equation}
  \label{eq:eq26}
  N_{ijk} = W_{:,i}^H Q F (P^H F)^{-1} P^H f(W_{:,j},W_{:,k}).
\end{equation}
Note that this expression does not require the nonlinear term to be point-wise. It is the bilinearity of the nonlinear operator that ensures an operation count of the order $p^3$.

In the case of a point-wise nonlinearity, the evaluation of the nonlinear term in the reduced order model can be achieved at a very small cost, that is the evaluation of the nonlinearity at $q$ points $ P^H f(Wz,Wz)= \tilde{f}(P^ HWz)$, where $\tilde{f}$ is a nonlinear operator taking a vector of size $q$ and giving back a vector of size $q$.
In such a case, the full reduced-order model is:
\begin{equation}
\frac{dz}{dt}=c+Lz+N_1\tilde{f}(N_2z),
\end{equation}
where $N_1=W^HQF(P^HF)^{-1}$ and $N_2=P^H W$ are matrices of size $(p,q) $ and $(q,p)$ respectively. Hence, the operation count scales as $\max(p^2,pq)$.
This implementation is consistent with
equation~(\ref{eq:eq26}).
If the nonlinearity is not pointwise, the  sparsity argument of \citep{chaturantabut2010nonlinear} or the introduction of auxiliary variables \citep{fosasde2016nonlinear}
may also lead to a reduction of the evaluation cost of the nonlinear term.

\begin{center}
  \begin{algorithm}[H]
    \KwData{nonlinear basis $F \in \mathbb{R}^{n\times q}$}
    \KwResult{row-selector matrix $P \in \mathbb{R}^{n\times q}$}
    $P$ = {\tt{zeros}}($n,q$)\;
    $n_{\max}$ = {\tt{argmax}}($\vert F_{:,1} \vert$)\;
    $P_{n_{\max},1}$ = $1$\;
    \For{$j$ = $2:q$}{
    $\hat{f}$ = $[P_{:,1:j-1} F_{:,1:j-1}]^{-1} [P_{:,1:j-1} F_{:,j}]$\;
    $r$ = $F_{:,j} - F_{:,1:j-1} \ \hat{f}$\;
    $n_{\max}$ = {\tt{argmax}}($\vert r \vert$)\;
    $P_{n_{\max},j}$ = $1$\;
    }
    {\tt{return}} $P$\;
    \caption{DEIM algorithm (adapted from \cite{chaturantabut2010nonlinear}).}
  \end{algorithm}
\end{center}

\subsection{Mathematical properties of the reduced-order models}
\label{sec:mathprop}

When comparing the stability, accuracy and robustness of various
model reduction techniques, it is imperative to introduce quality measures
and other mathematical properties to quantitatively assess their absolute
and relative performance. The kinetic perturbation energy will serve as
the quantity that will be monitored and compared for the full and the
reduced-order model.

The kinetic energy $K = z^H z/2$ of the reduced-order
model is governed by
\begin{equation}
  \label{eq:eq27}
  \frac{d K}{dt} = z^H c + z^H L z +
  \sum_{i,j,k} N_{ijk} z_i z_j z_k.
\end{equation}
If we define the symmetric and antisymmetric parts of $L_{ij}$
and $N_{ijk}$ as follows (with superscript ${}^S$ denoting the
symmetric and superscript ${}^A$ denoting the anti-symmetric
part)
\begin{subeqnarray}
  L_{ij}^S = \displaystyle{\frac{L_{ij} + L_{ji}}{2}}, \ 
  L_{ij}^A = \displaystyle{\frac{L_{ij} - L_{ji}}{2}}, \\
  N_{ijk}^S = \displaystyle{\frac{N_{ijk} + N_{ikj} + N_{jik} + N_{jki} +
  N_{kij} + N_{kji}}{6}},\ 
  N_{ijk}^A = \displaystyle{\frac{5 N_{ijk} - N_{ikj} - N_{jik} - N_{jki} -
  N_{kij} - N_{kji}}{6}}
\end{subeqnarray}
we can recast the evolution equation for the energy $K$ as
\begin{equation}
  \label{eq:eq32}
  \frac{d K}{dt} = z^H c + z^H L^S z
  + \sum_{i,j,k} N_{ijk}^S z_i z_j z_k.
\end{equation}
The energy evolution equation leads to the following tight bounds (i.e., there exist $ z$ that achieve the bound):
\begin{equation}
  \label{eq:eq33}
  \vert K \vert \leq \sqrt{2 K} \ \Vert c \Vert + 2 K \ \Vert L^S
  \Vert_F + \sqrt{(2K)^3} \ \Vert N^S \Vert_F
\end{equation}
where $\Vert \cdot \Vert_F$ stands for the Frobenius norm of a matrix
or a tensor, defined as
\begin{equation}
  \label{eq:eq34}
  \Vert L \Vert_F = \sqrt{\sum_{ij} L_{ij}^2}, \qquad \qquad \qquad
  \Vert N \Vert_F = \sqrt{\sum_{ijk} N_{ijk}^2}.
\end{equation}
The first term on the right-hand-side of equation (\ref{eq:eq33}) may
lead to an algebraic increase of $K$ of the form
\begin{equation}
  \label{eq:eq35}
  K \sim K_0 \left( 1 + \frac{\Vert c \Vert}
  {\sqrt{2 K_0}} \ t\right)^2,
\end{equation}
the second term to an exponential growth following
\begin{equation}
  \label{eq:eq36}
  K \sim K_0 \exp \left( 2 \Vert L^S \Vert_F \ t \right),
\end{equation}
and the third term to a finite-time blow-up according to
\begin{equation}
  \label{eq:eq37}
  K \sim \frac{1}{\left(\sqrt{1/K_0} - \sqrt{2}
  \Vert N^S \Vert_F \ t\right)^2}.
\end{equation}
The nonlinearities $F$ that we consider in this article are of purely
convective type and are therefore energy-preserving (under standard
boundary conditions). It is straightforward to show that
\begin{equation}
  \label{eq:38}
  \iint u \cdot (u \cdot \nabla u) \ dx \ dy = 0,
\end{equation}
which implies for the reduced-order model that
\begin{equation}
  \label{eq:eq39}
  N^S = 0.
\end{equation}
If this condition holds exactly, the finite-time singularity implied by
the third term in equation~(\ref{eq:eq33}) can be avoided; if, on the
other hand, the reduced-order model does not satisfy $N^S = 0,$ one
should expect a finite-time blow-up in energy for some initial
condition. The condition $N^S = 0$ thus constitutes an important and
effective test for the robustness and long-term stability of the reduced-order model.

\subsection{Quality measures of reduced-order models}
\label{sec:crit}

In this subsection, $w$ refers to either $w'$ if the BF formulation is chosen or $ w''$ if the MF formulation is selected. For a quantitative assessment of the quality of model reduction, we
introduce various error measures.

We first look for a fixed point $ z_b $ of the reduced-order model using a Newton method to solve:
\begin{equation}
c_i + \sum_j L_{ij} z_{b,j} +
  \sum_{j,k} N_{ijk} z_{b,j} z_{b,k}=0
\end{equation}
We evaluate the relative error between the predicted base-flow $Wz_{b}$ and the actual base-flow $ w_b $ ($w'_b$ with the BF formulation and $w''_b $ with the MF formulation):
\begin{equation}
   \epsilon_{w_b}= \sqrt{\frac{(Wz_{b}-w_b)^HQ(Wz_{b}-w_b)}{w_b^HQ{w_b}}}.
\end{equation}
We compute the leading eigenvalues/eigenvectors $ (\hat{\lambda}_i^{BF},\hat{z}_i^{BF} ) $ of the operator obtained by linearizing the equations governing the reduced-order model \eqref{eq:eq14} around $ z_b $.
The number of unstable eigenvalues is denoted $ \nu_\lambda^{BF}$ and should be equal to $ 2$ for any formulation.  We evaluate both the relative error between the leading eigenvalue $ \hat{\lambda}_{\mbox{max}}^{BF}$ and $\lambda^{BF}$ and the relative alignment error between the predicted leading eigenvector $W\hat{z}_{\mbox{max}}^{BF}$ and the actual one $ \hat{w}^{BF} $ following:
\begin{eqnarray}
  \label{eq:eq41}
  \epsilon_\lambda^{BF} &=& \frac{\vert \hat{\lambda}_{\mbox{max}}^{BF} - \lambda^{BF} \vert}
          {\vert \lambda^{BF} \vert} \\
    \epsilon_{\hat{w}}^{BF}&=&1-\frac{\left|(W\hat{z}_{\mbox{max}}^{BF})^HQ\hat{w}^{BF}\right|}{\sqrt{(W\hat{z}_{\mbox{max}}^{BF})^HQW\hat{z}_{\mbox{max}}^{BF}}\sqrt{\hat{w}^{BF,H}Q{\hat{w}^{BF}}}}.
\end{eqnarray}
We then compute the mean-flow $ \overline{z}$ by time-averaging the results of a temporal simulation of the reduced-order model \eqref{eq:eq14}:
\begin{equation}
\overline{z}=\lim_{T\rightarrow \infty} \int_0^T z(t) dt.
\end{equation}
In a similar way then before, we define the relative errors pertaining to the mean-flow and the leading eigenvalue / eigenvector of equations \eqref{eq:eq14} linearized around the mean-flow $ \overline{z} $:
\begin{eqnarray}
   \epsilon_{\overline{w}}&=& \sqrt{\frac{(W\overline{z}-\overline{w})^HQ(W\overline{z}-\overline{w})}{\overline{w}^HQ\overline{w}}} \\
  \epsilon_\lambda^{MF} &=& \frac{\vert \hat{\lambda}_{\mbox{max}}^{MF} - \lambda^{MF} \vert}
          {\vert \lambda^{MF} \vert} \\
    \epsilon_{\hat{w}}^{MF}&=&1-\frac{\left|(W\hat{z}_{\mbox{max}}^{MF})^HQ\hat{w}^{MF}\right|}{\sqrt{(W\hat{z}_{\mbox{max}}^{MF})^HQW\hat{z}_{\mbox{max}}^{MF}}\sqrt{\hat{w}^{MF,H}Q{\hat{w}^{MF}}}}.
\end{eqnarray}
The number of unstable eigenvalues $ \nu_\lambda^{MF}$ should again be equal to two.
The recovery of the least-damped eigenvalues $\lambda^{BF}$ and $ \lambda^{MF}$ is an important quality
measure for the model reduction procedure since, in the transient regime (TR),
the solution behaves like $\exp(\lambda^{BF} t)$, while in the limit-cycle
case (LC), $\lambda^{MF}$ displays near marginal stability properties with a frequency
close to the frequency of the limit-cycle~\cite{barkley2006linear}.

The remaining two errors, proposed for the assessment of model
reduction quality, are errors linked to the reconstruction of the TR, LC and MFTR trajectories.
In each case, the initial conditions for the
reduced-order models are set to those of
the large-scale simulation,
\begin{equation}
  \label{eq:eq40}
  z(t=t_0) = \tilde{z}(t=t_0),
\end{equation}
where $\tilde{z}$ denotes the projection of the snapshots of the unsteady simulation onto the POD modes representing the state: 
\begin{equation}
  \label{eq:eq44}
  \tilde{z} = W^H Q w.
\end{equation}
The mean truncation error $\epsilon_t$, which is the ratio between the integral energy (over time) of the retained $ p$ modes and the integral total energy, is given by
\begin{equation}
  \label{eq:eq42}
  \epsilon_t = \left( \frac{\sum_{j=0}^{T/\Delta t_S} \sum_{i \geq
      p+1} \left\Vert \tilde{z}_i(t_0 + j\Delta t_S) \right\Vert^2}
          {\sum_{j=0}^{T/\Delta t_S} \sum_{i\geq 1} \left\Vert \tilde{z}_i
  (t_0 + j\Delta t_S) \right\Vert^2}\right)^{1/2},
\end{equation}
and the relative mean model error $\epsilon_m$ by:
\begin{equation}
  \label{eq:eq43}
  \epsilon_m = \left( \frac{\sum_{j=0}^{T/\Delta t_S} \sum_{i=1}^p
    \left\Vert z_i(t_0 + j\Delta t_S) - \tilde{z}_i(t_0 + j\Delta
    t_S) \right\Vert^2} {\sum_{j=0}^{T/\Delta t_S} \sum_{i=1}^p
    \left\Vert \tilde{z}_i(t_0 + j\Delta
    t_S) \right\Vert^2}\right)^{1/2}.
\end{equation}
$\epsilon_t$ and $\epsilon_m,$ respectively, assess the error due to truncation of the projection basis to $ p$ POD modes and the error due to prediction of the time-evolution of the $ p $ retained states of the model. In \cite{alomar2020reduced}, it was shown that the total error $ \epsilon_{total}$ can be deduced from 
$ \epsilon_t$ and $ \epsilon_m $ following $ \epsilon_{total}=\epsilon_t+\epsilon_m(1-\epsilon_t)$.

Finally, we introduce the ratio $\epsilon_S$ between the Frobenius
norm of the symmetric part $N^S$ of the nonlinear term $N$ and the
Frobenius norm of $N$
\begin{equation}
  \label{eq:eq46}
  \epsilon_S = \frac{\Vert N^S \Vert_F}{\Vert N \Vert_F}.
\end{equation}
With the nonlinearities representing the convective term of the
Navier-Stokes equations, this ratio should be small (or zero). In
contrast, a large value of this quantity may indicate robustness
problems, since the presence of a non-zero $N^S$-term in the model may
induce a finite-time singularity for some initial conditions. We have therefore also evaluated the performance of the models when considering $ N_{ijk}^A$ in the reduced-order-model instead of $N_{ijk}$, that is when removing the symmetric part of the quadratic term. In this case, we denote the model error $ \epsilon_m'$.

\section{Kuramoto-Sivashinsky equation}
\label{sec:KuramotoS}

In this section, we assess the quality of the reduced-order models
 for the case of the
Kuramoto-Sivashinsky equation presented in section
\S~\ref{sec:KS}. After describing the POD bases for the state  and
the nonlinearities (section \S~\ref{sec:KS_POD}), we analyze
the different reduction techniques presented in section
\S~\ref{sec:ModelRed} (section \S~\ref{sec:KS_ModRed}).

\subsection{POD bases}
\label{sec:KS_POD}

In  the following subsections, we present the state and nonlinear POD modes associated with two different time-spans of the learning trajectory.

\subsubsection{Bases $W$ and $F$ determined from snapshots on the transient and the limit-cycle ($ 0\leq t \leq 150$)}

\begin{figure}[htbp]
  \begin{center}
    \begin{tabular}{cc}(a)&(b) \\
         \psfrag{n}{$i$}
     \psfrag{s}{$\mathrm{\Sigma}^{'2}_i,w'$}
     \includegraphics[width=0.45\textwidth]{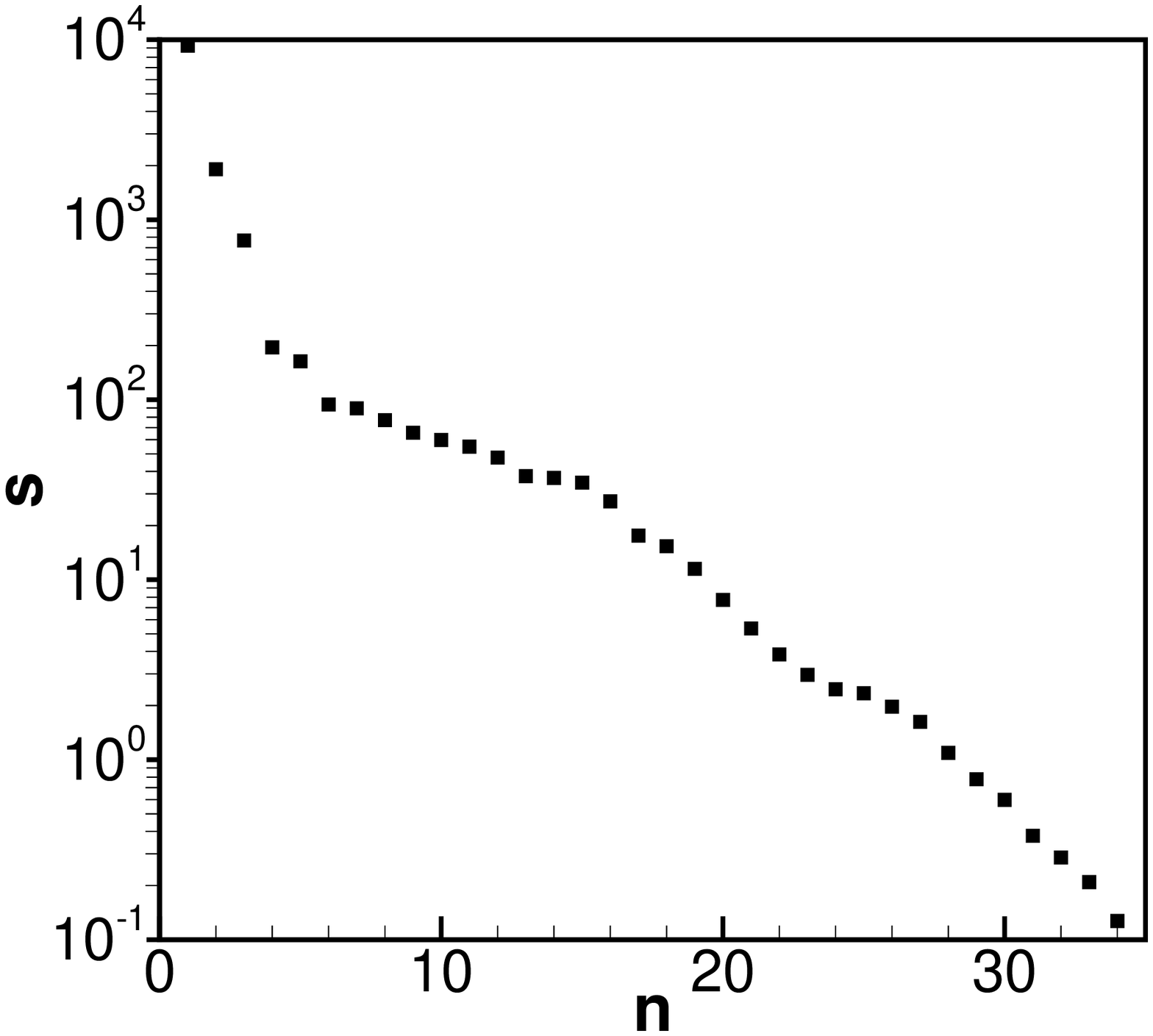} & 
              \psfrag{n}{$i$}
\psfrag{s}{$\mathrm{\Gamma}^{'2}_i,f(w',w')$}
     \includegraphics[width=0.45\textwidth]{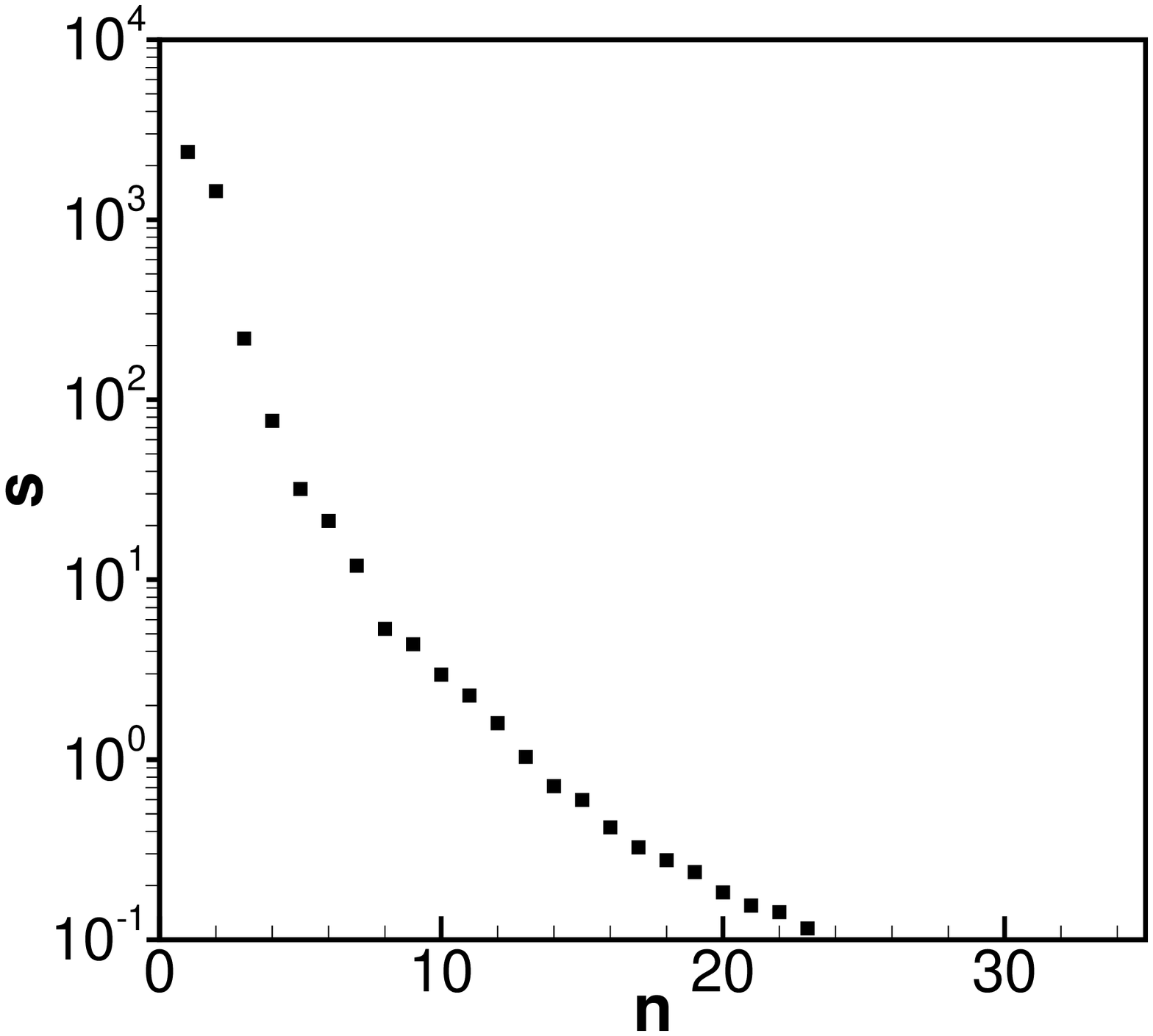} 
     \\ (c) & (d) \\
     \psfrag{a}{$W_{:,i=1}$}
     \psfrag{b}{$W_{:,i=2}$}
     \psfrag{c}{$W_{:,i=3}$}
     \psfrag{x}{$x$}
     \psfrag{w}{$u'$}
     \includegraphics[width=0.45\textwidth]{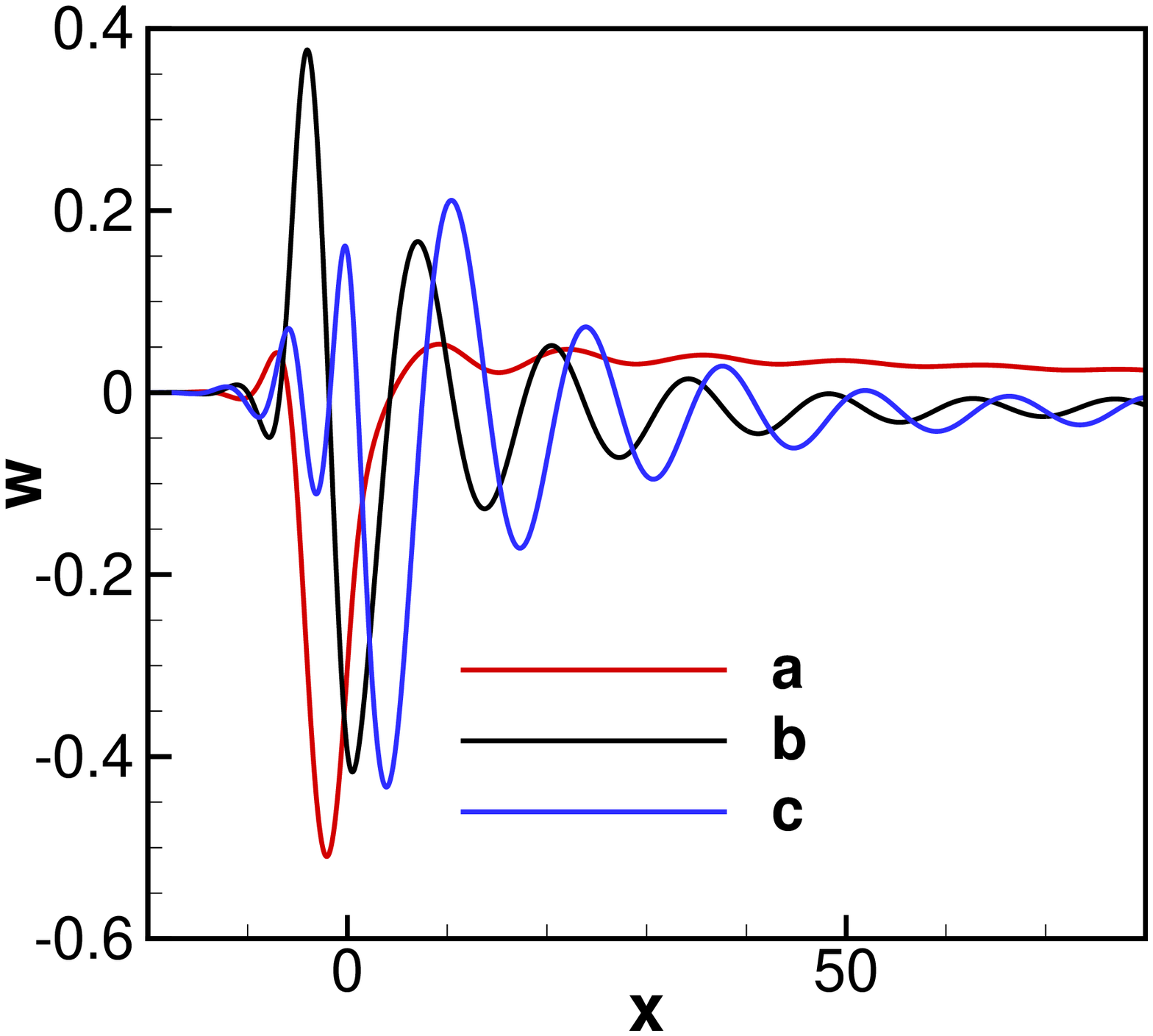} & 
     \psfrag{a}{$F_{:,i=1}$}
     \psfrag{b}{$F_{:,i=2}$}
     \psfrag{c}{$F_{:,i=3}$}
     \psfrag{x}{$x$}
     \psfrag{w}{$-u'\partial_x u'$}
  \includegraphics[width=0.45\textwidth]{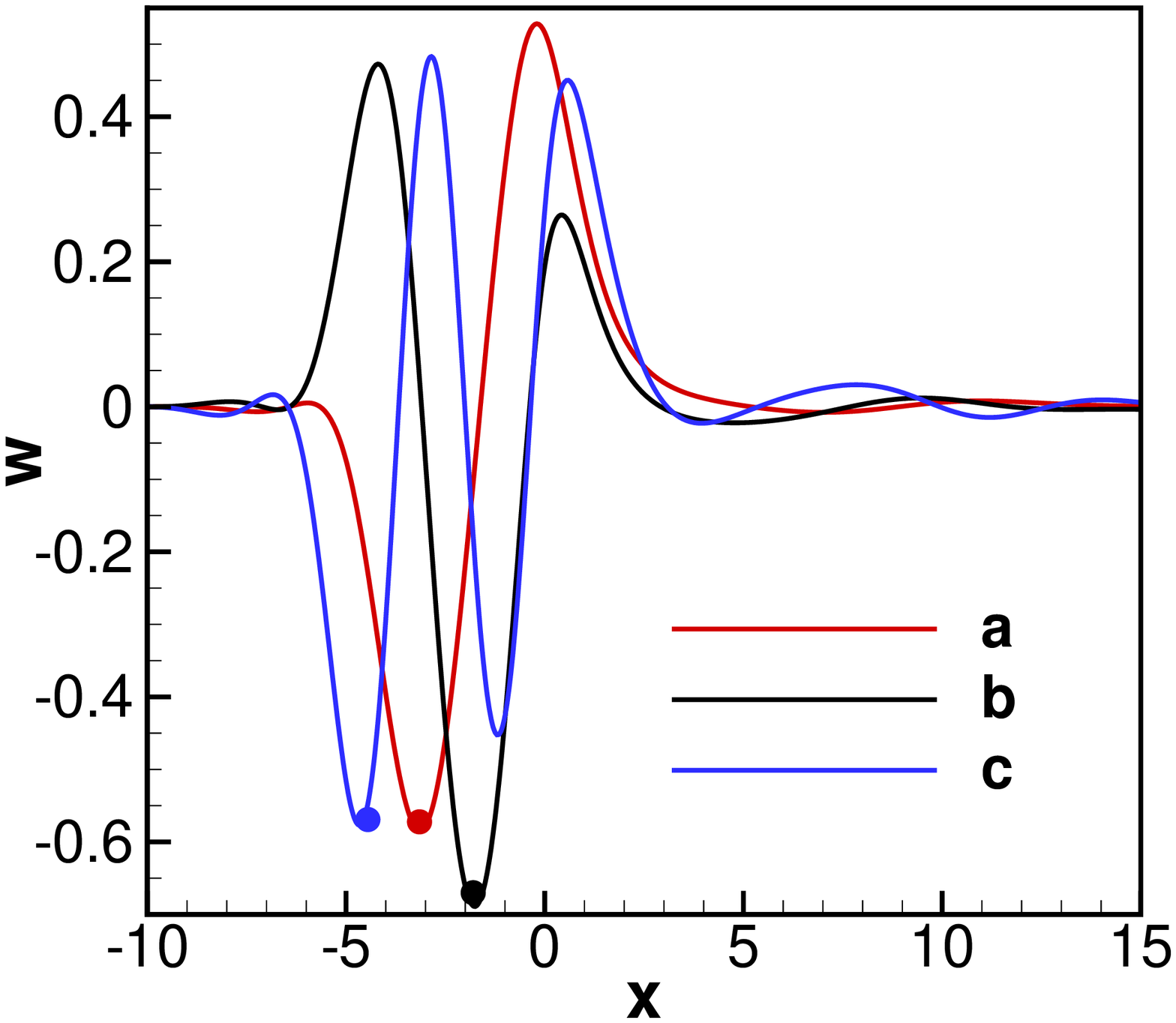}
     \end{tabular}
    \end{center}
    \caption{\label{fig:KSresults} Model reduction of Kuramoto-Sivashinsky equation with projection bases obtained from snapshots within $ 0 \leq t \leq 150 $ for the BF formulation. (a,b): Eigenvalues of the
    correlation matrices for the snapshots representing (a): the
    state and (b): the nonlinearity. (c,d): first three POD modes of (c): state
    and (d): nonlinearity. Colored circles show the corresponding DEIM points (only the
    $x$ position is relevant).}
\end{figure}
We take $751$ snapshots from the simulation shown in figures
\ref{fig:KS4}(c--e) over the time-span $ 0\leq t \leq 150$ to build representative bases $W$ and $F$ of the transient and the limit-cycle. The sampling time is $\Delta t_S = 150/750 =
0.2$ time units. Based on these snapshots, the leading eigenvalues of
the correlation matrices for the state and the nonlinearities
 are shown, for the BF formulation, in figures
\ref{fig:KSresults}(a,b) . We observe that the eigenvalues of the
correlation matrix for the nonlinearity snapshots drop off markedly
faster, indicating that the effective dimensionality of the space
spanned by nonlinearity snapshots is lower than the analogous
dimensionality formed by the state snapshots.

In figures~\ref{fig:KSresults}(c,d), we present the first three POD
modes of the state $w'$ and of the nonlinearity $f(w',w')$. We observe that, while the state POD modes are
non-zero over the entire region $x > -10,$ the support of the
nonlinearity POD modes is far more compact (within $-10 < x <
10$). This result is consistent with the pronounced drop-off in the
eigenvalues of the respective correlation matrices shown in
figure~\ref{fig:KSresults}(a,b).
It is also seen that the first state POD mode accounts for the mean-flow deformation $ \overline{w'}$ shown in fig. \ref{fig:KS4}(f). The other state POD modes allow us to represent, in an optimal way, the unstable growing mode shown in fig. \ref{fig:KS4}(b) with a black solid line as well as the time-oscillations depicted in \ref{fig:KS4}(e).
In figure~\ref{fig:KSresults}(d), we have displayed also the associated DEIM
points with symbols. As is evident from the plot, the DEIM points approximately
correspond to the location of maximum amplitude of the underlying POD
modes. 

\subsubsection{Bases $W$ and $F$ determined from snapshots on the limit-cycle ($ 150\leq t \leq 300$)}

\begin{figure}[htbp]
  \begin{center}
    \begin{tabular}{cc}(a)&(b) \\
         \psfrag{n}{$i$}
     \psfrag{s}{$\mathrm{\Sigma}^{''2}_i,w''$}
     \includegraphics[width=0.45\textwidth]{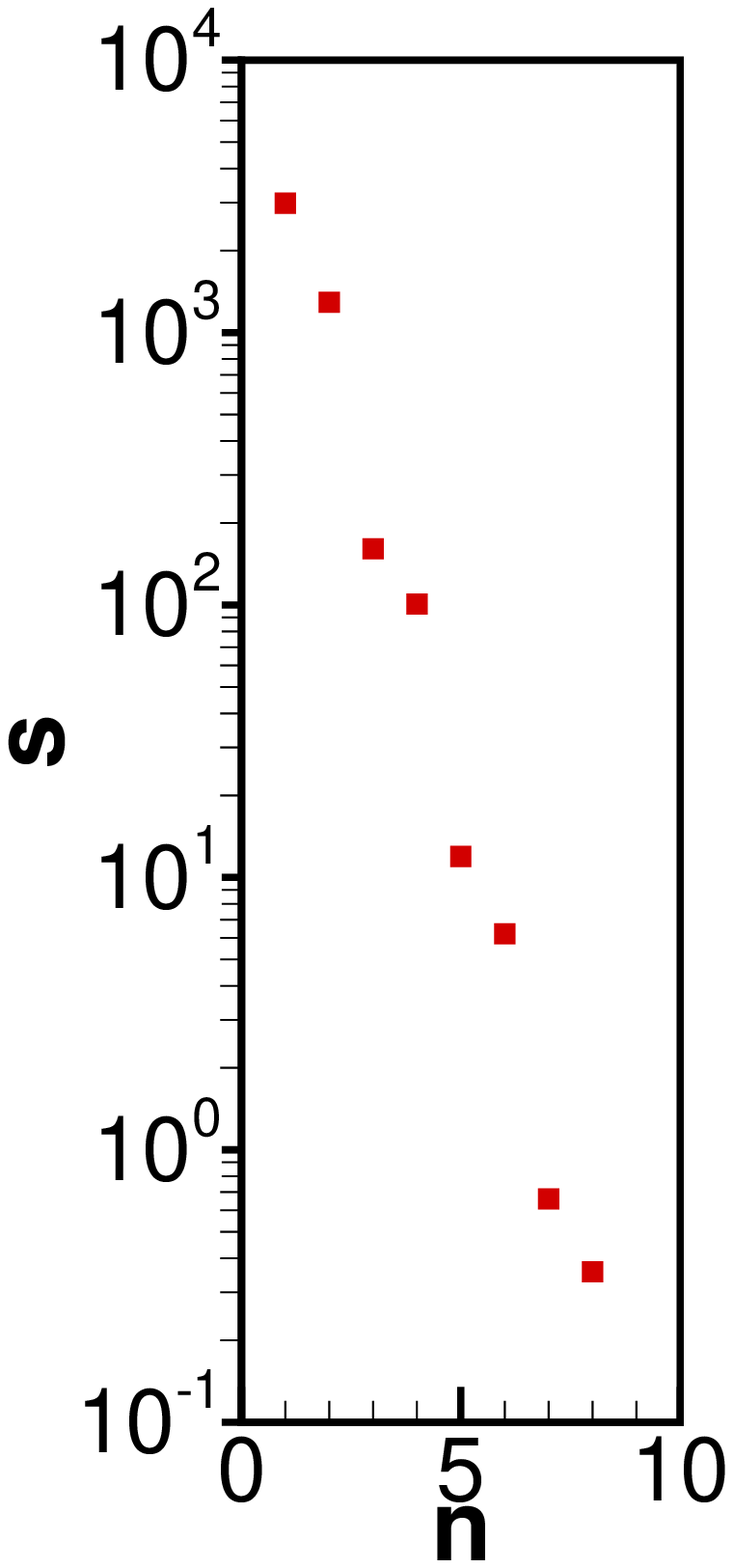} &
              \psfrag{n}{$i$}
\psfrag{s}{$\mathrm{\Gamma}^{''2}_i,f(w'',w'')$}
     \includegraphics[width=0.45\textwidth]{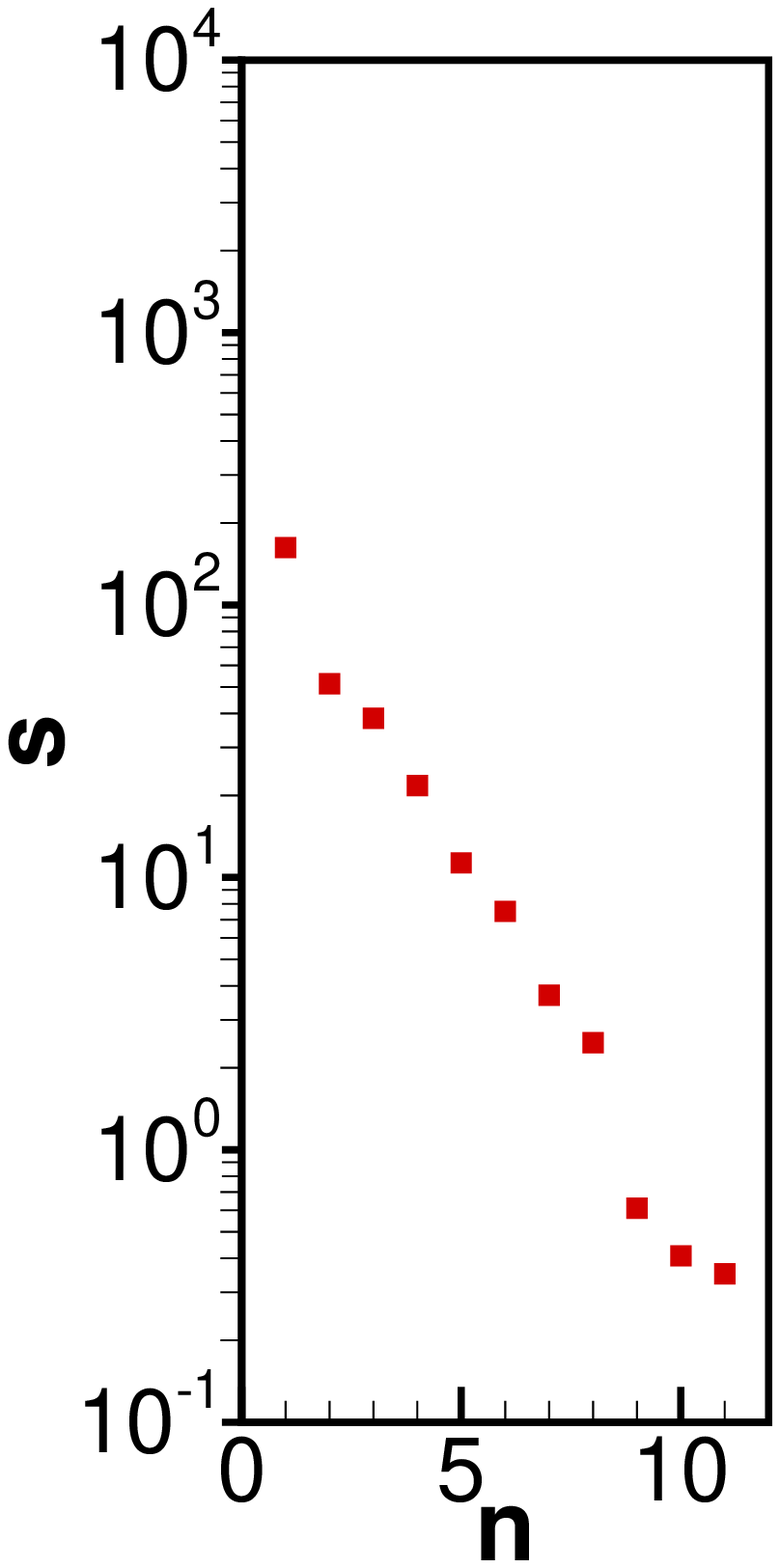} 
   \\ (c) & (d) \\
     \psfrag{a}{$W_{:,i=1}$}
     \psfrag{b}{$W_{:,i=2}$}
     \psfrag{c}{$W_{:,i=3}$}
     \psfrag{x}{$x$}
     \psfrag{w}{$u''$}
     \includegraphics[width=0.45\textwidth]{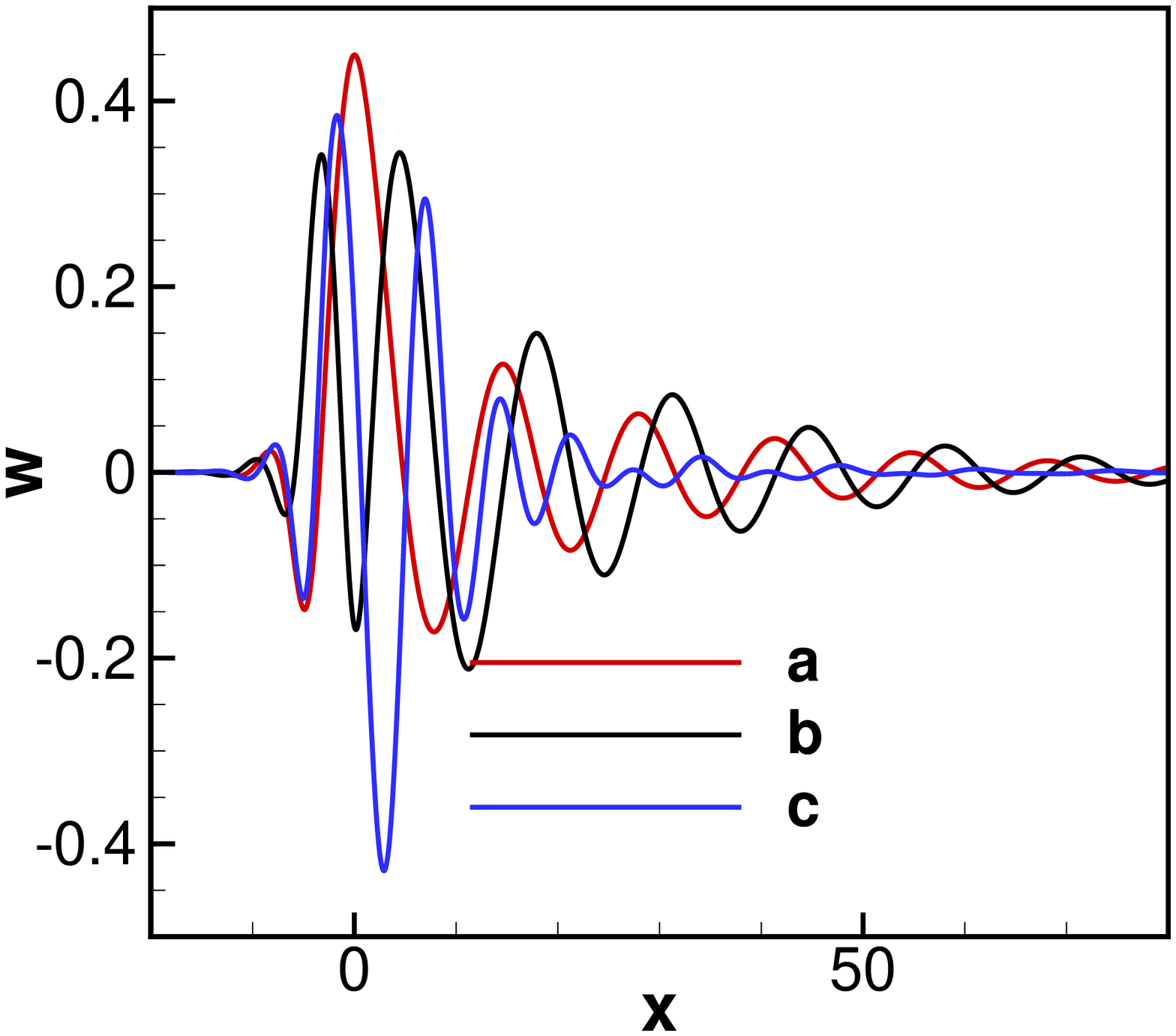} &
     \psfrag{a}{$F_{:,i=1}$}
     \psfrag{b}{$F_{:,i=2}$}
     \psfrag{c}{$F_{:,i=3}$}
     \psfrag{x}{$x$}
     \psfrag{w}{$-u''\partial_x u''$}
  \includegraphics[width=0.45\textwidth]{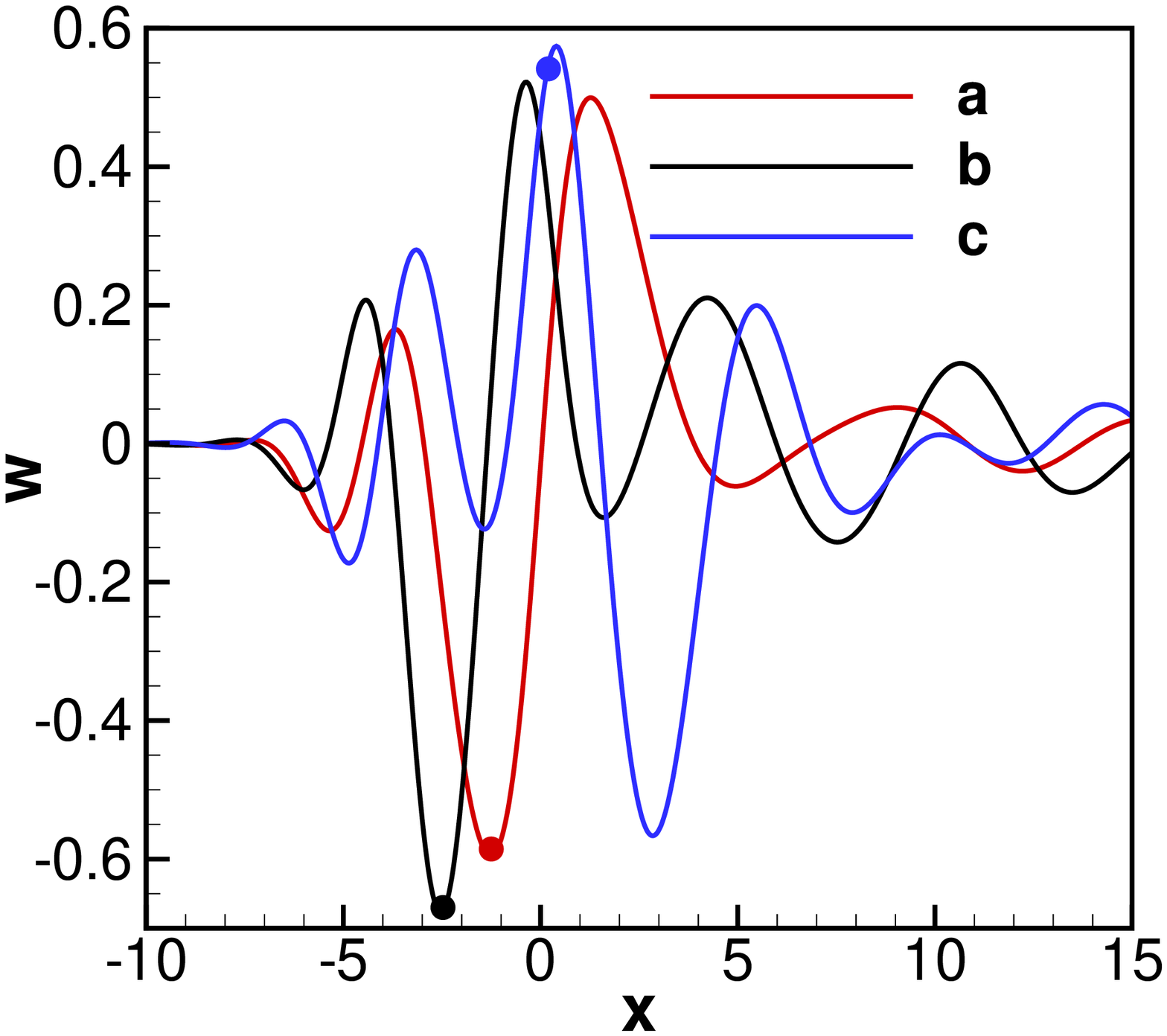}
       \end{tabular}
    \end{center}
    \caption{\label{fig:KSresults2} Model reduction of Kuramoto-Sivashinsky equation with projection bases obtained from snapshots within $ 150 \leq t \leq 300 $ for the MF formulation. Same caption as in fig. \ref{fig:KSresults}.
    }
\end{figure}

We take again $751$ snapshots from the simulation shown in figures
\ref{fig:KS4}(c--e) but over the time-span $ 150\leq t \leq 300$ (not represented in the figure),
which is only representative of the limit-cycling behavior. The sampling time remains unchanged, and the results for the MF formulation (which is consistent with the sole knowledge of the limit-cycle solution) are shown in fig. \ref{fig:KSresults2} in a similar way than in fig. \ref{fig:KSresults}. 
The dimensionality of the dynamics is drastically reduced in both cases, since only ten POD modes are required to achieve a decrease of four orders of magnitude of the eigenvalues (compared to approximately 30 POD modes in the previous section).
The eigenvalues for the state variable $w''$ approximately come in pairs and therefore represent spatio-temporal structures that are smoothly convected downstream.
The state-POD mode representing the mean-flow deformation has disappeared since this feature is not visible in the $w''$ variable over the time-span $ 150\leq t \leq 300$ of the learning trajectory. Only POD modes representing limit-cycle oscillations are seen in fig. \ref{fig:KSresults2}(c). 

\subsection{Model reduction}
\label{sec:KS_ModRed}

Based on the snapshot bases for the state and the nonlinearities, we
are now in a position to explore the various reduced-order models and compare them with the corresponding
large-scale unsteady simulations.
We will consider for that the TR, LC and MFTR trajectories presented in fig. \ref{fig:KS4}(c,d).
For
the first reduction method (Galerkin projection with a single basis), we choose a
number $p$ of POD modes to represent the state ($w'$ or $ w'' $ following the chosen formulation). For the second
method (Galerkin projection with two bases) and third method
(Galerkin projection for the linear term and DEIM for the nonlinear
term), we choose the number of POD modes for either basis, labelled
$p$ and $q$.

\begin{table}[htbp]
  \centering
  \begin{tabular}{|l|c|cccc|cccc|ccc|ccc|c|}
  \hline
  & & \multicolumn{4}{|c|}{BF} &  \multicolumn{4}{|c|}{MF} & \multicolumn{3}{|c|}{TR} & \multicolumn{3}{|c|}{LC} & \multicolumn{1}{|c|}{MFTR}\\
MF-$p$-$q$& $\epsilon_S$
& $\epsilon_{w_b}$ & $\nu_{\lambda}$ & $\epsilon_{\lambda}$ & $\epsilon_{\hat{w}}$
& $\epsilon_{\bar{w}}$ & $\nu_{\lambda}$ &$\epsilon_{\lambda}$ & $\epsilon_{\hat{w}}$ 
& $ \epsilon_t$ & $ \epsilon_m$ & $ \epsilon_m' $
& $ \epsilon_t$ & $ \epsilon_m$ & $ \epsilon_m'$
& $ \epsilon_m$
\\
\hline
1B-60& \ora{12} & $\grn{0.00}$ & \grn{2} & \grn{0.1} & \grn{0.00} & \grn{2} & \grn{2}&\grn{1}&\grn{0.1} & \grn{0.01}&\grn{0.5}&\grn{0.5} &  \grn{0.00} &\grn{0.1}&\grn{0.3} & \grn{1}\\
1B-50& \ora{13} & $\grn{0.00}$ & \grn{2} & \grn{0.5} & \grn{0.00} & \grn{2} & \grn{2}&\grn{1}&\grn{0.1} & \grn{0.03}&\grn{3}&\grn{3} &  \grn{0.02} &\grn{0.2}&\grn{0.3} &\grn{3}\\
1B-40& \ora{14} & $\grn{0.00}$ & \grn{2} & \grn{2} & \grn{0.03} & \grn{2} & \grn{2}&\grn{1}&\grn{0.2} & \grn{0.2}&\ora{13}&\ora{13} &  \grn{0.1} &\grn{3}&\grn{1} &\ora{38}\\
1B-30& \ora{12} & $\grn{0.00}$ & \grn{2} & \grn{2} & \grn{10} & \grn{0.3} & \grn{2}&\grn{2}&\grn{2} & \grn{1}&\red{85}&\red{85} &  \grn{1} &\grn{3}&\grn{4} &\ora{49}\\
1B-20& \grn{4} & $\grn{0.00}$ & \red{8} & \ora{45} & \ora{15} & \grn{4} & \grn{2}&\grn{10}&\grn{3} & \grn{4}&\red{103}&\red{103} &  \grn{4} &\grn{9}&\grn{9} &\ora{27}\\
1B-10& \grn{0.3} & $\grn{0.00}$ & \red{4} & \ora{45} & \ora{15} & \ora{29} & \grn{2}&\ora{28}&\grn{10} & \ora{19}&\red{100}&\red{100} &  \grn{7} &\red{75}&\red{75} &\red{81}\\
1M-60& \ora{12} & $\grn{0.01}$ & \grn{2} & \grn{0.1} & \grn{0.00} & \grn{2} & \grn{2}&\grn{1}&\grn{0.1} & \grn{0.01}&\grn{2}&\grn{8} &  \grn{0.01} &\grn{0.1}&\grn{0.1} &\grn{1}\\
1M-40& \ora{14} & $\grn{2}$ & \grn{2} & \grn{3} & \grn{0.04} & \grn{2} & \grn{2}&\grn{1}&\grn{0.2} & \grn{0.3}&\red{119}&\red{117} &  \grn{0.2} &\grn{4}&\grn{4} &\red{69}\\
1M-20& \grn{5} & $\ora{28}$ & \red{6} & \red{74} & \ora{17} & \grn{1} & \grn{2}&\grn{9}&\grn{3} & \grn{5}&\red{140}&\red{140} &  \grn{7} &\ora{23}&\ora{23} &\ora{49}\\
\hline
2B-60-60& \ora{14} & $\grn{0.00}$ & \grn{2} & \grn{0.1} & \grn{0.00} & \grn{2} & \grn{2}&\grn{1}&\grn{0.1} & \grn{0.01}&\grn{1}&\ora{12} &  \grn{0.00} &\grn{0.1}&\ora{13} &\grn{1}\\
2B-60-40& \ora{20} & $\grn{0.00}$ & \grn{2} & \grn{0.1} & \grn{0.00} & \grn{2} & \grn{2}&\grn{1}&\grn{0.2} & \grn{0.01}&\grn{3}&\ora{15} &  \grn{0.00} &\grn{1}&\ora{35} &\grn{4}\\
2B-60-20& \ora{24} & $\grn{0.00}$ & \grn{2} & \grn{0.1} & \grn{0.00} & \grn{2} & \grn{2}&\grn{1}&\grn{1} & \grn{0.01}&\ora{12}&\ora{18} &  \grn{0.00} &\grn{4}&\ora{26} &\grn{10}\\
2B-40-40& \ora{16} & $\grn{0.00}$ & \grn{2} & \grn{2} & \grn{0.03} & \grn{2} & \grn{2}&\grn{1}&\grn{0.2} & \grn{0.2}&\ora{13}&\ora{14} &  \grn{0.1} &\grn{2}&\ora{13} &\ora{41}\\
2B-40-20& \ora{18} & $\grn{0.00}$ & \grn{2} & \grn{2} & \grn{0.03} & \grn{2} & \grn{2}&\grn{1}&\grn{1} & \grn{0.2}&\ora{17}&\ora{18} &  \grn{0.1} &\grn{5}&\ora{14} &\ora{45}\\
2B-20-20& \grn{8} & $\grn{0.00}$ & \red{8} & \ora{45} & \ora{15} & \grn{4} & \grn{2}&\grn{7}&\grn{3} & \grn{4}&\red{104}&\red{103} &  \grn{4} &\ora{12}&\ora{11} &\ora{25}\\
2B-10-10& \ora{19} & $\grn{0.00}$ & \red{4} & \ora{45} & \ora{15} & \ora{35} & \red{0}&\ora{25}&\grn{7} & \ora{19}&\red{100}&\red{100} &  \grn{7} &\red{88}&\red{106} &\red{71}\\
2M-60-60& \ora{14} & $\grn{0.03}$ & \grn{2} & \grn{0.1} & \grn{0.00} & \grn{2} & \grn{2}&\grn{1}&\grn{0.1} & \grn{0.01}&\grn{3}&\red{124} &  \grn{0.01} &\grn{1}&\grn{4} &\grn{4}\\
\hline
3B-60-60& \red{62} & $\grn{0.00}$ & \grn{2} & \grn{0.1} & \grn{0.00} & \grn{1} & \grn{2}&\grn{1}&\grn{0.1} & \grn{0.01}&\grn{1}&\grn{10} &  \grn{0.00} &\grn{1}&\ora{26} &\grn{5}\\
3B-60-40& \red{65} & $\grn{0.00}$ & \grn{2} & \grn{0.1} & \grn{0.00} & \grn{2} & \grn{2}&\grn{1}&\grn{0.2} & \grn{0.01}&\grn{3}&\grn{7} &  \grn{0.00} &\grn{1}&\ora{28} &\grn{6}\\
3B-60-20& \red{67} & $\grn{0.00}$ & \grn{2} & \grn{0.1} & \grn{0.00} & \grn{2} & \grn{2}&\grn{1}&\grn{1} & \grn{0.01}&\ora{16}&\red{61} &  \grn{0.00} &\grn{3}&\ora{47} &\ora{12}\\
3B-40-40& \red{58} & $\grn{0.00}$ & \grn{2} & \grn{2} & \grn{0.03} & \grn{2} & \grn{2}&\grn{1}&\grn{0.2} & \grn{0.2}&\ora{14}&\ora{13} &  \grn{0.1} &\grn{2}&\grn{8} &\ora{42}\\
3B-40-20& \red{60} & $\grn{0.00}$ & \grn{2} & \grn{2} & \grn{0.03} & \grn{2} & \grn{2}&\grn{1}&\grn{1} & \grn{0.2}&\ora{21}&\ora{29} &  \grn{0.1} &\grn{4}&\ora{14} &\ora{43}\\
3B-20-20& \ora{30} & $\grn{0.00}$ & \red{8} & \ora{45} & \ora{15} & \grn{4} & \grn{2}&\grn{7}&\grn{3} & \grn{4}&\red{103}&\red{103} &  \grn{4} &\ora{13}&\ora{31} &\ora{23}\\
3B-10-10& \ora{23} & $\grn{0.00}$ & \red{4} & \ora{45} & \ora{15} & \ora{39} & \red{0}&\red{97}&\red{75} & \ora{19}&\red{100}&\red{100} &  \grn{7} &\red{79}&\red{96} &\red{80}\\
3M-60-60& \red{64} & $\grn{0.03}$ & \grn{2} & \grn{0.1} & \grn{0.00} & \grn{2} & \grn{2}&\grn{1}&\grn{0.1} & \grn{0.01}&\grn{4}&\red{114} &  \grn{0.01} &\grn{1}&\grn{5} &\grn{3}\\
\hline
\end{tabular}
  \caption{\label{tab:tab1} Error analysis of various reduced-order
    modelling techniques for the Kuramoto-Sivashinsky equation with projection bases $W$ and $F$ based on
    snapshots taken from the transient and the limit-cycle. M, refers to methods 1, 2 and 3, F to base-flow (B) and mean-flow (M) formulations, $p$ and $q$ to the number of
    POD modes representing the state and the nonlinearity (when applicable),
    $\epsilon_S$ to the
    ratio between $\Vert N^S \Vert_F$ and $\Vert N \Vert$.
    The four next columns concern the recovery of the base-flow (BF) properties: $\epsilon_{w_b}$ is the relative error with respect to the true base-flow solution, $ \nu_\lambda$ the number of unstable eigenvalues of the operator linearized around the base-flow,
    $\epsilon_\lambda$ the relative error with respect to the most unstable eigenvalue $\lambda^{BF}$,
    $\epsilon_{\hat{w}}$ the relative alignment error with respect to the most unstable eigenvector $\hat{w}^{BF}$. The next four columns give analogous results for the mean-flow properties (mean-flow solution $\overline{w}$, number of unstable eigenvalues, eigenvalue $ \lambda^{MF}$, eigenvector $\hat{w}^{MF}$). The next three columns deal with the recovery of the TR trajectory:
    $\epsilon_t$, $\epsilon_m$ and $\epsilon'_m$ are the mean truncation error, model
    error and model error when setting to zero the symmetric part of the reduced-order model.
    The next three columns provide the same information for the MF trajectory, while the last column deals with the recovery of the MFTR trajectory (transient initialized with the mean-flow solution).
    We have shown in green
    reduced-order simulations that achieve a relative error
    less than $10\%$, in orange those achieving an error between $10\%$ and $50\%$, and in red the remaining ones. All $\epsilon $ quantities are given in percentage.}
\end{table}

All results from the simulations are summarized in
table~\ref{tab:tab1} for bases $W$ and $F$ determined from the transient and limit-cycle behavior (fig. \ref{fig:KSresults}) and in table~\ref{tab:tab1b} for bases solely obtained from the limit-cycle trajectory (fig. \ref{fig:KSresults2}).
These tables provide all parameters of the tested
models and the associated errors. We display in green errors $ \epsilon $ that are less than
$10\%$, in orange those between 
$10\%$ and $50\%$ and in red those who exceed $ 50\%$ or simulations that diverge. We have also indicated in green or red whether the model recovers or not the right number of unstable eigenvalues of the equations linearized around the fixed-point and time-averaged solutions. This is an important feature for subsequent use of the models in a flow control strategy, especially for the base-flow eigenvalue. 

From table~\ref{tab:tab1}, which deals with the case of POD bases built with snapshots in the transient and on the limit-cycle,  we can
draw the following conclusions :
\begin{itemize}
\item Comparing 1B-60, 2B-60-60 and 3B-60-60, all methods manage to achieve excellent and nearly equivalent
  results, if the number of POD modes, $p$ and $q,$ are sufficiently
  high and close. In this case, we have $\epsilon_m \leq 1\%$. There
  thus seems to be no apparent gain in using the more elaborate two-bases strategies (methods 2 and 3).
\item Comparing 1B-60 to 1M-60, 2B-60-60 to 2M-60-60 and 3B-60-60 to 3M-60-60, it seems slightly better to use BF formulations than MF formulations, whichever method is used.
\item Comparing 1B-60 to 1B-50, the number of POD modes required for high precision for the TR trajectory, i.e.,
  $\epsilon_m \leq 1\%,$ is rather large ($p \approx q \approx
  60$). In particular, the recovery of the unstable eigenvalue/eigenvector
  ($\epsilon_\lambda^{BF} \ll 1$ and $\epsilon_{\hat{w}}^{BF} \ll 1$) requires large values of $p$, typically $ p \approx 50 $.
\item On both TR and LC trajectories, the truncation error $\epsilon_t$ is always small compared to
  the model error $\epsilon_m$.
\item Considering the column $ \epsilon_S $, the most robust models are those provided by the first method and the second method with $ p=q $. The models deduced by the DEIM method are less energy-preserving with always high values of $\epsilon_S$. However, a high
  value of $\epsilon_S$ does not necessarily prevent the model from
  being accurate ($\epsilon_m \ll 1$), see line 3B-60-60. With the third technique, decreasing the number $q$ of POD modes
  for the representation of the nonlinearities always alters the
  energy-preserving property. This is not the case, however, for the second method.
\item Errors are consistently smaller on the LC trajectory than on the TR trajectory.
\item Considering the recovery of the mean-flow properties (four sub-columns of MF column), it is seen that both the mean-flow solution and the linear properties around it are easily reproduced with low values of $ \epsilon$, even for $ p, q \approx 30 $.
\item Considering the columns $ \epsilon'_m$, it is seen that removing the symmetric part of the quadratic term generally always decreases the accuracy of the model. Hence, although the symmetric part is not energy preserving, it is important for the precision of the model.
\item Considering the column MFTR, it is seen that all modelling strategies manage to well recover the trajectory MFTR for high values of $p$ and $q$. We remind the reader that none of the snapshots along this trajectory have been considered for building the bases $ W$ and $F$: the states explored along the trajectory MFTR are contained in the subspace spanned by the snapshots along the trajectory initialized from the base-flow ($0 \leq t \leq 150$).  
\end{itemize}

\begin{table}[htbp]
  \centering
  \begin{tabular}{|l|c|cccc|cccc|ccc|ccc|c|}
  \hline
  & & \multicolumn{4}{|c|}{BF} &  \multicolumn{4}{|c|}{MF} & \multicolumn{3}{|c|}{TR} & \multicolumn{3}{|c|}{LC} & \multicolumn{1}{|c|}{MFTR}\\
MF-$p$-$q$& $\epsilon_S$
& $\epsilon_{w_b}$ & $\nu_{\lambda}$ & $\epsilon_{\lambda}$ & $\epsilon_{\hat{w}}$
& $\epsilon_{\bar{w}}$ & $\nu_{\lambda}$ &$\epsilon_{\lambda}$ & $\epsilon_{\hat{w}}$ 
& $ \epsilon_t$ & $ \epsilon_m$ & $ \epsilon_m' $
& $ \epsilon_t$ & $ \epsilon_m$ & $ \epsilon_m'$
& $ \epsilon_m$
\\
\hline
1B-12& \grn{0.02} & $\grn{0}$ & \grn{2} & \ora{25} & \grn{4} & \grn{2} & \grn{2}&\grn{6}&\grn{6} & \grn{10} &\red{112}&\red{112} &\grn{9} &\grn{3}&\grn{3} &\red{97}\\
1B-10& \grn{0.02} & $\grn{0}$ & \grn{2} & \ora{29} & \grn{6} & \grn{1} & \grn{2}&\grn{9}&\grn{6} &\ora{14}&\red{89}&\red{89} &\grn{9}&\grn{2}&\grn{2} &\red{92}\\
1M-12& \grn{0.00} & $\red{94}$ & \grn{2} & \ora{41} & \ora{20} & \grn{1} & \grn{2}&\grn{8}&\grn{6} &\ora{42}&\red{147}&\red{147} &\ora{17}&\grn{4}&\grn{4} &\red{107}\\
1M-10& \grn{0.00} & $\red{94}$ & \grn{2} & \ora{42} & \ora{20} & \grn{1} & \grn{2}&\grn{10}&\grn{6}&\ora{46} &\red{148}&\red{148} &\ora{17}&\grn{3}&\grn{3} &\red{92}\\
1M-6& \grn{0.00} & $\red{96}$ & \grn{2} & \ora{45} & \ora{22} & \grn{1} & \grn{2}&\ora{11}&\grn{7} &\red{51}&\red{149}&\red{149} 
&\ora{17}&\grn{3}&\grn{3} &\red{84}\\
\hline
2B-10-10& \ora{22} & $\grn{0}$ & \grn{2} & \ora{29} & \grn{6} & \grn{1} & \grn{2}&\ora{11}&\grn{6}&\ora{14} &\red{91}&\red{85} &\grn{9}&\grn{4}&\ora{53} &\red{82}\\
2M-10-10& \ora{22} & $\red{94}$ & \grn{2} & \ora{42} & \ora{21} & \grn{1} & \grn{2}&\grn{10}&\grn{6}&\ora{46} &\red{148}&\red{147}&\ora{17} &\grn{2}&\ora{14} &\red{96}\\
2M-10-6& \ora{29} & $\red{94}$ & \grn{2} & \ora{42} & \ora{22} & \grn{2} & \grn{2}&\grn{10}&\grn{6} &\ora{46}&\red{149}&\red{147} &\ora{17}&\ora{12}&\ora{16} &\red{90}\\
\hline
3B-10-10& \red{66} & \grn{0} & \grn{2} & \ora{29} & \grn{6} & \grn{1} & \grn{2}&\ora{13}&\grn{6}&\ora{13} &\red{116}&\red{83} &\grn{9}&\grn{4}&\red{62} &\red{82}\\
3M-10-10& \red{70} & \red{94} & \grn{2} & \ora{42} & \ora{21} & \grn{1} & \grn{2}&\grn{10}&\grn{6} &\ora{46}&\red{147}&\red{138} &\ora{17}&\grn{6}&\ora{47} &\red{90}\\
3M-10-6& \red{67} & \red{94} & \grn{2} & \ora{42} & \ora{22} & \grn{2} & \grn{2}&\grn{10}&\grn{6} &\ora{46}& \red{149}&\red{153} &\ora{17}&\grn{8}&\ora{16} &\red{91}\\
3M-6-6& \red{52} & \red{96} & \grn{2} & \ora{44} & \ora{22} & \grn{3} & \grn{2}&\ora{11}&\grn{7} &\red{51 }&\red{156}&\red{154} &\ora{17}&\ora{13}&\ora{19} &\red{84}\\
\hline
\end{tabular}
  \caption{\label{tab:tab1b} Error analysis of various reduced-order
    modelling techniques for the Kuramoto-Sivashinsky equation with projection bases $W$ and $F$ obtained from snapshots on the limit-cycle. Same caption as in tab. \ref{tab:tab1}.}
\end{table}

From table~\ref{tab:tab1b}, which deals with the case of bases $W$ and $F$ solely determined from snapshots on the limit-cycle, we can draw the following conclusions:
\begin{itemize}
    \item Considering the column BF and its four sub-columns, it is seen that the model is not able to predict the base-flow properties with the MF formulation. The BF formulation is slightly better, since it accounts for explicit knowledge of the base-flow solution. Yet, the stability properties are poorly captured.
    \item the TR trajectory is poorly recovered both in terms of truncation and modelling errors. This shows that the dynamics along the LC cycle misses an important part of the subspace spanned during the transient initialized by the base-flow. The same conclusion holds for the transient initialized from the mean-flow solution (column MFTR).
    \item Comparing 1B-10, 2B-10-10 and 3B-10-10, the limit-cycle trajectory is equally well captured by the three methods for sufficiently large $p$ and $q$.
    \item The mean-flow properties (MF column and 4 sub-columns) are easily recovered by all models.
\end{itemize}

\section{Flow past a cylinder with projection bases $W$ and $F$ based on snapshots from the transient and the limit-cycle}
\label{sec:cyl_TR}

We proceed
analogously to the above Kuramoto-Sivashinsky simulations. Simulation snapshots from the transient and the limit-cycle are first processed in a POD analysis in section
\S~\ref{sec:TR_POD}, whereas the assessment of the reduced-order
models is carried out in section \S~\ref{sec:TR_ModRed}.

\subsection{POD bases}
\label{sec:TR_POD}

\begin{figure}[htbp]
  \begin{center}
    \begin{tabular}{cc} (a)&(b) \\
     \psfrag{n}{$i$}
     \psfrag{w}{$\mathrm{\Sigma}^{2}_i,w'$}
     \psfrag{x}{$x$}
     \psfrag{y}{$y$}
     \psfrag{u}{$u'$}
     \psfrag{v}{$v'$}
 \includegraphics[width=0.45\textwidth]{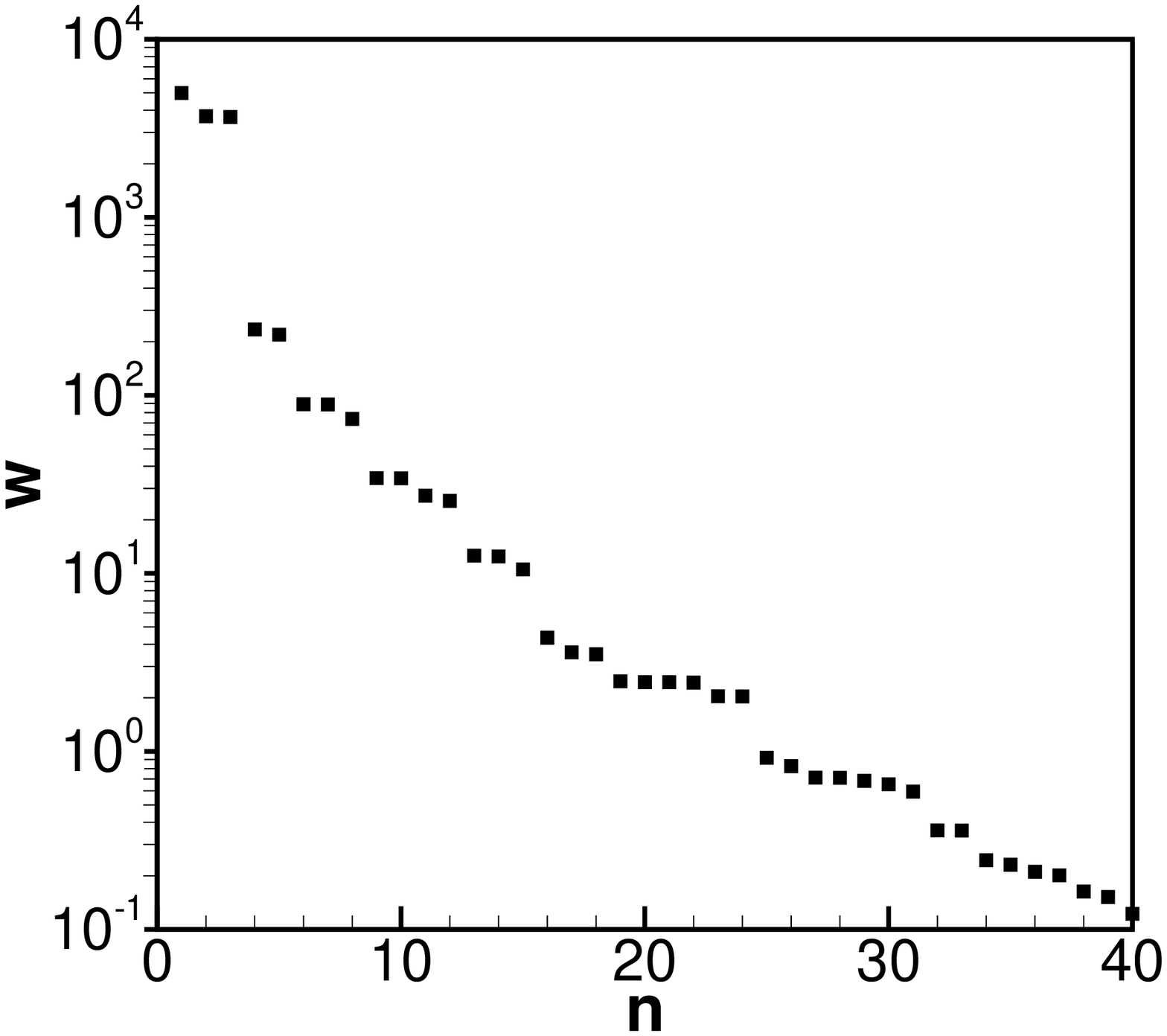} &  \psfrag{n}{$i$}
     \psfrag{w}{$\mathrm{\Gamma}^{2}_i,f(w',w')$}
\psfrag{x}{$x$}
     \psfrag{y}{$y$}
     \psfrag{u}{$u'$}
     \psfrag{v}{$v'$}
 \includegraphics[width=0.45\textwidth]{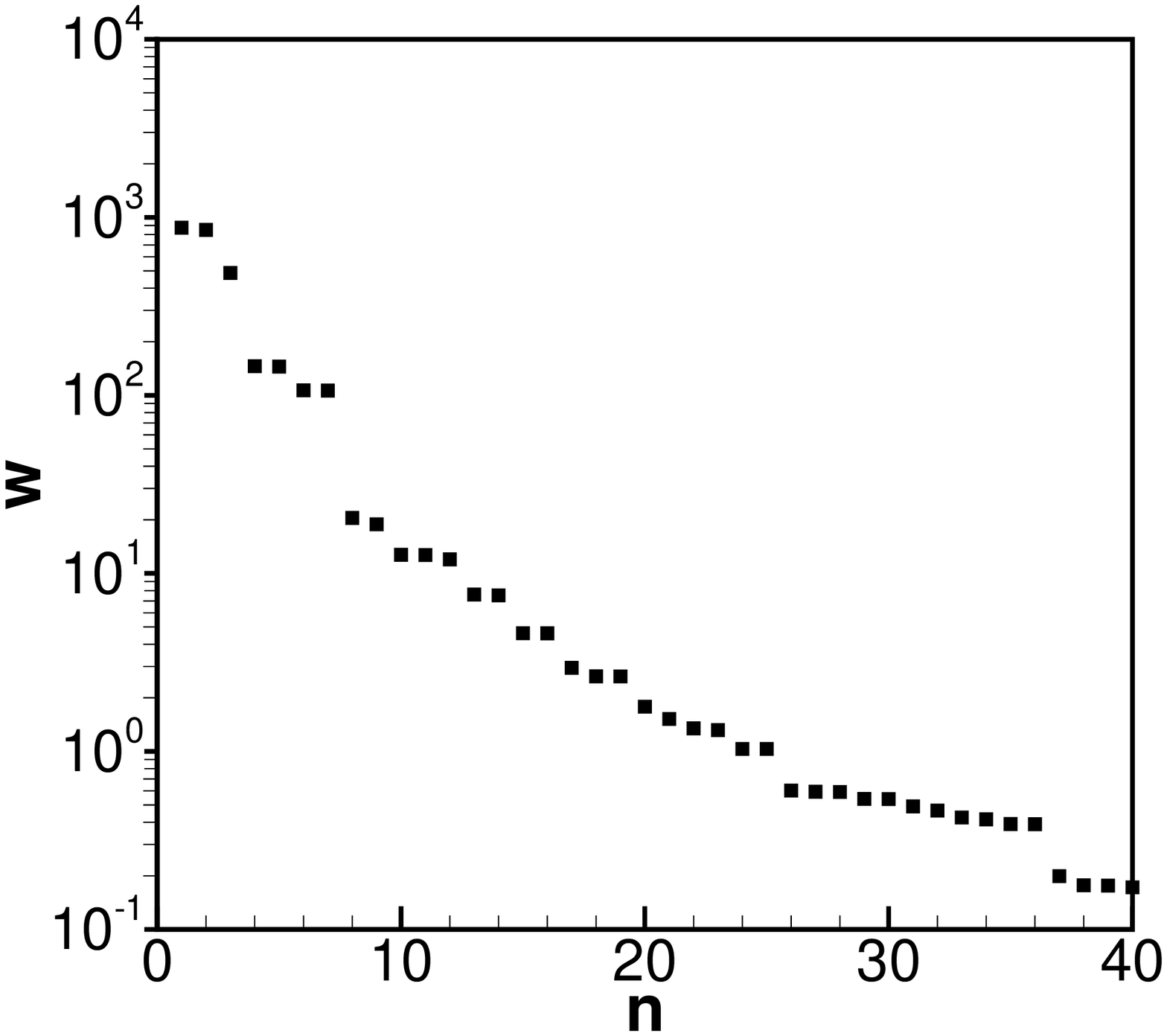} \\ (c) & (d) \\
     \psfrag{n}{$i$}
     \psfrag{a}{$\mathrm{\Sigma}^{'2}_i,w'$}
     \psfrag{b}{$\mathrm{\Gamma}^{'2}_i,f(w',w')$}
     \psfrag{s}{Eigenvalues}
     \psfrag{x}{$x$}
     \psfrag{y}{$y$}
     \psfrag{u}{$u'$}
     \psfrag{v}{$v'$}
      \includegraphics[width=0.5\textwidth]{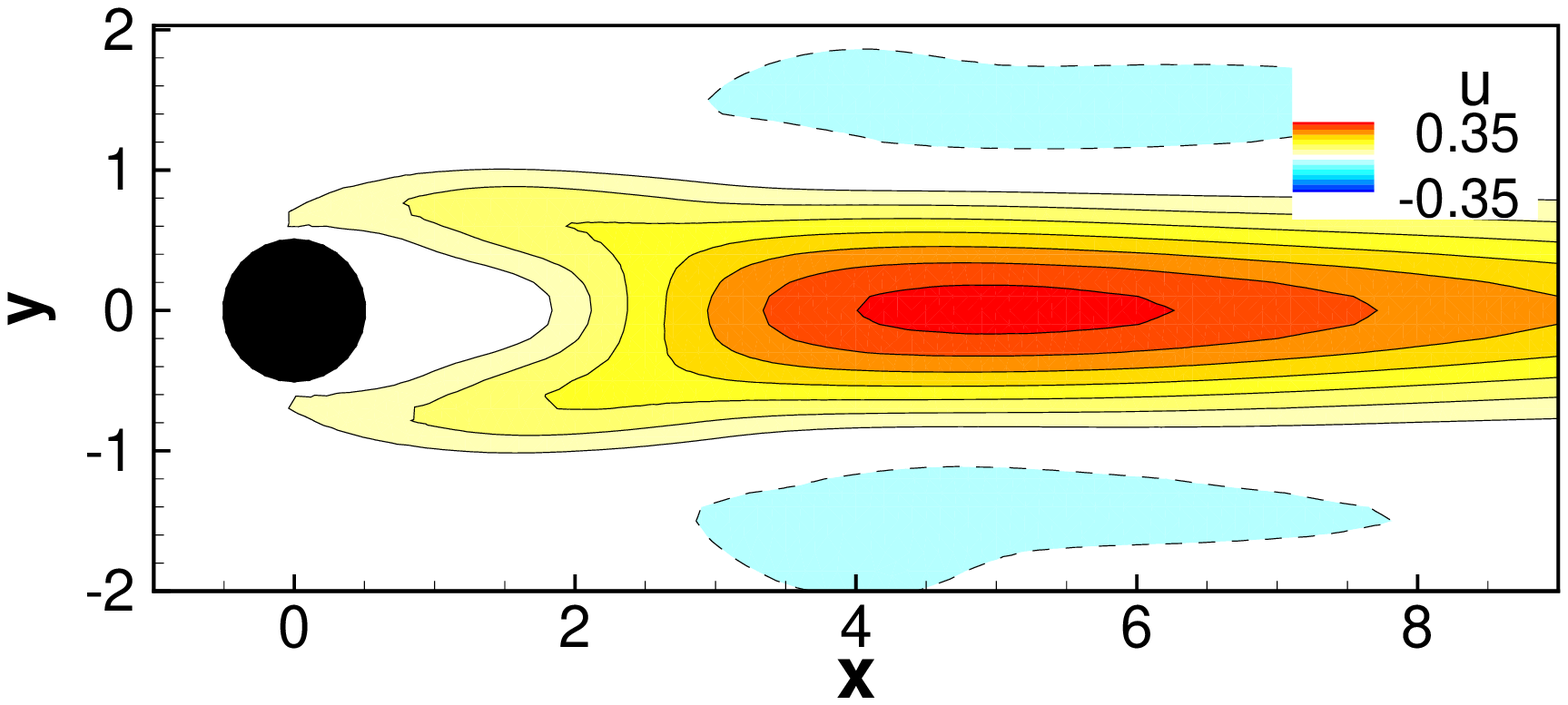} &
     \psfrag{n}{$i$}
     \psfrag{a}{$\mathrm{\Sigma}^{2}_i,w'$}
     \psfrag{b}{$\mathrm{\Gamma}^{2}_i,f(w',w')$}
     \psfrag{s}{Eigenvalues}
     \psfrag{x}{$x$}
     \psfrag{y}{$y$}
     \psfrag{u}{$u'$}
     \psfrag{v}{$v'$}
      \includegraphics[width=0.5\textwidth]{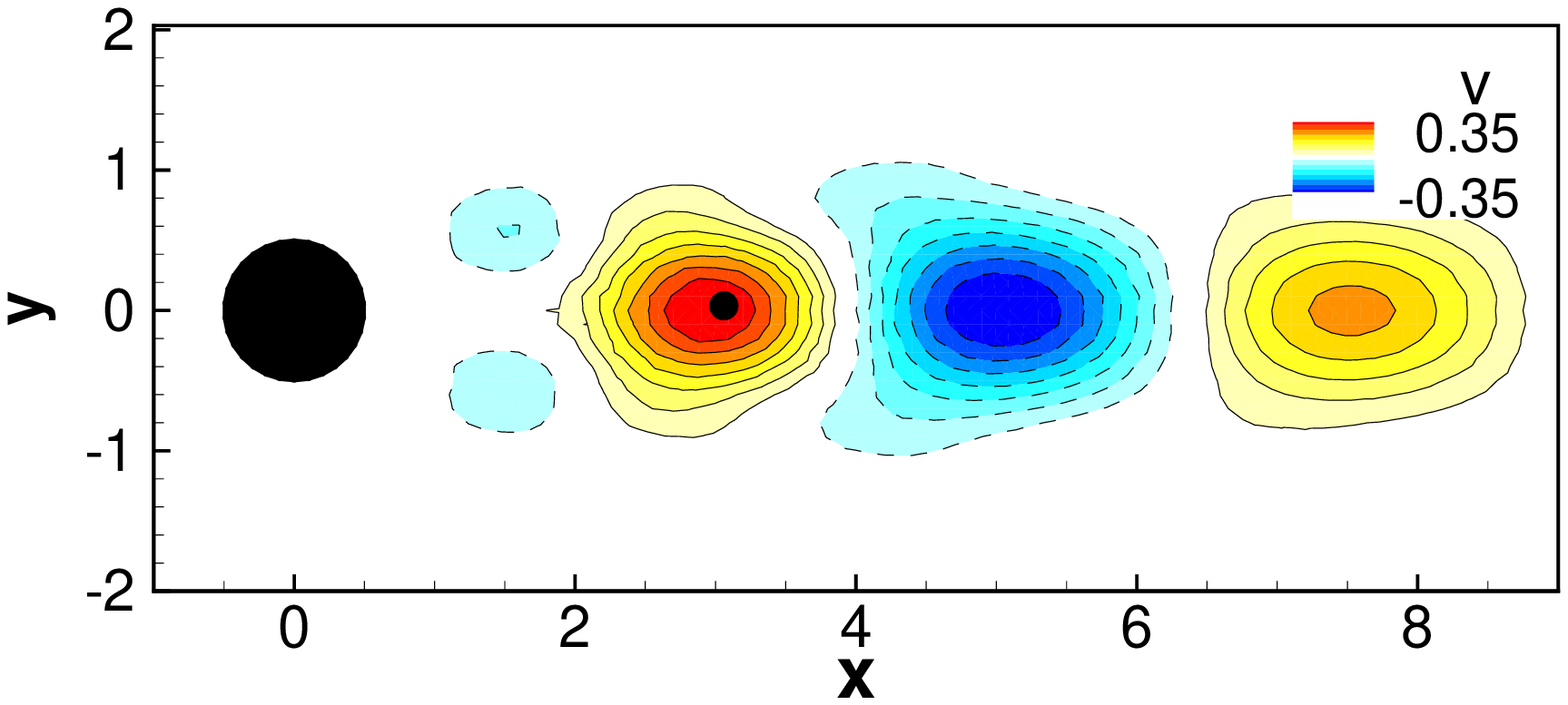} \\ (e) & (f) \\
     \psfrag{n}{$i$}
     \psfrag{a}{$\mathrm{\Sigma}^{2}_i,w'$}
     \psfrag{b}{$\mathrm{\Gamma}^{2}_i,f(w',w')$}
     \psfrag{s}{Eigenvalues}
     \psfrag{x}{$x$}
     \psfrag{y}{$y$}
     \psfrag{u}{$u'$}
     \psfrag{v}{$v'$}
      \includegraphics[width=0.5\textwidth]{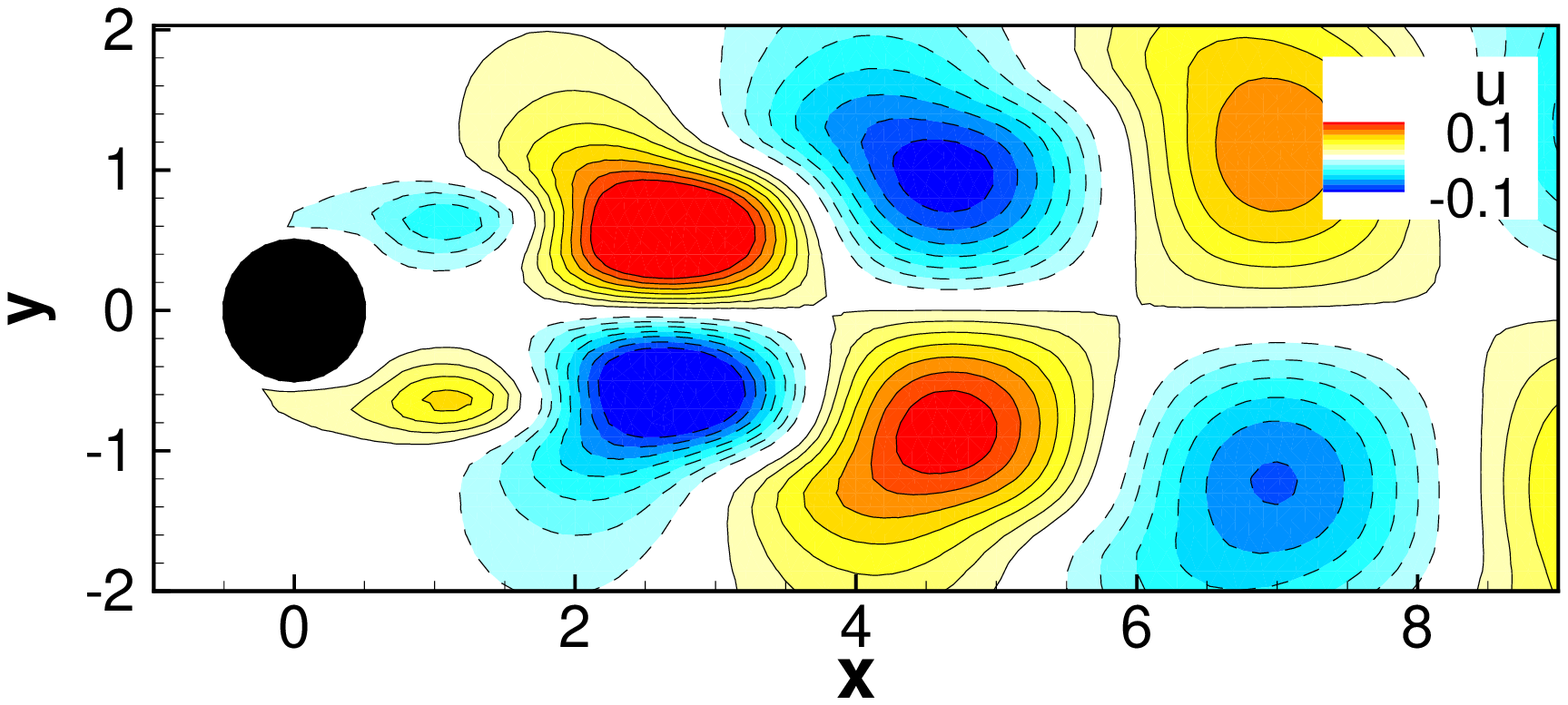} &
     \psfrag{n}{$i$}
     \psfrag{a}{$\mathrm{\Sigma}^{'2}_i,w'$}
     \psfrag{b}{$\mathrm{\Gamma}^{'2}_i,f(w',w')$}
     \psfrag{s}{Eigenvalues}
     \psfrag{x}{$x$}
     \psfrag{y}{$y$}
     \psfrag{u}{$u'$}
     \psfrag{v}{$v'$}
      \includegraphics[width=0.5\textwidth]{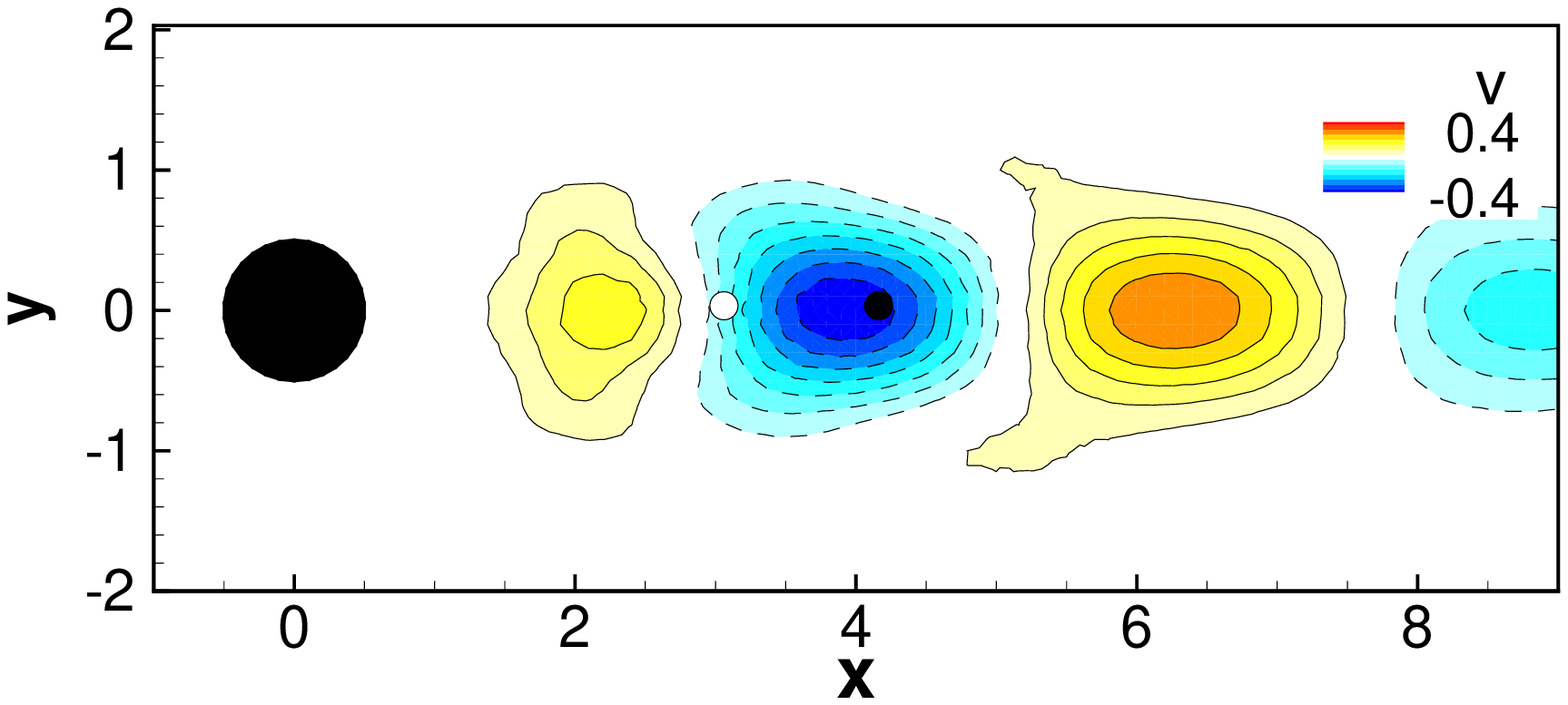} \\ (g) & (h) \\
      \psfrag{n}{$i$}
     \psfrag{a}{$\mathrm{\Sigma}^{'2}_i,w'$}
     \psfrag{b}{$\mathrm{\Gamma}^{'2}_i,f(w',w')$}
     \psfrag{s}{Eigenvalues}
     \psfrag{x}{$x$}
     \psfrag{y}{$y$}
     \psfrag{u}{$u'$}
     \psfrag{v}{$v'$}
      \includegraphics[width=0.5\textwidth]{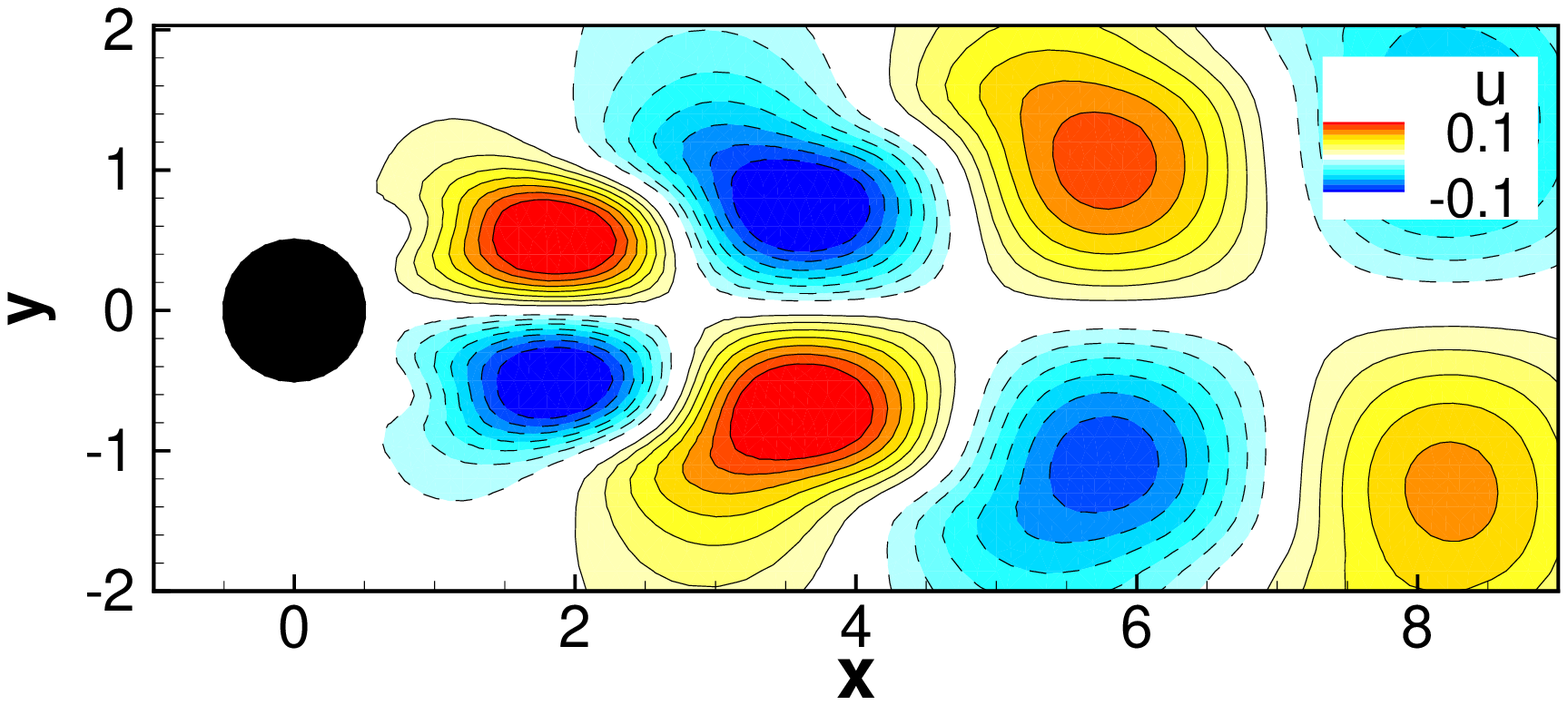} &
      \psfrag{n}{$i$}
     \psfrag{a}{$\mathrm{\Sigma}^{'2}_i,w'$}
     \psfrag{b}{$\mathrm{\Gamma}^{'2}_i,f(w',w')$}
     \psfrag{s}{Eigenvalues}
     \psfrag{x}{$x$}
     \psfrag{y}{$y$}
     \psfrag{u}{$u'$}
     \psfrag{v}{$v'$}
      \includegraphics[width=0.5\textwidth]{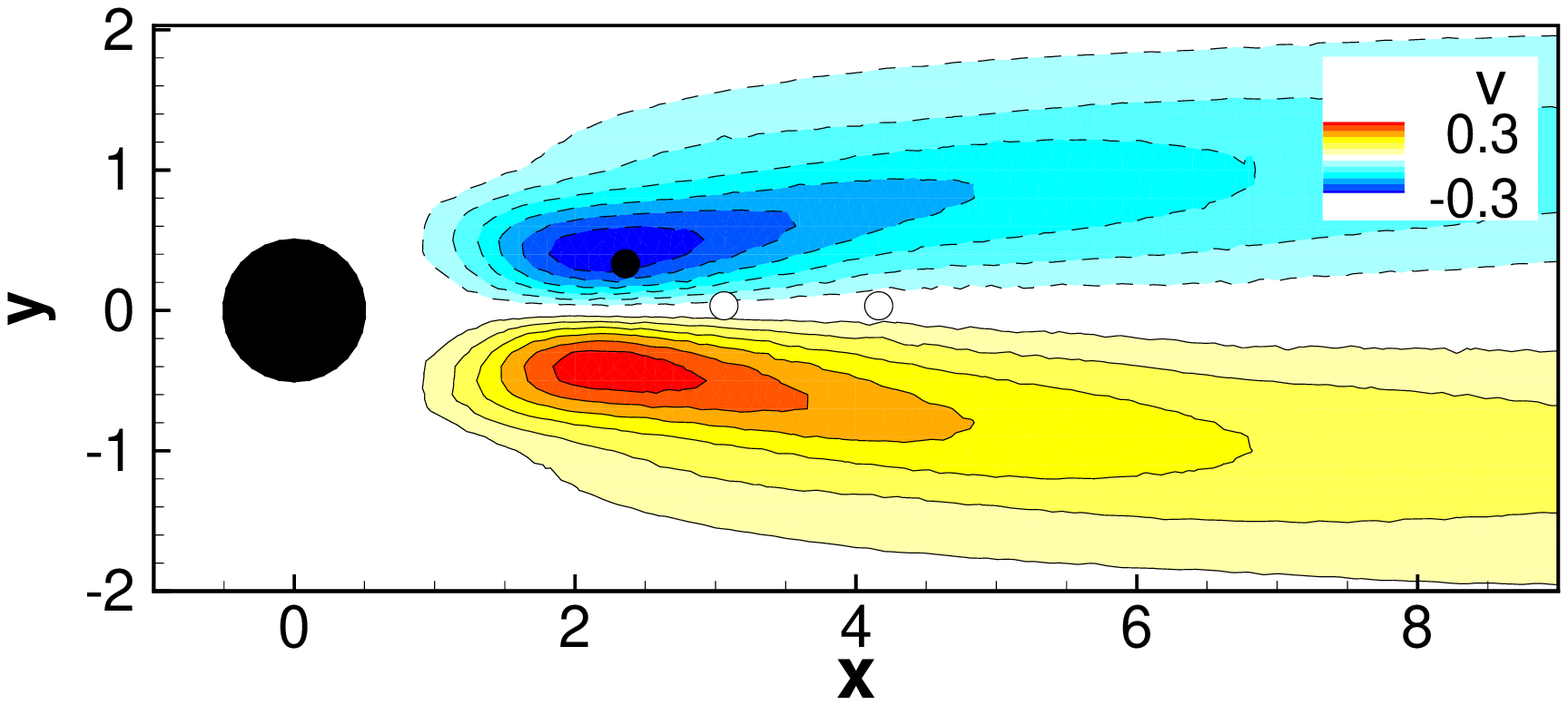}
    \end{tabular}
  \end{center}
  \caption{\label{fig:TRresults} Model reduction of cylinder flow at $Re=100$ with projection bases obtained from snapshots in the transient and the limit-cycle. (a,b): Eigenvalues of the
    correlation matrices for the snapshots representing (a): the
    state and (b): the nonlinearity. (c-h): first three POD modes of (c,e,g): state
    and (d,f,h): nonlinearity. Colored circles show the corresponding DEIM points.}
  \end{figure}

The snapshots for the POD analysis have been gathered from the
unsteady simulations shown in figure~\ref{fig:evoEnergy}(a,b) in the
time range $0 \leq t \leq 150.$ The sampling time is equal to $\Delta
t_S = 0.2$ which results in $751$ snapshots. We consider the snapshots
for the state and the nonlinearity for both the BF and the MF formulation. The eigenvalues of the correlation matrices
for the state and the nonlinearity are
represented in figures~\ref{fig:TRresults}(a,b) for the BF formulation (state $w'$ and nonlinearity $f(w',w')$). We observe that the
eigenvalues either appear in pairs or singly. The eigenvalues
appearing in pairs represent the downstream advection of coherent structures.

The first three POD modes representing the state in the BF formulation are shown in
figures~\ref{fig:TRresults}(c,e,g). We notice that the first POD mode
(linked to a single eigenvalue of the correlation matrix) represents
the mean flow generated by the growth of the unstable global mode (it
also corresponds to the shift-mode discussed in~\cite{noack2003hierarchy}),
while the second and third POD modes represent vortex shedding (linked
to the unstable global mode).

The first three nonlinear POD modes are shown in
figure~\ref{fig:TRresults}(d,f,h) with the associated DEIM points (a black
symbol representing the most recent DEIM point, and the white symbols
showing the previous ones).
We have represented the component that is selected by the DEIM procedure in the following way: if the row selector matrix $ P^H$ designates a streamwise (resp. cross-stream) velocity component, we have represented the $u'$ (resp. $v'$) component. It is seen here that the first three DEIM points are linked to cross-stream $v'$ components.
Again, it appears that the DEIM points
correspond to the large values in amplitude of the associated POD
modes. We also observe that the first two POD modes appear as a pair,
indicating that they represent a downstream travelling structure for
$f(w',w')$. Also, these structures are anti-symmetric ($v'$ is symmetric) and therefore stem from the interaction of a symmetric mode (first state POD mode) with an anti-symmetric one (second and third state POD modes). The third POD mode corresponds to a
single eigenvalue and is symmetric: it represents the nonlinear interaction between the second and third state anti-symmetric POD modes. It actually features a forcing term responsible for the mean-flow deformation (shortening of the recirculation bubble in the presence of high-amplitude vortex shedding modes).

\subsection{Model reduction}
\label{sec:TR_ModRed}

As in section \S~\ref{sec:KS_ModRed}, we present the characteristics
of the models and the analysis of the errors in a single table, see
tab.~\ref{tab:tab2}. Again, for the three reduction methods, the two formulations and various values of  $ p $ and $q$, we assess the robustness of the reduced-order models linked to the energy-preservation of the quadratic term ($ \epsilon_S$), we evaluate the recovery of the base-flow (BF) and mean-flow (MF) properties, and the performance of the models to predict the transient (TR), limit-cycle (LC) and transient from mean-flow (MFTR) trajectories. 

\begin{table}[htbp]
  \centering
  \begin{tabular}{|l|c|cccc|cccc|ccc|ccc|c|}
  \hline
  & & \multicolumn{4}{|c|}{BF} &  \multicolumn{4}{|c|}{MF} & \multicolumn{3}{|c|}{TR} & \multicolumn{3}{|c|}{LC} & \multicolumn{1}{|c|}{MFTR}\\
MF-$p$-$q$& $\epsilon_S$
& $\epsilon_{w_b}$ & $\nu_{\lambda}$ & $\epsilon_{\lambda}$ & $\epsilon_{\hat{w}}$
& $\epsilon_{\bar{w}}$ & $\nu_{\lambda}$ &$\epsilon_{\lambda}$ & $\epsilon_{\hat{w}}$ 
& $ \epsilon_t$ & $ \epsilon_m$ & $ \epsilon_m' $
& $ \epsilon_t$ & $ \epsilon_m$ & $ \epsilon_m'$
& $ \epsilon_m$
\\
\hline
1B-60& \grn{7} & $\grn{0.00}$ & \red{4} & \grn{0.02} & \grn{0.04} & \grn{1} & \grn{2} &\grn{2}&\grn{5} & \grn{1}&\grn{1}&\grn{3} &  \grn{0.2} &\grn{0.4}&\grn{3} &\red{76}\\
1B-50& \grn{6} & $\grn{0.00}$ & \red{4} & \grn{0.1} & \grn{0.1} & \grn{2}  & \red{0}&\grn{2}&\grn{5} & \grn{1}&\grn{2}&\grn{2} &  \grn{0.3} &\grn{1}&\grn{3} &\red{78}\\
1B-40& \grn{5} & $\grn{0.00}$ & \grn{2} & \grn{0.2} & \grn{0.1} & \grn{2} & \red{0}&\grn{2}&\grn{5} & \grn{1}&\grn{3}&\grn{4} &  \grn{0.4} &\grn{4}&\grn{2}&\red{76}\\
1B-30& \grn{4} & $\grn{0.00}$ & \grn{2} & \grn{0.3} & \grn{0.4} & \grn{2} & \red{0}&\grn{1}&\grn{6} & \grn{3}&\grn{5}&\grn{6} &  \grn{1} &\grn{4}&\grn{6} &\red{76}\\
1B-20& \grn{4} & $\grn{0.00}$ & \grn{2} & \grn{2} & \grn{1} & \grn{3} & \red{0}&\grn{1}&\grn{6} & \grn{5}&\red{53}&\red{54} &  \grn{2} &\grn{4}&\grn{5} &\red{84}\\
1B-10& \grn{0.4} & $\grn{0.00}$ & \grn{2} & \grn{5} & \ora{19} & \grn{4} & \grn{2}&\grn{1}&\grn{9} & \ora{16}&\ora{40}&\ora{40} &  \grn{4} &\grn{4}&\grn{5} &\red{86}\\
1M-60& \grn{7} & $\grn{0.2}$ & \red{4} & \grn{0.1} & \grn{0.04} & \grn{1} & \grn{2}&\grn{2}&\grn{5} & \grn{1}&\grn{2}&\grn{4} &  \grn{0.3} &\grn{1}&\grn{2} &\red{97}\\
1M-40& \grn{5} & $\grn{0.5}$ & \red{4} & \grn{0.2} & \grn{0.1} & \grn{2} & \red{0}&\grn{2}&\grn{5} & \grn{1}&\grn{4}&\grn{6} & \grn{1} &\grn{6}&\grn{2} &\red{97}\\
1M-20& \grn{4} & $\grn{2}$ & \grn{2} & \grn{2} & \grn{1} & \grn{3} & \red{0}&\grn{1}&\grn{6} & \grn{5}&\red{54}&\red{55} &  \grn{3} &\grn{6}&\grn{5} &\red{107}\\
\hline
2B-60-60& \ora{31} & $\grn{0.00}$ & \red{4} & \grn{0.02} & \grn{0.04} & \grn{2} & \grn{2}&\grn{2}&\grn{8} & \grn{1}&\grn{1}&\red{87} &  \grn{0.2} &\grn{1}&\red{118} &\red{89}\\
2B-60-40& \ora{34} & $\grn{0.00}$ & \red{4} & \grn{0.02} & \grn{0.04} & \grn{2} & \grn{2}&\grn{2}&\grn{9} & \grn{1}&\grn{4}&\red{88} &  \grn{0.2} &\grn{1}&\red{122} &\red{85}\\
2B-60-20& \ora{40} & $\grn{0.00}$ & \red{4} & \grn{0.02} & \grn{0.04} & \grn{3} & \grn{2}&\grn{2}&\grn{10} & \grn{1}&\ora{25}&\red{85} &  \grn{0.2} &\grn{5}&\red{116} &\red{89}\\
2B-40-40& \ora{31} & $\grn{0.00}$ & \grn{2} & \grn{0.2} & \grn{0.1} & \grn{2} & \grn{2}&\grn{1}&\grn{8} & \grn{1}&\grn{4}&\red{71} &  \grn{0.4} &\grn{2}&\red{114} &\red{83}\\
2B-40-20& \ora{31} & $\grn{0.00}$ & \grn{2} & \grn{0.2} & \grn{0.1} & \grn{5} & \grn{2}&\grn{2}&\grn{9} & \grn{1}&\ora{18}&\red{57} &  \grn{0.4} &\ora{18}&\red{124} &\red{83}\\
2B-20-20& \ora{30} & $\grn{0.00}$ & \grn{2} & \grn{2} & \grn{1} & \grn{8} & \grn{2}&\grn{3}&\grn{9} & \grn{5}&\red{100}&\red{143} &  \grn{2} &\grn{10}&\red{122} &\red{122}\\
2B-10-10& \ora{21} & $\grn{0.00}$ & \grn{2} & \grn{5} & \ora{19} & \ora{32} & \grn{2}&\grn{7}&\ora{12} & \ora{16}&\red{147}&\red{128} &  \grn{4} &\red{97}&\red{79} &\red{134}\\
2M-60-60& \ora{35} & $\grn{0.3}$ & \red{4} & \grn{0.1} & \grn{0.04} & \grn{2} & \grn{2}&\grn{2}&\grn{5} & \grn{1}&\grn{2}&\ora{32} &  \grn{0.3} &\grn{1}&\red{84} &\red{105}\\
2M-40-40& \ora{37} & $\grn{1}$ & \red{8} & \grn{0.3} & \grn{1} & \grn{2} & \red{0}&\grn{2}&\grn{5} & \grn{1}&\grn{4}&\ora{34} &  \grn{1} &\grn{2}&\red{117} &\red{112}\\
2M-20-20& \ora{40} & $\grn{4}$ & \red{6} & \grn{6} & \ora{17} & \grn{3} &\red{0} &\grn{2}&\grn{6} & \grn{5}&\red{128}&\red{65} &  \grn{3} &\grn{3}&\red{134} &\red{90}\\
\hline
3B-60-60& \red{58} & $\grn{0.00}$ & \red{4} & \grn{0.02} & \grn{0.04} & \red{$\infty$} & && & \grn{1}&\red{$\infty$}&\red{95} &  \grn{0.2} &\red{$\infty$}&\red{115} &\red{91}\\
3B-60-40& \red{58} & $\grn{0.00}$ & \red{4} & \grn{0.02} & \grn{0.04} & \red{$\infty$} & && & \grn{1}&\red{$\infty$}&  \red{108} & \grn{0.2} &\red{$\infty$}&\red{62} &\red{$\infty$}\\
3B-60-20& \red{59} & $\grn{0.00}$ & \red{4} & \grn{0.02} & \grn{0.04} & \red{$\infty$} & && & \grn{1}&\red{54}&\red{97} &  \grn{0.2} &\ora{37}&\red{92} &\red{$\infty$}\\
3B-40-40& \red{57} & $\grn{0.00}$ & \grn{2} & \grn{0.2} & \grn{0.1} & \red{$\infty$} & && & \grn{1}&\red{$\infty$}&\red{80} &  \grn{0.4} &\red{$\infty$}&\red{116} &\red{$\infty$}\\
3B-40-20& \red{57} & $\grn{0.00}$ & \grn{2} & \grn{0.2} & \grn{0.1} & \red{$\infty$} & && & \grn{1}&\red{86}&\red{98} &  \grn{0.4} &\ora{24}&\red{92} &\red{$\infty$}\\
3B-20-20& \red{56} & $\grn{0.00}$ & \grn{2} & \grn{2} & \grn{1} & \red{$\infty$} & && & \grn{5}&\red{$\infty$}&\red{140} &  \grn{2} &\red{$\infty$}&\red{115} &\red{141}\\
3B-10-10& \ora{41} & $\grn{0.00}$ & \grn{2} & \grn{5} & \ora{19} & \red{57} & \grn{2}&\ora{14}&\ora{16} & \ora{16}&\red{153}&\red{100} &  \grn{4} &\red{102}&\red{83} &\red{154}\\
3M-60-60& \red{59} & $\grn{2}$ & \red{8} & \red{222} & \red{96} & \red{$\infty$} & && & \grn{1}&\red{$\infty$}&\red{110} &  \grn{0.3} &\grn{8}&\red{118} &\red{156}\\
3M-60-40& \red{59} & $\grn{3}$ & \red{6} & \red{258} & \red{76} & \grn{2} & \grn{2}&\grn{2}&\grn{5} & \grn{1}&\red{$\infty$}&\red{90} &  \grn{0.3} &\grn{3}&\red{115} &\red{97}\\
3M-40-40& \red{59} & $\grn{2}$ & \red{6} & \red{302} & \red{85} & \grn{2} & \red{0}&\grn{2}&\grn{5} & \grn{1}&\ora{11}&\red{97} &  \grn{1} &\grn{1}&\red{148} &\red{105}\\
3M-40-20& \red{59} & $\ora{22}$ & \red{7} & \grn{3} & \grn{10} & \red{$\infty$} & && & \grn{1}&\ora{33}&\red{105} &  \grn{1} &\grn{8}&\red{91} &\red{103}\\
3M-30-30& \red{59} & $\grn{4}$ & \red{6} & \grn{1} & \grn{1} & \grn{5} & \red{0}&\grn{3}&\grn{5} & \grn{3}&\ora{19}&\red{97} &  \grn{1} &\ora{12}&\red{126} &\red{115}\\
3M-20-20& \red{56} & $\ora{14}$ & \red{6} & \grn{4} & \grn{9} & \red{$\infty$} & && & \grn{5}&\red{66}&\red{108} &  \grn{3} &\ora{18}&\red{143} &\red{97}\\
\hline
\end{tabular}
  \caption{\label{tab:tab2} Error analysis of various reduced-order
    modelling techniques for cylinder flow at $Re=100$ with projection bases $W$ and $F$ based on
    snapshots taken from the TR trajectory. Same caption as in tab. \ref{tab:tab1}.}
\end{table}

With the results listed in
table~\ref{tab:tab2}, we can state that
the conclusions are overall the same than for the Kuramoto-Sivashinsky equation with bases obtained from snapshots in the transient and in the limit-cycle. We will therefore not reproduce these conclusions here but only stress the differences:
\begin{itemize}
\item Only the Galerkin methods (methods 1 and 2) achieve accurate
  results. The
  accuracy is approximately the same for the TR and for
  the LC trajectories.
\item Method 3 either diverges (most of the time) or yields
  inaccurate results (with model errors $\epsilon_m > 10\%$).
\item Method 1 (traditional Galerkin method) with $p = 30$ is more
  accurate than method 2 (double-base) with $p = q = 30.$
\item The energy-preserving criterion is best met by method 1 with $\epsilon_S \leq 5\%$ in any case. Method 2 produces models with $\epsilon_S \approx 30\%$, while method 3 yields values $\epsilon_S \approx 60$ in most of the cases.
\item Method 3 generates models that are not robust and diverge most of the time or are inaccurate on the TR trajectory. On the LC trajectory, they provide accurate results only with the MF formulation (see for example 3M-40-40 case). Yet,
  the level of $\epsilon_S$ is not significantly higher than in the
  case of the Kuramoto-Sivashinsky simulations.
  \item Considering columns $ \epsilon'_m$, we observe that, removing the non-preserving energy part of the quadratic term, prevents the models from diverging (as expected), but does never generate an accurate model.
   \item Considering the column labelled MFTR, it is seen that none of the reduced order models is able to precisely reproduce the mean-flow transient trajectory. This indicates that the subspace explored on this trajectory is significantly different from the one characterizing the base-flow transient.
   \item Looking at 1B-60, it is seen that four unstable eigenvalues might exist (even for large $p$) for the linearized operator near the fixed-point. This is detrimental for flow control since a non-physical unstable eigenmode exists on top of the physical one (which is well captured). Yet, if one considers only $p=40$ modes (1B-40), the two unstable eigenvalues are recovered. In such a case, it is also seen that the linearized operator around the mean-flow only exhibits stable eigenvalues, while the large-scale operator exhibits two slightly unstable modes.  
\end{itemize}

\section{Flow past a cylinder with projection bases $W$ and $F$ based on snapshots from the limit-cycle} \label{sec:CYL_LC}

Here we consider snapshots only on the limit-cycle to build the bases $W$ and $F$. We proceed as in the
previous section: the POD
analysis is performed in section \S~\ref{sec:LC_POD}, and the
assessment of the reduced-order models is carried out in section
\S~\ref{sec:LC_ModRed}.

\subsection{POD bases}
\label{sec:LC_POD}

\begin{figure}[htbp]
  \begin{center}
    \begin{tabular}{cc}
     (a) & (b) \\
     \psfrag{n}{$i$}
     \psfrag{w}{$\mathrm{\Sigma}^{2}_i,w''$}
     \psfrag{x}{$x$}
     \psfrag{y}{$y$}
     \psfrag{u}{$u'$}
     \psfrag{v}{$v'$}
 \includegraphics[width=0.45\textwidth]{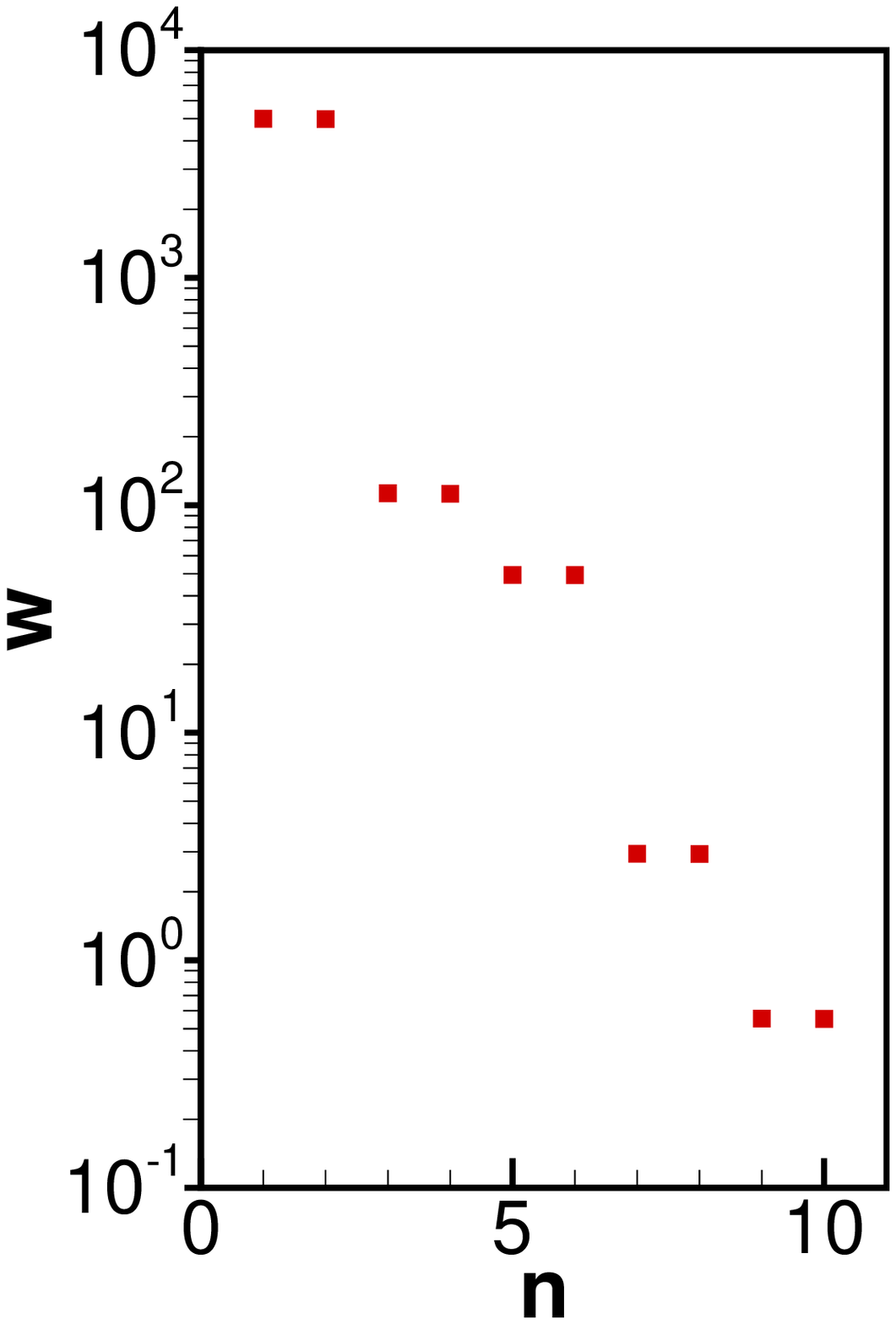} &     \psfrag{n}{$i$}
     \psfrag{w}{$\mathrm{\Gamma}^{2}_i,f(w'',w'')$}
\psfrag{x}{$x$}
     \psfrag{y}{$y$}
     \psfrag{u}{$u'$}
     \psfrag{v}{$v'$}
 \includegraphics[width=0.45\textwidth]{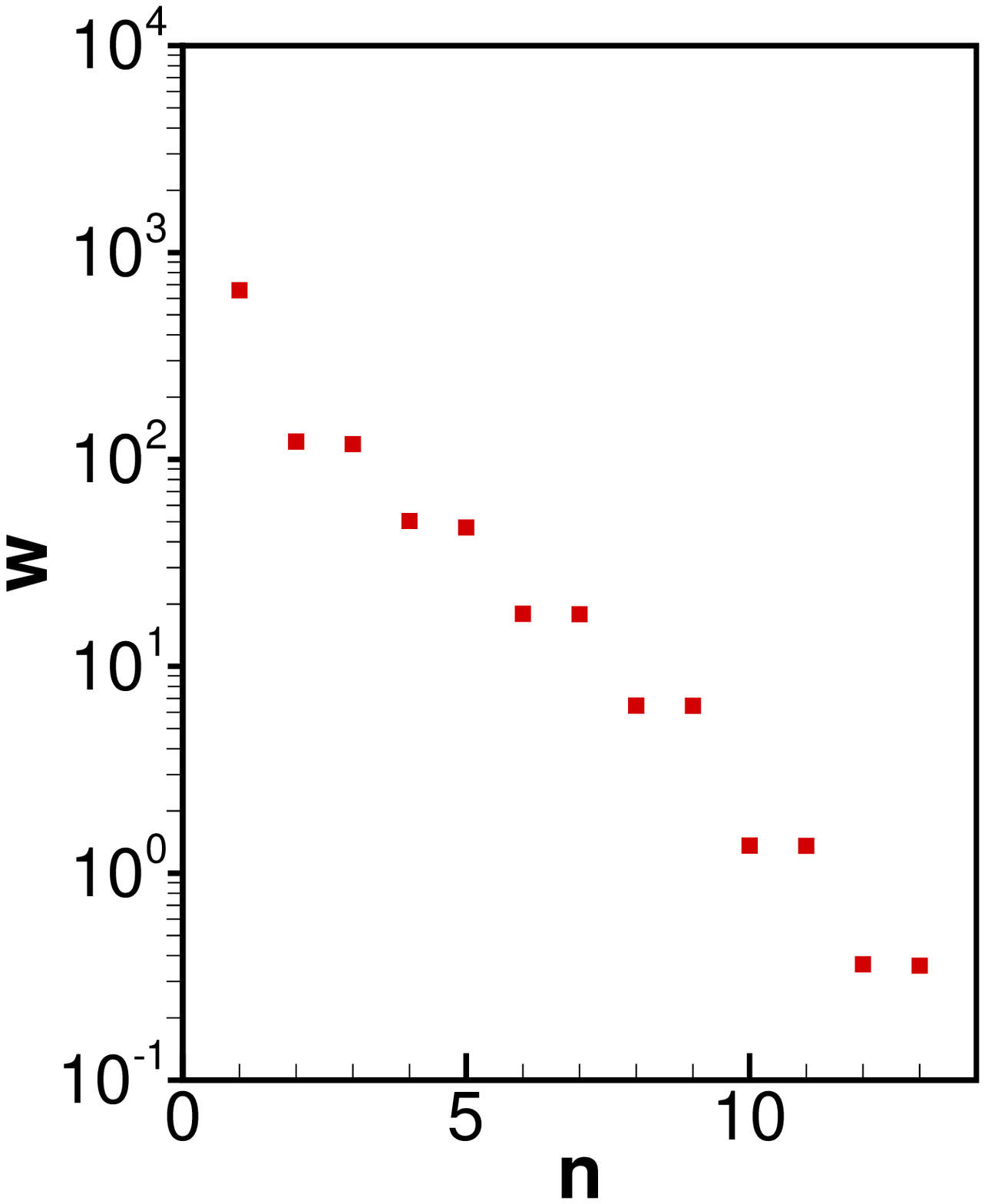} \\ (c) & (d) \\
           \psfrag{x}{$x$}
     \psfrag{y}{$y$}
     \psfrag{u}{$u''$}
     \psfrag{v}{$v''$}
      \includegraphics[width=0.5\textwidth]{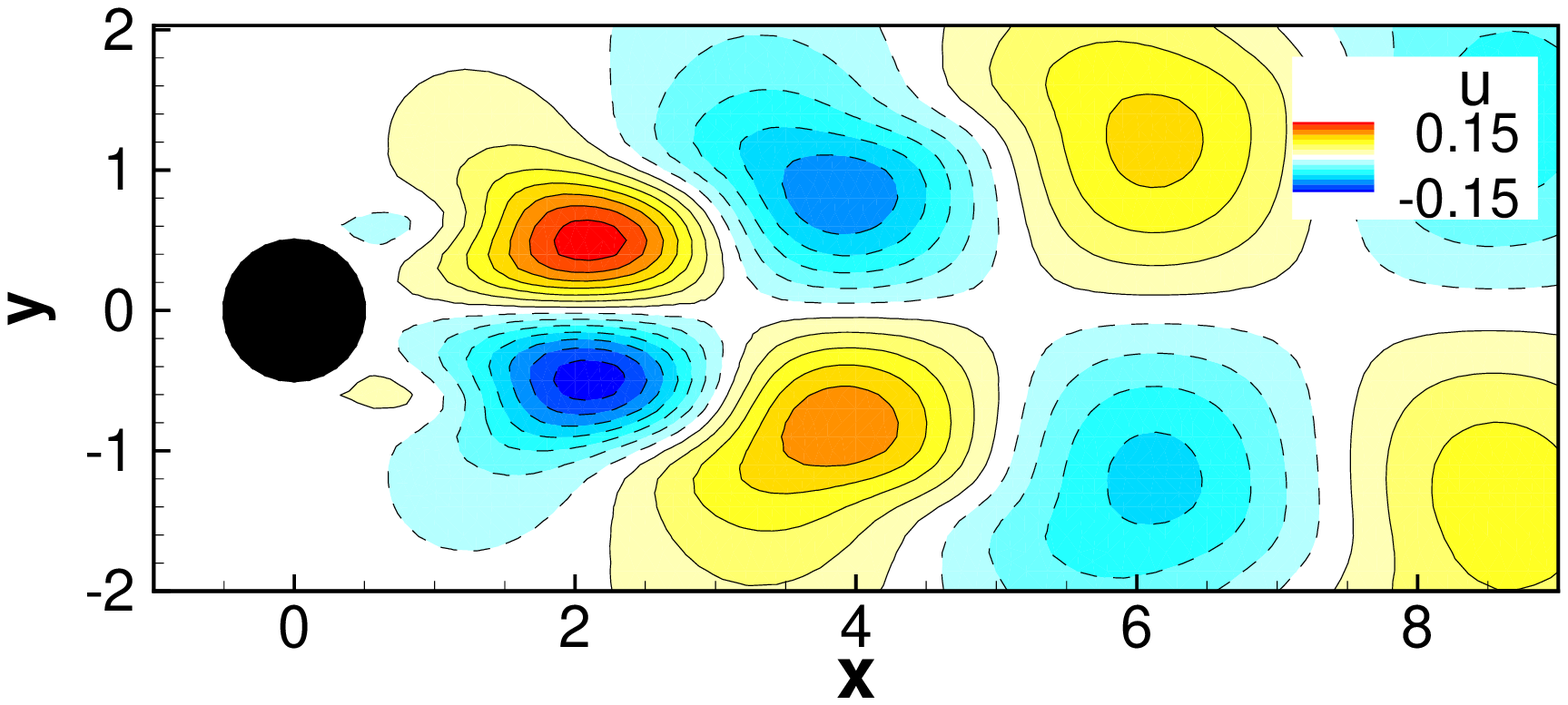} &
           \psfrag{x}{$x$}
     \psfrag{y}{$y$}
     \psfrag{u}{$u''$}
     \psfrag{v}{$v''$}
      \includegraphics[width=0.5\textwidth]{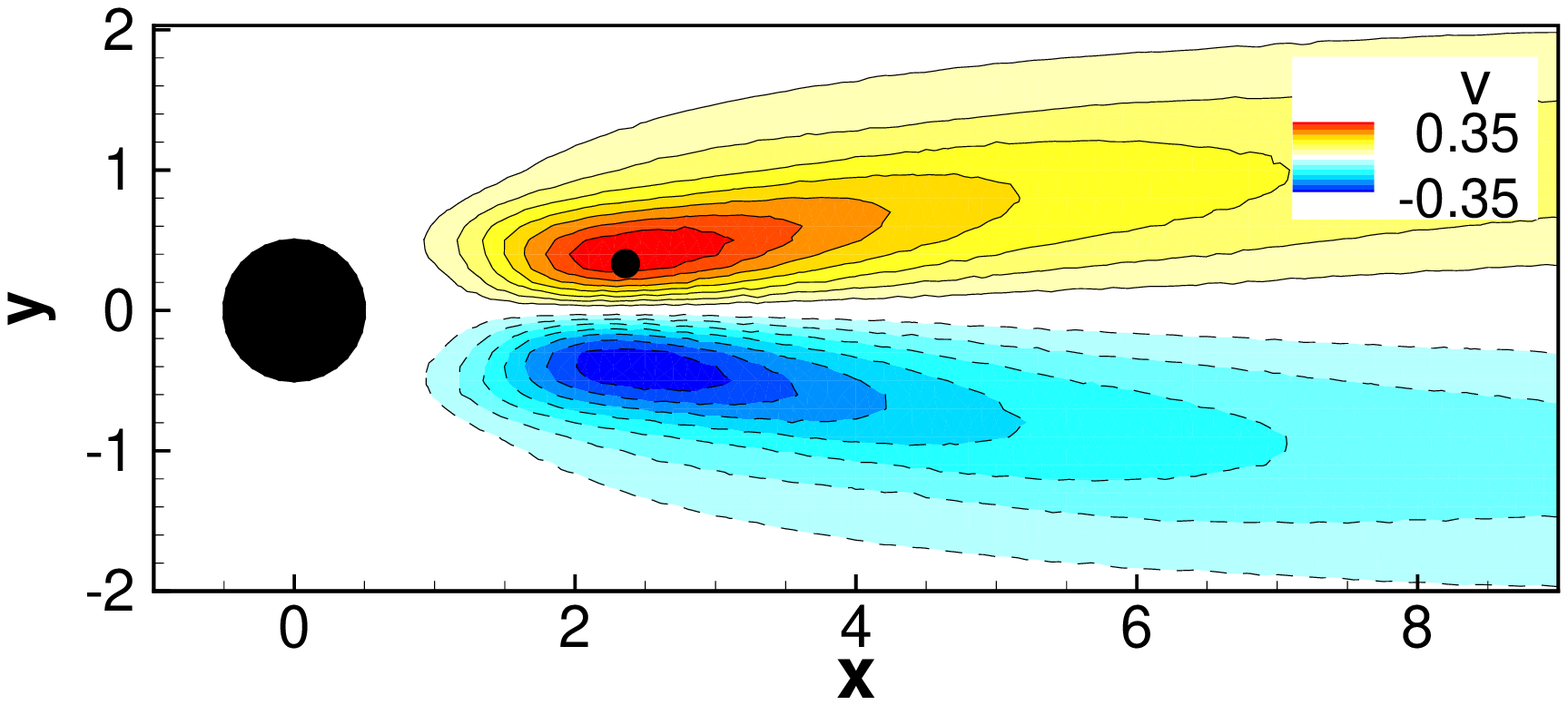} \\ (e) & (f) \\
           \psfrag{x}{$x$}
     \psfrag{y}{$y$}
     \psfrag{u}{$u''$}
     \psfrag{v}{$v''$}
      \includegraphics[width=0.5\textwidth]{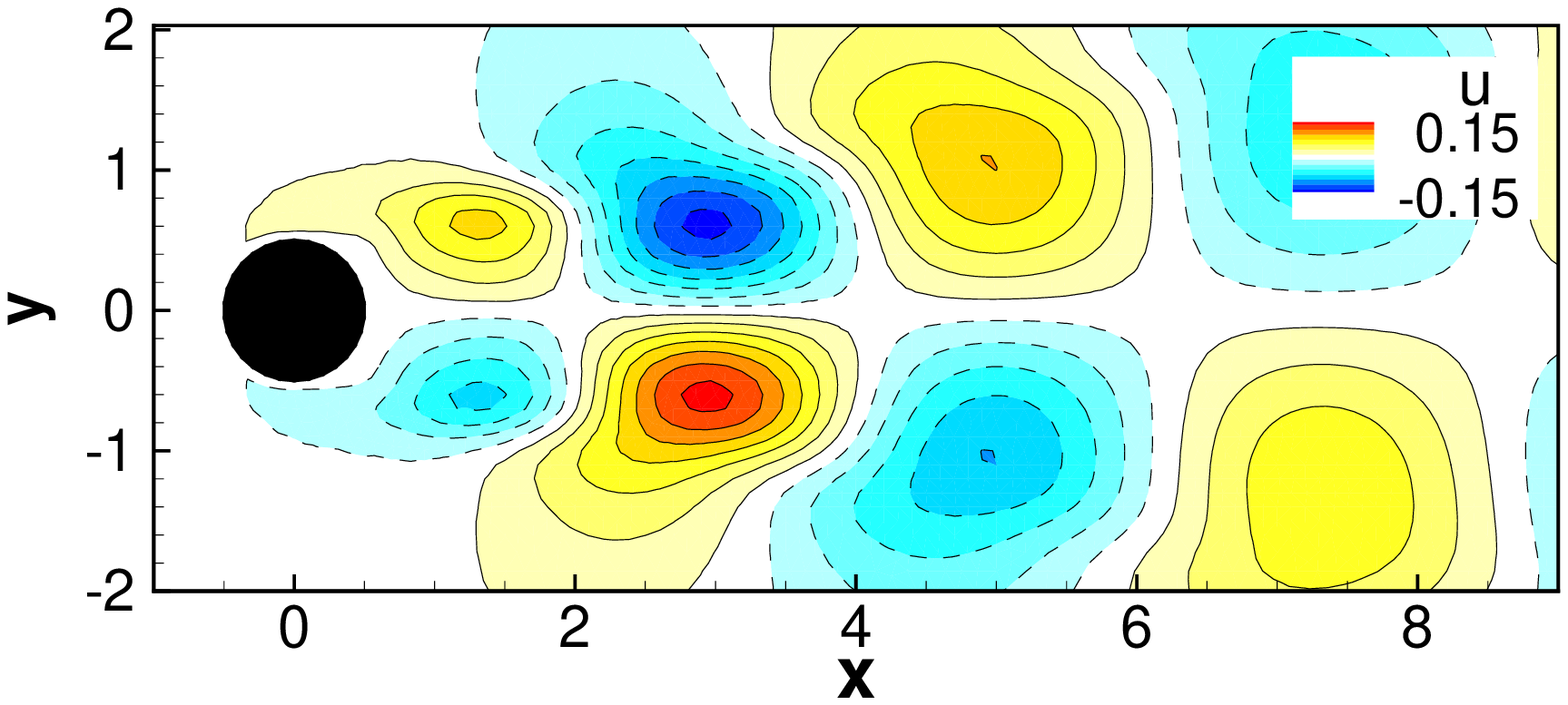} &
           \psfrag{x}{$x$}
     \psfrag{y}{$y$}
     \psfrag{u}{$u''$}
     \psfrag{v}{$v''$}
      \includegraphics[width=0.5\textwidth]{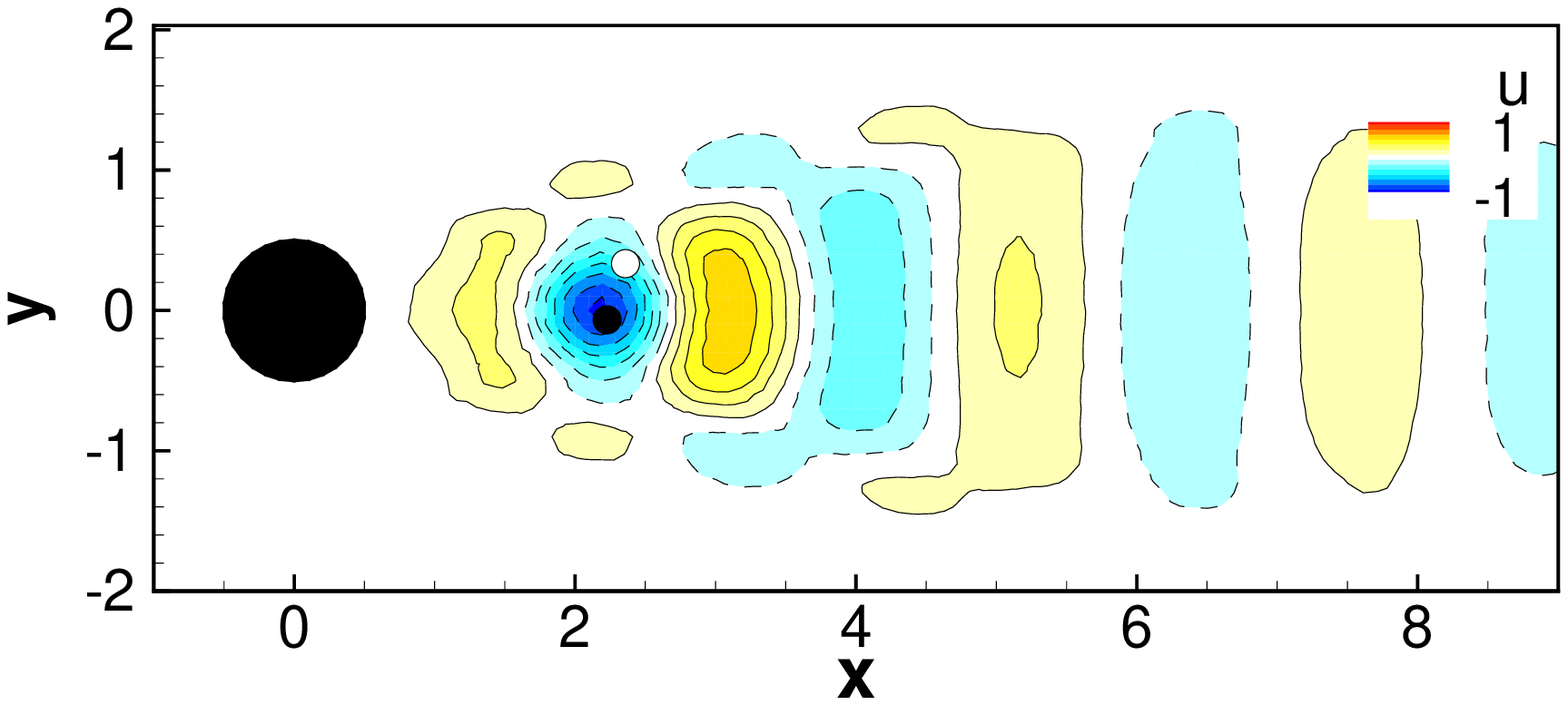} \\ (g) & (h) \\
           \psfrag{x}{$x$}
     \psfrag{y}{$y$}
     \psfrag{u}{$u''$}
     \psfrag{v}{$v''$}
      \includegraphics[width=0.5\textwidth]{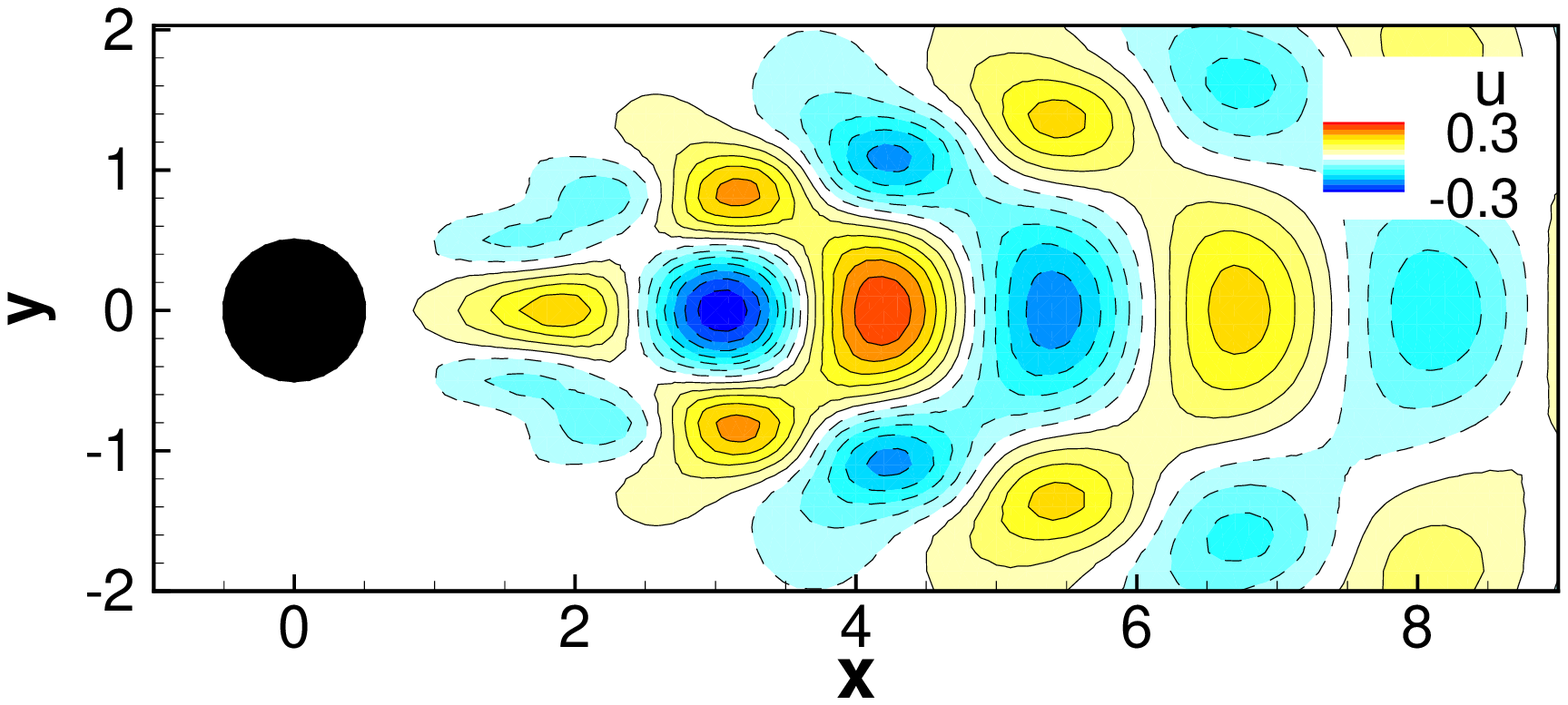} &
           \psfrag{x}{$x$}
     \psfrag{y}{$y$}
     \psfrag{u}{$u''$}
     \psfrag{v}{$v''$}
      \includegraphics[width=0.5\textwidth]{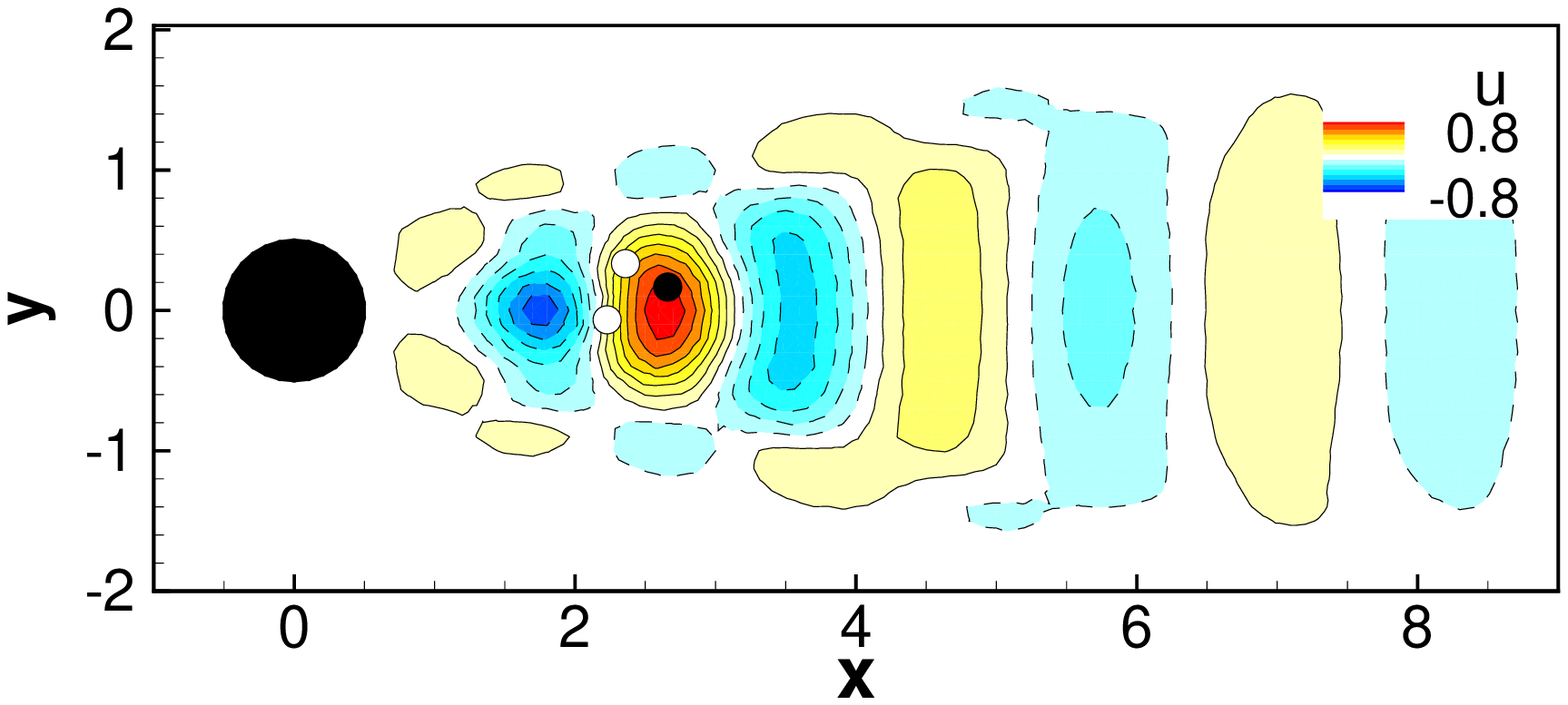}
    \end{tabular}
  \end{center}
 \caption{\label{fig:LCresults} Model reduction of cylinder flow at $Re=100$ with projection bases obtained from the LC trajectory with the MF formulation. Same caption as in fig. \ref{fig:TRresults}.}
\end{figure}

The snapshots for the POD analysis have been taken from the unsteady
simulation presented in figures~\ref{fig:evoEnergy}(a,b) in the time range
$150 \leq t \leq 300$ (not shown in the figure). The sampling time is equal to
$\Delta t_S = 0.2$ which results in $751$ snapshots. The eigenvalues of the correlation matrices in the case of the MF formulation (which is the most natural here because it does not require explicit knowledge of the base-flow as only limit-cycle snapshots are considered for the building of the $W$ and $F$ bases) are displayed in
figures~\ref{fig:LCresults}(a,b) for the sate $w''$ and the nonlinearity $f(w'',w'')$. The eigenvalues decay much faster than in the previous section (see figures~\ref{fig:TRresults}(a,b)), indicating
that the dimensionality of the space spanned by the limit-cycle
snapshots is markedly lower than the analogous dimensionality for the
combined transient and limit-cycle behavior. All eigenvalues appear as pairs; furthermore, the mean-flow
distortion linked to the growth of the unstable global modes has
disappeared, as this component is already accounted for by the variable $ w'' $, which is a perturbation around the
mean-flow $\overline{w}$.

The first three POD modes representing the state $w''$ are depicted in
figure~\ref{fig:LCresults}(c,e,g). We observe that the two first POD modes
form a pair representing the vortex shedding phenomenon (i.e., the
global mode). In contrast, the first three nonlinear POD modes are
shown in figure~\ref{fig:LCresults}(d,f,h), together with the
corresponding DEIM points (black symbol representing the most recent
DEIM point, and the white symbols indicating the earlier DEIM
points). Again, the DEIM points correspond to large values in
amplitude of the associated nonlinear POD mode. The first nonlinear POD-mode is the same as the third non-linear POD-mode of the previous section (see fig. \ref{fig:TRresults}(h)). It is a symmetric mode and is linked to the interaction of the two first state POD modes. The next two nonlinear POD modes capture a structure that is convected downstream (the eigenvalues appear in pairs). It is seen that the row selector matrix $P^H$ selects here streamwise velocity components. These modes represent symmetric forcings and may result from interactions between two symmetric or two antisymmetric state POD modes.

\subsection{Model reduction}
\label{sec:LC_ModRed}

As in section \S~\ref{sec:TR_ModRed}, the parameters of the various
models and their resulting error values are listed in a single table,
see tab.~\ref{tab:tab2b}.  An
analysis of the numbers allows us to infer that the overall conclusions of the Kuramoto-Sivanshinski equations with bases $W$ and $F$ obtained from snapshots on the limit-cycle also hold here:
\begin{itemize}
\item All considered model-reduction techniques 
  show excellent and nearly equivalent results for the LC trajectory as long as the number
  of POD modes is sufficiently high (about $p \sim q \sim 10$).
\item The energy-preserving criterion is most accurately satisfied by
  method 1 ($\epsilon_S < 1\%$), while methods 2 and 3
  exhibit significant values for $\epsilon_S$ ($40-50\%$ for method 2,
  $60-70\%$ for method3). Yet, despite these high values of $\epsilon_S,$
  these models are nonetheless reasonably accurate for the LC trajectory ($\epsilon_m <
  10\%$).
  \item All models fail with the TR and MFTR trajectory.
  \item The MF properties are well reproduced by all models.
  \item The base-flow properties are poorly recovered as in the analogous Kuramoto-Sivashinski case.
\end{itemize}
The excellent results of the DEIM technique on the LC trajectory agree with the results
given in~\cite{fosasde2016nonlinear}. This is in stark contrast to the results of the case discussed in the previous section where DEIM failed to reproduce the dynamics of the TR trajectory although both the transient and the limit-cycle were considered for the building of the bases $W$ and $F$. A distinctive feature is that the dynamical system evolves in that case from the  
base flow solution, where the linearized dynamics is characterized by an unstable
eigenvalue, to a limit cycle, where the mean flow dynamics 
displays marginal stability. This fact suggests that the linear
dynamics significantly changes along the trajectory of the system.
It should be noted that the POD-DEIM technique involves different
approximations for the linear and nonlinear terms and, in the case of
the Navier--Stokes equations, this split strongly depends on the reference
field that is considered. 
As mentioned in~\cite{chaturantabut_state_2012}, the presence of
linear dynamics in the nonlinear term can result in a poor 
approximation.
In the present case, there is no decomposition  
that results in a nonlinear term which does not contain any linear driving all along the trajectory of the system: in the case of the BF (resp. MF) formulation, the nonlinear term exhibits a linear driving on the limit-cycle (resp. close to the base-flow).
This fact
could in principle explain why the DEIM technique results in an 
accurate model only when considering flow on the limit cycle while failing in the transient
simulation.

\begin{table}[htbp]
  \centering
  \begin{tabular}{|l|c|cccc|cccc|ccc|ccc|cc|}
  \hline
  & & \multicolumn{4}{|c|}{BF} &  \multicolumn{4}{|c|}{MF} & \multicolumn{3}{|c|}{TR} & \multicolumn{3}{|c|}{LC} & \multicolumn{1}{|c|}{MFTR}\\
MF-$p$-$q$& $\epsilon_S$
& $\epsilon_{w_b}$ & $\nu_{\lambda}$ & $\epsilon_{\lambda}$ & $\epsilon_{\hat{w}}$
& $\epsilon_{\bar{w}}$ & $\nu_{\lambda}$ &$\epsilon_{\lambda}$ & $\epsilon_{\hat{w}}$ 
& $ \epsilon_t$ & $ \epsilon_m$ & $ \epsilon_m' $
& $ \epsilon_t$ & $ \epsilon_m$ & $ \epsilon_m'$
& $ \epsilon_m$
\\
\hline
1B-14& \grn{0.2} & $\grn{0.00}$ & \grn{2} & \ora{20} & \ora{50} & \grn{4} & \grn{2}&\grn{2}&\grn{8} & \grn{0.1}&\red{102}&\red{102} &  \grn{0.1} &\grn{1}&\grn{2} &\red{95}\\
1B-12& \grn{0.1} & $\grn{0.00}$ & \grn{2} & \ora{20} & \ora{50} & \grn{4} & \grn{2}&\grn{1}&\grn{8} & \grn{0.2}&\red{102}&\red{102} &  \grn{0.3} &\grn{1}&\grn{1} &\red{95}\\
1B-10& \grn{0.1} & $\grn{0.00}$ & \grn{2} & \ora{20} & \ora{50} & \grn{4} & \grn{2}&\grn{2}&\grn{8} & \grn{1}&\red{102}&\red{102} &  \grn{1} &\grn{2}&\grn{3} &\red{95}\\
1M-14& \grn{0.4} & $\red{100}$ & \grn{2} & \ora{35} & \ora{49} & \grn{0.4} & \grn{2}&\grn{2}&\grn{8} & \ora{30}&\red{100}&\red{100} &  \grn{0.1} &\grn{1}&\grn{1} &\red{100}\\
1M-12& \grn{0.3} & $\red{100}$ & \grn{2} & \ora{35} & \ora{49} & \grn{0.4} & \grn{2}&\grn{2}&\grn{8} & \ora{30}&\red{100}&\red{100} &  \grn{0.2} &\grn{2}&\grn{1} &\red{100}\\
1M-10& \grn{0.2} & $\red{100}$ & \grn{2} & \ora{35} & \ora{49} & \grn{0.4} & \grn{2}&\grn{2}&\grn{8} & \ora{30}&\red{100}&\red{100} &  \grn{0.5} &\grn{2}&\grn{2} &\red{100}\\
1M-8& \grn{0.2} & $\red{100}$ & \grn{2} & \ora{35} & \ora{49} & \grn{0.4} & \grn{2}&\grn{2}&\grn{8} & \ora{30}&\red{100}&\red{100} &  \grn{1} &\grn{2}&\grn{2} &\red{100}\\
1M-6& \grn{0.1} & $\red{100}$ & \grn{2} & \ora{35} & \ora{49} & \grn{0.4} & \grn{2}&\grn{2}&\grn{8} & \ora{30}&\red{100}&\red{100} &  \grn{3} &\grn{4}&\grn{4} &\red{100}\\
1M-4& \grn{0.7} & $\red{100}$ & \grn{2} & \ora{35} & \ora{49} & \grn{0.4} & \grn{2}&\grn{2}&\grn{8} & \ora{31}&\red{100}&\red{100} &  \grn{10} &\ora{37}&\ora{36} &\red{100}\\
\hline
2B-10-10& \ora{26} & $\grn{0}$ & \grn{2} & \ora{20} & \ora{49} & \grn{4} & \grn{2}&\grn{1}&\grn{8} & \grn{1}&\red{102}&\red{102} &  \grn{1} 
&\grn{2}&\ora{35} &\red{94}\\
2M-10-14& \ora{43} & $\red{100}$ & \grn{2} & \ora{35} & \ora{50} & \grn{0.4} & \grn{2}&\grn{2}&\grn{8} & \ora{30}&\red{100}&\red{100} &  \grn{0.5} &\grn{2}&\ora{25} &\red{100}\\
2M-10-10& \ora{45} & $\red{100}$ & \grn{2} & \ora{35} & \ora{50} & \grn{0.4} & \grn{2}&\grn{2}&\grn{8} & \ora{30}&\red{100}&\red{100} &  \grn{0.5} &\grn{2}&\ora{25} &\red{100}\\
2M-10-6& \ora{56} & $\red{100}$ & \grn{2} & \ora{35} & \ora{50} & \grn{0.4} & \grn{2}&\grn{2}&\grn{8} & \ora{30}&\red{100}&\red{100} &  \grn{0.5} &\grn{7}&\ora{26} &\red{100}\\
2M-8-8& \ora{51} & $\red{100}$ & \grn{2} & \ora{35} & \ora{50} & \grn{0.4} & \grn{2}&\grn{2}&\grn{8} & \ora{30}&\red{100}&\red{100} &  \grn{1} &\grn{2}&\ora{25} &\red{100}\\
2M-4-4& \ora{61} & $\red{100}$ & \grn{2} & \ora{35} & \ora{50} & \grn{0.4} & \grn{2}&\grn{2}&\grn{8} & \ora{31}&\red{100}&\red{100} &  \grn{10} &\grn{9}&\ora{25} &\red{100}\\
\hline
3B-10-10& \red{60} & $\grn{0}$ & \grn{2} & \ora{20} & \ora{49} & \ora{36} & \grn{2}&\grn{3}&\grn{8} & \grn{1}&\red{102}&\red{102} &  \grn{1} 
&\ora{24}&\red{68} &\red{95}\\
3M-10-14& \red{67} & $\red{100}$ & \grn{2} & \ora{35} & \ora{49} & \grn{0.4} & \grn{2}&\grn{2}&\grn{8} & \ora{30}&\red{100}&\red{100} &  \grn{0.5} &\grn{2}&\ora{26} &\red{100}\\
3M-10-10& \red{62} & $\red{100}$ & \grn{2} & \ora{35} & \ora{49} & \grn{0.4} & \grn{2}&\grn{2}&\grn{8} & \ora{30}&\red{100}&\red{100} &  \grn{0.5} &\grn{2}&\ora{25} &\red{100}\\
3M-10-6& \red{61} & $\red{100}$ & \grn{2} & \ora{35} & \ora{49} & \grn{0.4} & \grn{2}&\grn{2}&\grn{8} & \ora{30}&\red{100}&\red{100} &  \grn{0.5} &\grn{8}&\ora{27} &\red{100}\\
3M-8-8& \red{67} & $\red{100}$ & \grn{2} & \ora{35} & \ora{49} & \grn{0.4} & \grn{2}&\grn{2}&\grn{8} & \ora{30}&\red{100}&\red{100} &  \grn{1} &\grn{2}&\ora{25} &\red{100}\\
3M-4-4& \red{64} & $\red{100}$ & \grn{2} & \ora{35} & \ora{49} & \grn{0.4} & \grn{2}&\grn{2}&\grn{8} & \ora{30}&\red{100}&\red{100} &  \grn{10} &\grn{8}&\ora{26} &\red{100}\\
\hline
\end{tabular}
  \caption{\label{tab:tab2b} Error analysis of various reduced-order
    modelling techniques for cylinder flow at $Re=100$ with projection bases $W$ and $F$ based on
    snapshots taken from the LC trajectory. Same caption as in tab. \ref{tab:tab1}.}
\end{table}

\section{Conclusions \label{sec:Concl}}

Three different model reduction techniques have been applied to a
model equation and to the Navier-Stokes equations, and the results
have been compared: (i) Galerkin projection onto a POD basis
representing the state variable, (ii) Galerkin projection onto two
distinct POD bases (one basis for the state variable, a second basis
for the nonlinear convection term) and (iii) the Discrete Empirical
Interpolation Method (DEIM) for the evaluation of the projected
nonlinear terms. The two test cases included: a non-parallel version
of the Kuramoto-Sivashinski equation and flow past a 
cylinder at a supercritical Reynolds number of $Re=100$.
For each case, we considered three trajectories:
the base-flow transient (TR) initialized by a solution close to the base-flow, the limit-cycle solution (LC) and a mean-flow transient (MFTR) initialized from the mean-flow solution. 
The
first test case is a one-dimensional model problem that displays all
characteristics of the more complex flow past a
cylinder: the nonlinearity is of the same nature (convection), the
solution exhibits advection, diffusion and instability, and the system
undergoes a Hopf bifurcation, after which oscillatory behavior
prevails. We show that (single base) Galerkin projections provide excellent results
 (if the number of POD modes is sufficiently large), with a reduced nonlinear term that very well preserves
energy. The use of a second basis to represent the nonlinearities with
Galerkin projection also provides good results. Yet, the
energy-preserving criterion is significantly altered. Finally, the
DEIM model-reduction technique is successful for the case of the
transient Kuramoto-Sivashinski solution and in the case of flow past a
cylinder in the limit-cycle regime. However, it fails for the case of
flow past a cylinder in the transient regime, producing reduced-order
models that exhibit a finite-time singularity. Tentative explanations for this behavior have been provided.  Further investigations are necessary to come to more
definitive and predictive conclusions about the accuracy, stability
and robustness of projection- and interpolation-based model-reduction
techniques.




\bibliographystyle{elsarticle-num}
\bibliography{biblio}


\end{document}